%% file: main.tex
\documentclass[fleqn,usenatbib]{mnras}

\usepackage{newtxtext,newtxmath}

\usepackage[T1]{fontenc}

\DeclareRobustCommand{\VAN}[3]{#2}
\let\VANthebibliography\thebibliography
\def\thebibliography{\DeclareRobustCommand{\VAN}[3]{##3}\VANthebibliography}

\usepackage{graphicx}	
\usepackage{amsmath}	
\usepackage{orcidlink}  
\usepackage{multicol}  
\usepackage{multirow}  
\usepackage{color,soul} 
\usepackage{booktabs} 

\usepackage[version=4]{mhchem} 
\usepackage{adjustbox}

\defcitealias{ShenTurner08}{ST08}
\defcitealias{kipping2013}{K13}
\newcommand*{\rom}[1]{\expandafter\@slowromancap\romannumeral #1@}

\title[RV eccentricity priors]{RV-exoplanet eccentricities: Good, Beta, and Best}

\author[A. T. Stevenson et al.]{
A. T. Stevenson$^{1}$\thanks{E-mail: adam.stevenson@open.ac.uk}\orcidlink{0000-0003-2399-7619},
C. A. Haswell$^{1}$\orcidlink{0000-0002-8050-1897},
J. P. Faria$^{2}$\orcidlink{0000-0002-6728-244X},
J. R. Barnes$^{1}$\orcidlink{0000-0001-6105-2902},
J. K. Barstow$^{1}$\orcidlink{0000-0003-3726-5419},
H. Dickinson$^{1}$\orcidlink{0000-0003-0475-008X},
\newauthor{~M. R. Standing$^{3}$}\orcidlink{https://orcid.org/0000-0002-7608-8905}
\\
$^{1}$School of Physical Sciences, The Open University, Milton Keynes MK7 6AA, UK \\ 
$^{2}$Observatoire Astronomique de l’Université de Genève, Chemin Pegasi 51, CH-1290 Versoix, Switzerland \\
$^{3}$European Space Agency (ESA), European Space Astronomy Centre (ESAC), Camino Bajo del Castillo s/n, E-28692 Villanueva de la Ca\~nada, Madrid, Spain
}
\date{Accepted 2025 March 25. Received 2025 March 7; in original form 2024 September 20}

\pubyear{\the\year{}}

\begin{document}
\label{firstpage}
\pagerange{\pageref{firstpage}--\pageref{lastpage}}
\maketitle

\begin{abstract} 
We examine the eccentricity distribution(s) of radial velocity detected exoplanets. Previously, the eccentricity distribution was found to be described well by a Beta distribution with shape parameters $a=0.867, b=3.03$. Increasing the sample size by a factor of 2.25, we find that the CDF regression method now prefers a mixture model of Rayleigh + Exponential distributions over the Beta distribution, with an increase in Bayesian evidence of $\Delta\ln{Z}\sim 77$ ($12.6\,\sigma$). Using PDF regression, the eccentricity distribution is best described by a Gamma distribution, with a Rayleigh + Exponential mixture a close second. The mixture model parameters, $\alpha = 0.68\pm0.05, \lambda=3.32\pm0.25, \sigma =0.11\pm0.01$, are consistent between methods. We corroborate findings that exoplanet eccentricities are drawn from independent parent distributions when splitting the sample by period, mass, and multiplicity. Systems with a known outer massive companion provide no positive evidence for an eccentricity distribution distinct from those without. We quantitatively show M-dwarf hosted planets share a common eccentricity distribution with those orbiting FGK-type stars. We release our python code, \textsc{eccentriciPy}, which allows bespoke tailoring of the input archive to create more relevant priors for particular problems in RV planet discovery and characterisation. We re-characterised example planets using either traditional Beta, or updated priors, finding differences for recovery of low-amplitude multi-signal systems. We explore the effects of a variety of prior choices. The accurate determination of small but non-zero eccentricity values has wide-ranging implications for modelling the structure and evolution of planets and their atmospheres due to the energy dissipated by tidal flexing.
\end{abstract}
%



\section{Introduction}
 
Recent advancements in both observational hardware and analysis methods used for planet-hunting have helped spur on the steadily increasing rate of exoplanetary detections, which now number in the several thousands. The increasing number of planets facilitates more powerful statistical inferences from their demographics, relating observed trends to implications for formation and evolution \citep[e.g. reviews from][]{WF2015,WeiSubo2021}. This helps place our own Solar System's history in context of a wider Galactic population. 

The orbital eccentricity is one of the most important parameters to measure for exoplanets. Eccentricities have formation and evolutionary processes imprinted upon them, and determine the rate of heating due to tidal flexing. The Solar System planets exhibit nearly circular orbits (0.007 -- 0.206 for Mercury) whereas radial velocity (RV) detected exoplanets are usually found on far more elliptical orbits. \citet{BMJ2021} suggested that the Solar System planets may belong to the tail of a continuous distribution, and the eccentricities are not uncommonly low, but instead that the number of planets is uncommonly high. 

Eccentricities can also play a significant role in shaping the potential habitability of exoplanets, and influencing atmospheric compositions. As the stellar insolation will vary over the course of an elliptical orbit, such a planet may be unable to retain surface water over the course of its year \citep[e.g.][and references therein]{Bolmont2016}, or may freeze completely \citep{Dressing2010}. The water-loss rate will be increased (by up to a factor of $\sim3$) leading to stronger \ce{H2O} and \ce{CO2} spectral absorption features, provided the water-world can retain an ocean over long time-scales \citep{Liu2023}. A study by \citet{Dressing2010} found that larger eccentricity could, in some cases, increase the semi-major axis range where orbits could be considered `habitable'.

Finite eccentricity of an exoplanet will change which species we expect to observe in the atmosphere, emphasizing the need to accurately constrain its value. Even small eccentricities can be enough to tidally induce surface volcanism, demonstrated in our Solar System on the Jovian moon Io ($e=0.004$). This can lead to the production of molecules bearing elements such as sulfur in higher quantities than would be otherwise expected \citep{Seligman2024,Banerjee2024} -- though these could also be initiated by photochemical forcing on a highly eccentric orbit \citep{Tsai2023}. 

\citet{Seligman2024} suggested L98-59\,d as a promising candidate for observing an eccentric orbit driving Io-like volcanism, creating a \ce{SO2} atmosphere. This exoplanet orbits close to a 2:4:7 resonance with eccentricity of $e=0.074$. Signs of sulfur species have recently been identified \citep{Banerjee2024,Gressier2024}, corroborating the link between a slightly eccentric orbit and observable features in the atmospheric composition. Tidal heating is therefore an important process to account for when considering atmospheric chemistry and interior structure inferences \citep{Welbanks2024}, even when eccentricity is small. This will affect most super-Earth to Saturn-mass planets -- such as the archetypal example of WASP-107\,b \citep{Sing2024,Welbanks2024}, with eccentricity of $e=0.06\pm0.04$ \citep{Piaulet2021}.

Dynamical effects such as tides, or planet-planet scattering will modulate eccentricities \citep{Rasio1996,RasioFord1996,Juric2008,Chatterjee2008,Carrera2019,2024AJ....168..115B}, helping us diagnose chaotic histories of exoplanetary systems. To ascertain eccentricities during planet detection and characterisation via Bayesian inference, where the data often are not entirely informative due to the noise and sampling, we require prior estimates for fitting models to the observational data. Previously, \citet[][henceforth \citetalias{kipping2013}]{kipping2013} found that the observed RV-planet eccentricities could be well-modelled by a Beta distribution, providing an informative prior modulating our expectations for newly-discovered systems. Before this, observers had typically adopted uninformative uniform priors on eccentricity over $0\leq e<1$ \citepalias{kipping2013}. Users of modern parameter sampling algorithms advocate for use of a better justified prior that favours lower $e$, but can still explore high $e$ if the data allows \citep{Standing2022}.

The Beta distribution has been extensively used as a prior for parameter space exploration, or been considered as a functional form for others studying the eccentricity distribution of exoplanets \citep[e.g.][and so on...]{VEA2015,FM+2016,Delisle2018,VE19,Bowler2020}, be it for transit, RV, or direct imaging studies. Works that use Beta for a prior generally do so with the derived parameters from \citetalias{kipping2013}, i.e. Beta\,($0.867,3.03$). Many studies using the \textsc{kima} RV analysis framework for exoplanet detection \citep{kima-joss,kima-ascl} use the more flexible Kumaraswamy distribution \citep{kumaraswamy} to provide a close match to \citetalias{kipping2013}'s Beta \citep[such as][]{Faria2020,Faria2022,Standing2022,AnnaJohn2023,Baycroft2023b,Barnes2023}.

In this paper we investigate how the RV discoveries of the past $\sim11$~years affect the eccentricity distribution of RV-discovered exoplanets. Whilst most exoplanets have been discovered with the transit method \citep[primarily from \textit{Kepler} and \textit{TESS},][]{Kepler,TESS}, the number of RV-planets has still more than doubled since \citetalias{kipping2013}. For transiting planets, recent works have calculated eccentricities from \textit{Kepler} light curves, where accurate measures of transit duration, impact parameter, and stellar density are available \citep{VEA2015,VE19,Sagear2023,GilbertPetiguraESS}. However, RVs are lagging behind. There is no uniform study of a population of RV planets, analogous to \textit{Kepler}, and eccentricity is often hard to determine from RVs if poor sampling causes gaps in the phase-folded orbit.

This work is a first step towards updating the eccentricity distribution of RV exoplanets. Our Bayesian analysis method closely follows that of \citetalias{kipping2013}, to see if and how an increased sample size changes the preferred eccentricity parameterisation. In Section~\ref{sec:sample} we discuss the sample we used, and any modifications we have made to create this. Section~\ref{sec:methods} describes the methods for regression, either to the empirical distribution function (EDF) as in \citetalias{kipping2013}, or directly to the probability density function (PDF). Section~\ref{sec:functions} introduces and describes the trial distributions, where we use those from \citetalias{kipping2013} and other works based on observed qualities of the eccentricity population. In Section~\ref{sec:EDF} we outline our results for the EDF method: first for function regression to the global sample; then in local cases as a test of similarity between distributions, where the sample is split based on characteristics such as orbital period, number of planets in the system, the presence of massive outer companions, or the type of star hosting each planet. Section~\ref{sec:PDF} describes results for an alternative method, where the eccentricity values directly inform distribution PDF regression. Section~\ref{sec:discussion} discusses the results; compares the two trialled methods; establishes the need for generating bespoke eccentricity priors on a case-by-case basis; tests how parameter inference is altered by changing the prior used; and considers some of the biases that may be present in an RV planet sample. We conclude our findings in Section~\ref{sec:conclusion}.

\section{RV exoplanet sample}\label{sec:sample}

There are now $\sim1100$ RV-detected exoplanets, of which $917$ could be used for eccentricity analysis after performing the same cuts as in \citetalias{kipping2013}. The original catalogue used in 2013 is no longer updated, so we have elected to use the comprehensive and up to date \href{https://exoplanetarchive.ipac.caltech.edu/}{NASA Exoplanet Archive} \citep{NEA}. Samples are derived from catalogue version downloaded on 12 June 2024, and we initially use the `default' parameter set for each planet. We ensure that planets have valid eccentricities, periods, semi-amplitudes, and impose that $K/\sigma_{K}>5$ to select good signal-to-noise detections. With $891$ exoplanets (see Section~\ref{subsec:fixed-ecc} below), our sample is now $2.25$ times larger than the original sample of $396$. The increased sample size now makes a comparison of different populations viable

\subsection{Eccentricity Errors}\label{sub:e_errors}
The eccentricities being used are simply either the maximum likelihood ``best'' estimates or posterior median values obtained from the NASA exoplanet archive.
The choice of maximum likelihood or posterior median may not always be consistent, depending on how discovery papers report the parameters for each planet. It is outside the scope of this work to assess all $>900$ planets or check their archival values, though it would be beneficial if in future authors could release posterior distributions -- and if these could be readily available on an archive to avoid this problem.

Using eccentricities in the manner we do in this work \citepalias[as was also done in][]{kipping2013} does not incorporate the uncertainty in each value (see Sections~\ref{subsec:edf} and~\ref{subsec:pdf}), which can be quite large for RV studies with poorly constrained orbits. 
Many of these eccentricity estimates may also not be significantly different from zero. \citet{LucySweeney1971} stated that to be $95$\,per~cent confident that an eccentricity is non-zero, it must be more than $2.45\sigma$ away from the hard boundary at $e=0$. A fraction of our sample will be consistent with a circular orbit interpretation, yet provide a non-zero eccentricity value for either EDF or PDF methods (discussed in Sections~\ref{subsec:edf} and~\ref{subsec:pdf}, respectively). This often applies to planets with small but non-zero eccentricities, such as WASP-107\,b mentioned earlier in this paper ($e=0.06\pm0.04)$. This is not ideal, though there is no way to account for this without a complete re-analysis of all currently known RV exoplanets.

The ideal approach would be to avoid using summary statistics \citep{KippingWang2024}, and instead use hierarchical Bayesian models (HBMs) to incorporate eccentricity errors \citep*{Hogg2010}. The HBM method has been employed in multiple works analysing eccentricity demographics recently \citep[e.g.][]{VE19,Sagear2023,Nagpal2023}, and requires the eccentricity posterior for each planet in the data-set. HBMs have not been applied to RV datasets yet, with recent literature focusing on either photometric or direct imaging data.

By simulating RVs for sets of either 30 or 300 example exoplanets, \citet{Hogg2010} found that the inferred eccentricity distribution created with a hierarchical probabilistic method using each planet's posterior as input describes the underlying distribution more effectively than a naive histogram of maximum likelihood eccentricity ($\Tilde{e}_{\rm n}$) estimates. They also showed that the mean $\Tilde{e}_{\rm n}$ overestimated the true mean of the inferred distribution, and that this discrepancy was not removed by increasing the number of simulated planets in their sample.

The HBM method successfully describes the underlying distribution for either 30 or 300 planets, displaying no significant improvement when sample size is increased.
However, the same cannot be said for the $\Tilde{e}_{\rm n}$ histogram: it becomes closer in shape to the imposed ground truth distribution when the number of simulated planets is increased from 30 to 300 (see figures~2 and~3 of \citealt{Hogg2010}). The mean is overestimated, and probability density incorrectly approximated, due to a large number of detections at $\Tilde{e}_{\rm n}\sim1$. \citet{Hogg2010} simulated planet periods drawn from a distribution $p(P)\propto 1/P$, between 2 and 2000\,d. The 30 RV measurements for each planet were simulated at uniform intervals over a baseline of 1000\,d. It is possible that some of the synthetic planet orbits were not sampled well, and the fitting routines have recovered the wrong parameters (or alternatively, the maximum likelihood solution overestimates eccentricity as the orbit is so poorly constrained). Orbit-fitting pipelines can produce a pile up at $e\sim0.9$ when periods are longer than the observation baseline, and the Keplerian fit fails \citep{ShenTurner08}. For real RV data, one would be wary of such a solution, and require additional observations before claiming to have discovered a new planet. 
Without these $\Tilde{e}_{\rm n}\sim1$ detections, the histogram and inferred distribution would look far more alike. We see no such excess of high-eccentricity orbits in our sample histogram (Figure~\ref{fig:ecc-hist}). Additionally, one can see that $\Tilde{e}_{\rm n}$ is overestimated for the simulated planets in \citet{Hogg2010} because there is an under-representation of very small eccentricities ($e\sim0$) in the recovered distribution, when compared to the true distribution.

As our sample data comes from an archive, we do not have access to the eccentricity posteriors, and creating these for $>900$ planets is beyond the scope of this work. HBM analysis would be a possible extension to the present study, but we choose to proceed with the quoted archival values here to explicitly test how increasing the size of the RV planet sample affects the results of \citetalias{kipping2013}. Moreover, in \citet{VE19}, the authors compared the posteriors of the HBM method and a histogram of best-estimate eccentricities, finding the probability distributions were consistent between the full treatment and a more simplistic analysis.

\subsection{Fixed eccentricities}\label{subsec:fixed-ecc}

In RV analyses, single or multiple Keplerian models are fit to the RV data. Where eccentricity ($e$) is significantly non-zero and constrained by the data, a value is determined for $e$. Often, for small $e$, the data are insufficient to constrain the value meaningfully and a circular orbit is fit with $e$ fixed to zero.
To use any methods involving a probability density function for parameterising the eccentricity distribution, we may need to make a minor modification to the sample. When using the PDF directly, some distributions may not be well defined at $e=0$ for a subset of parameter values. 
Another issue arises with the individual planets orbital parameters: generally when the archive reports $e=0$, this has been fixed in the analysis process, usually to make the problem simpler if the eccentricity is not significantly different from zero. 
Planets with small but non-zero eccentricity are particularly important so it is imperative that we are clear about how we treat these cases.

We considered simply disregarding any planets where $e=0$. This would reduce the sample from 917 to 867 exoplanets. Blindly removing these $50$ planets would bias the eccentricity distribution upwards, away from low values. The eccentricity being fixed to zero generally implies the true value is too small to be determined by the extant data. On the other hand, if we leave the $e=0$ fixed values in the sample, there may be problems with the fit, as well as potentially biasing the sample towards completely circular values. We therefore addressed this for as many of the $50$ planets with eccentricity fixed to zero, by inspecting the entries of the \href{https://exoplanetarchive.ipac.caltech.edu/}{NASA Exoplanet Archive}. As a compromise, we have searched the archive for other entries besides the default solution that may have a measured eccentricity value, with estimated uncertainty. Provided the remainder of the parameters are consistent (confirming we are dealing with the same planet), we accept the most recent value with highest precision. The 24 planets assigned a non-zero eccentricity in this manner are listed in Table~\ref{tab:updated_eccs}. This allows us to use the maximum possible number of \textit{measured} eccentricities, excluding those lacking error estimates \citep[cf.][]{Weldon2024}.

\begin{table}
    \centering
    \begin{tabular}{lcc}
    \hline
    Planet     & Non-default $e$ value & Reference  \\
    \hline \\
    55 Cnc b    & 0.0048 & \citet{Rosenthal2021} \\
    CoRoT-7 c & 0.026 & \citet{Faria2016} \\
    GJ 581 b & 0.022 & \citet{Trifonov2018} \\
    GJ 581 c & 0.0870 & \citet{Trifonov2018} \\
    GJ 581 e & 0.125 & \citet{Trifonov2018} \\
    HD 102195 b & 0.167 & \citet{Paredes2021} \\
    HD 109749 b &  0.01 & \citet{Fischer2006} \\
    HD 116029 b & 0.054 & \citet{Johnson2011} \\
    HD 136352 b & 0.079  & \citet{Kane2020}  \\
    HD 136352 c & 0.037  & \citet{Kane2020}  \\
    HD 136352 d & 0.075  & \citet{Kane2020}  \\
    HD 142245 b & 0.09 & \citet{Johnson2011} \\
    HD 149026 b & 0.051 & \citet{Ment2018} \\
    HD 156668 b & 0.235 & \citet{Rosenthal2021} \\
    HD 189733 b & 0.027 & \citet{Rosenthal2021} \\
    HD 20794 b &  0.27  & \citet{Feng2017} \\
    HD 20794 c &  0.17  & \citet{Feng2017} \\
    HD 20794 d &  0.25  & \citet{Feng2017} \\
    HD 209458 b & 0.01 & \citet{Rosenthal2021} \\
    HD 219134 b & 0.0630 & \citet{Rosenthal2021} \\
    HD 2638 b  & 0.187 & \citet{Paredes2021} \\
    HD 285968 b & 0.16 & \citet{Rosenthal2021} \\
    HD 50499 c & 0.241 & \citet{Rosenthal2021} \\
    YZ Cet c & 0.04 & \citet{Astudillo-Defru2017} \\
        \hline \\
    \end{tabular}
    \caption{A list of all exoplanets in the sample from Section~\ref{sec:sample} where the ``default'' archival solution has $e=0$ (\textit{fixed}), but other solutions provide an eccentricity estimate. The most recent value has been used, for all planets with consistent solutions, in an attempt to populate the very-low-$e$ section of parameter space. References are provided for each eccentricity.}
    \label{tab:updated_eccs}
\end{table}

\begin{figure}
    \centering
    \includegraphics[width=0.95\linewidth]{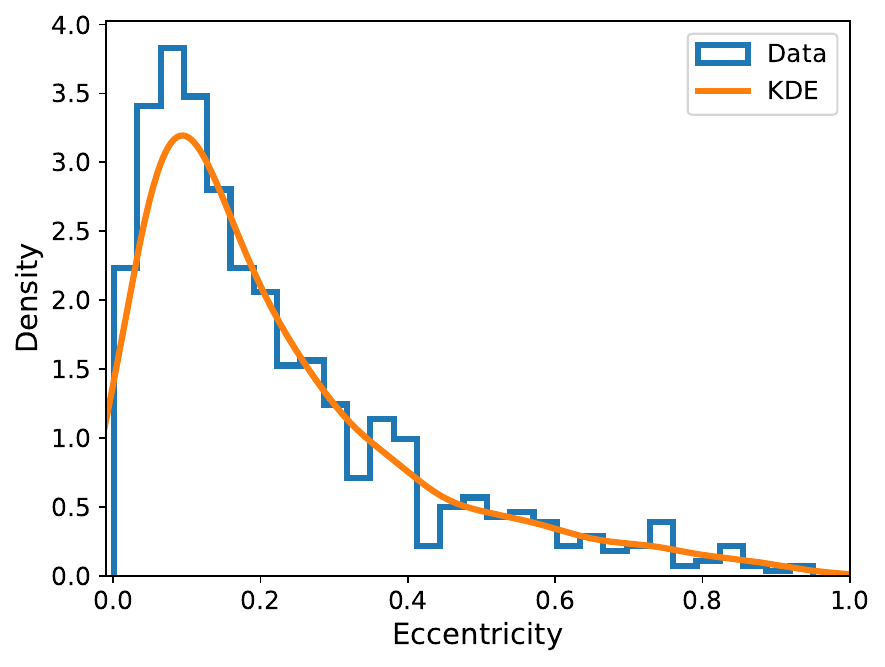}
    \caption{A histogram plot showing eccentricities of the RV exoplanet sample. The number of eccentricity bins is set equal to the rounded value of $\sqrt{N}$, with linear spacing between (0,1). The KDE of the PDF is over-plotted, generated using Gaussian kernels and automatic bandwidth determination \citep{Scott1992}.}
    \label{fig:ecc-hist}
\end{figure}

The resulting sample contains 891 non-zero eccentricities (Fig.~\ref{fig:ecc-hist}), meaning we only lose a small number (26) that are fixed at zero and cannot be estimated without performing a re-analysis of the radial velocities. These planets will likely have essentially undetermined non-zero eccentricities due to insufficient RV data. They therefore would not add much weight in a full treatment, so can be discarded with minimal impact to the derived results. The shape of the distribution is very similar in either case, but with a slight increase at $e=0$ if these planets were allowed to remain. The peak of the eccentricity histogram is located at $e>0$ in both scenarios, whether these 26 planets are included or are removed. Our sample of 891 planets includes 86 planets with $e<0.1$, sufficient to constrain the probability density shape for the lowest eccentricity region. In future, it would be beneficial if all planet discovery papers were to list eccentricity values for models where it is a free parameter, or better yet, provide the entire posterior samples/ MCMC chains used in the analysis, allowing for a more accurate treatment of eccentricities and their errors with HBMs \citep{Hogg2010,KippingWang2024}.

Figure~\ref{fig:ecc-hist} shows a Gaussian kernel density estimate (KDE) of the probability function. This is a non-parametric method to estimate the PDF, where kernels are used as weights to smooth out the histogram. The sample is more than twice as large as the sample from \citetalias{kipping2013}, and we use it to investigate changes in the optimum parameterisation of the exoplanet eccentricity distribution(s).

\section{Parameterisation methods}\label{sec:methods}
In this work we use two different methods to parameterise the eccentricity distribution. The first was introduced by \citetalias{kipping2013}, and we begin by reproducing their analysis to investigate the impact of an updated sample. We then trial an alternative method that does not require any binning of the sample.

\subsection{EDF method}\label{subsec:edf}

Following \citetalias{kipping2013}, we construct an empirical cumulative distribution function (EDF, or eCDF less commonly) associated with the measurement of a sample with $n$ data points. This EDF distribution, $\hat{F}(x)$, is the CDF that increases by 1/n at each data point. This can be simply expressed as:
\begin{equation}
    \hat{F}(x) = \frac{\textrm{number of elements in sample} \leq x}{n}.
    \label{eqn:EDF}
\end{equation}
\noindent This amounts to counting the elements less than or equal to a certain eccentricity value. For our $x$ array in Equation~\ref{eqn:EDF}, we simply take all unique eccentricities from the sample \citepalias{kipping2013}. 

As we are trying to estimate the underlying distribution our data are drawn from, the `true' CDF, $F(x)$, we need to quantify how constrained our EDF is by assigning uncertainties. Confidence intervals allow us to determine how good our distribution is, e.g. how far $\hat{F}(x)$ is from $F(x)$. 

The Dvoretzky-Kiefer-Wolfowitz (DKW; \citealt*{DKW}) inequality provides a method to estimate the probability of agreement between EDF and CDF by generating a confidence band that contains the true CDF at some specific confidence level. This is generally described in terms of $\alpha$, where 
the probability that $\hat{F}(x)$ is more than $\epsilon$ away from $F(x)$ is given by $1-\alpha$, so to generate a $\sim1\sigma$ DKW confidence band one sets $\alpha=0.32$ to get the 68$^{\textrm{th}}$ percentile. The DKW inequality states
\begin{equation}
    P(|\hat{F}(x)-F(x)| > \epsilon) \leq 2e^{-2n\epsilon^{2}} \textrm{~for every~} \epsilon >0, 
\end{equation}
\noindent where $\epsilon$ is given by:
\begin{equation}
    \epsilon = \sqrt{\frac{1}{2n}\log{\left(\frac{2}{\alpha}\right)}}.
\end{equation}
\noindent The interval that contains $F(x)$ with probability $1-\alpha$ is then:
\begin{equation}
    \hat{F}(x) - \epsilon \leq F(x) \leq \hat{F}(x) + \epsilon.
\end{equation}

\begin{figure}
    \centering
    \includegraphics[width=0.9\linewidth]{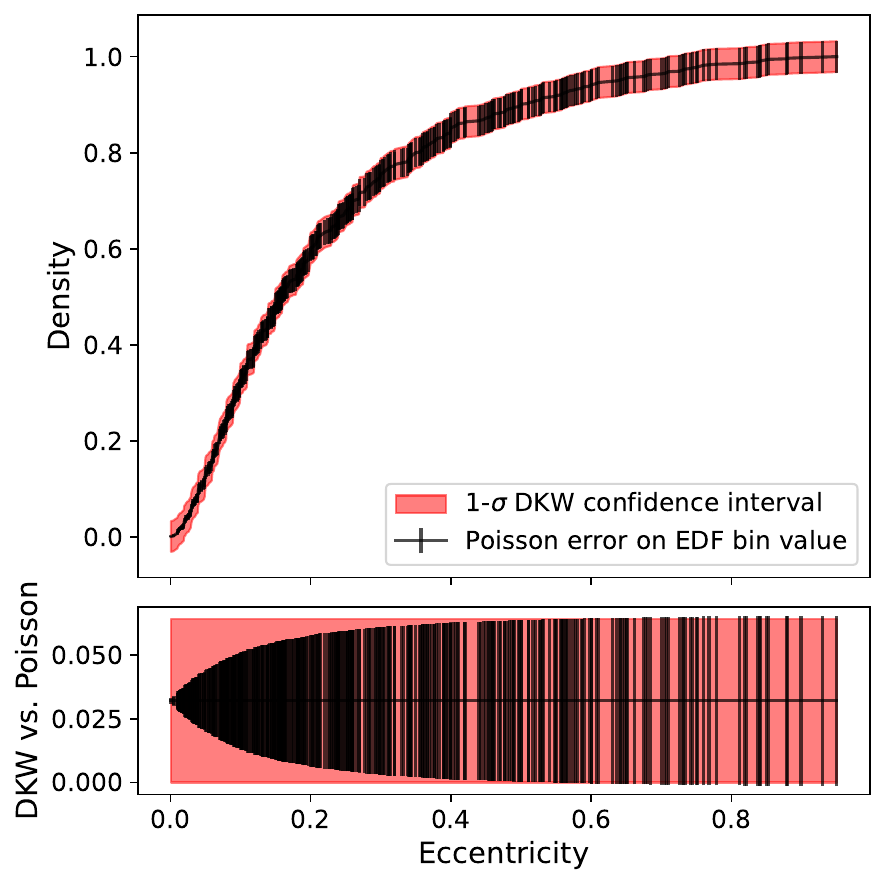}
    \caption{EDF created for 891 exoplanet eccentricities, showing both 1-$\sigma$ DKW intervals (red confidence band) and Poisson-derived errors on individual bins. At eccentricities below $\sim0.3$, the differences between error estimation methods becomes noticeable.}
    \label{fig:DKW}
\end{figure}

In Figure~\ref{fig:DKW} we show the computed EDF for all exoplanet eccentricities, alongside the $\sim1\sigma$ DKW confidence band, calculated by setting $\alpha=0.32$. We have over-plotted error bars derived via Poisson counting statistics on each EDF bin, the method used in \citetalias{kipping2013}. Towards higher eccentricities -- at larger EDF densities -- these error bars are a very good approximation for the DKW bounds, but there is a departure from this similarity towards low eccentricity/EDF densities (below $e\sim0.3$) where $\sqrt{N}$ errors are underestimating the uncertainty. The DKW confidence interval is more correct, as can be readily seen if we considered instead an EDF of `circularity'\,$=1-e$: this alternative EDF would have Poisson counting errors which are largest at small $e$.

\subsection{PDF method}\label{subsec:pdf}

Alternatively, we could instead perform a more direct method of computing parameters for each model, avoiding the need for E(C)DFs. Taking a distribution, such as Beta for example, we have priors on the ($a,b$) shape parameters and a Beta likelihood, and we can use the eccentricity measurements of the sample as observed values. The model parameters are then modified in light of the data -- the actual eccentricity values rather than each EDF bin -- to create the best-fit parameterisation and perform model comparisons.

\section{Distributions}\label{sec:functions}
We trial a number of distributions, to assess how well they represent the RV eccentricity population. Four of these were used in \citetalias{kipping2013}, and we include two additional distributions where models either share observed qualities with the sample, or are thought to reflect underlying physical processes affecting the eccentricities.

\subsection{Beta and Kumaraswamy}
The Beta distribution, $P_{\mathcal{B}}(e; a,b)$, was proposed by \citetalias{kipping2013} as a match to the observed eccentricity distribution, and as a potential prior for fitting exoplanet orbital models. It is defined and normalised over the required range of $0\leq e \leq 1$.  Despite being defined in terms of only
two parameters ($a,b$), it can successfully model a plethora of potential probability distributions \citepalias[see Fig.~\ref{fig:pdfs}, inspired by fig.~1 in][]{kipping2013}. The functional form is given below, either in terms of Gamma ($\Gamma$) or Beta ($B$) functions (not to be confused with the Gamma or Beta \textit{distributions}) for the normalisation:

\begin{align}
    P_{\mathcal{B}}(e;a,b) &= \textrm{constant}~\cdot~e^{a-1} (1-e)^{b-1}, \\ 
    &= \frac{\Gamma (a+b)}{\Gamma (a)\Gamma (b)}e^{a-1} (1-e)^{b-1}, \\
    &= \frac{1}{B(a,b)}e^{a-1} (1-e)^{b-1}.
\end{align}

\noindent The CDF follows, given in \citetalias{kipping2013} as
\begin{equation}
    C_{\mathcal{B}}(e;a,b) = \frac{B(e;a,b)}{B(a,b)}.
\end{equation}

The Beta distribution ($\mathcal{B}$\footnote{We assign symbols to the Beta distribution ($\mathcal{B}$) and Beta function ($B$) in different fonts to make the distinction between them clear.}) PDF and CDFs are available in many programming languages, making Beta convenient for use as a prior distribution. For applications of exoplanet parameter inference in light of RV data (using e.g. \textsc{kima}; \citealt{kima-ascl}), a Kumaraswamy distribution \citep{kumaraswamy} is often used in place of the Beta distribution as the eccentricity prior \citep[][]{Faria2022,Standing2023,AnnaJohn2023,Baycroft2023b}. The Kumaraswamy ($\mathcal{K}$) PDF also takes two shape parameters\footnote{We adopt the convention that $a,b$ are used for the parameters of the Beta distribution, and $\alpha,\beta$ for Kumaraswamy.}, $\alpha$ and $\beta$, and is is generally very similar to the Beta distribution. The Kumaraswamy PDF for a variety of parameters is represented pictorially in Figure~\ref{fig:pdfs}, and is: \begin{equation}
    P_{\mathcal{K}}(e;\alpha,\beta) =  \alpha\beta e^{\alpha-1} (1-e^{\alpha})^{\beta-1}.
\end{equation}
\noindent The Kumaraswamy distribution is simpler to use in simulations and manipulate numerically \citep{Faria2020,Faria2022}, as the PDF and CDF can be expressed in closed form making it often more tractable computationally \citep{jones2009,widemann2011}.

Typically, shape parameters from \citetalias{kipping2013}'s Beta distribution are used in the Kumaraswamy distribution, with the erroneous assertion of $a=\alpha=0.867,~b=\beta=3.03$. The resulting Kumaraswamy closely resembles Beta, but we find that using the exact same parameters does not provide a perfect fit. To reproduce $\mathcal{B}(0.867, 3.03$), for example, a more precise match can be achieved by using $\mathcal{K}(0.881,2.878)$. The parameters are often very close, but not usually identical. The Beta and Kumaraswamy distributions are both only valid for eccentricities between 0 and 1, and shape parameters $(a,b)\,,\,(\alpha,\beta) >0$. 

\begin{figure*}
    \centering
    \includegraphics[width=0.90\linewidth]{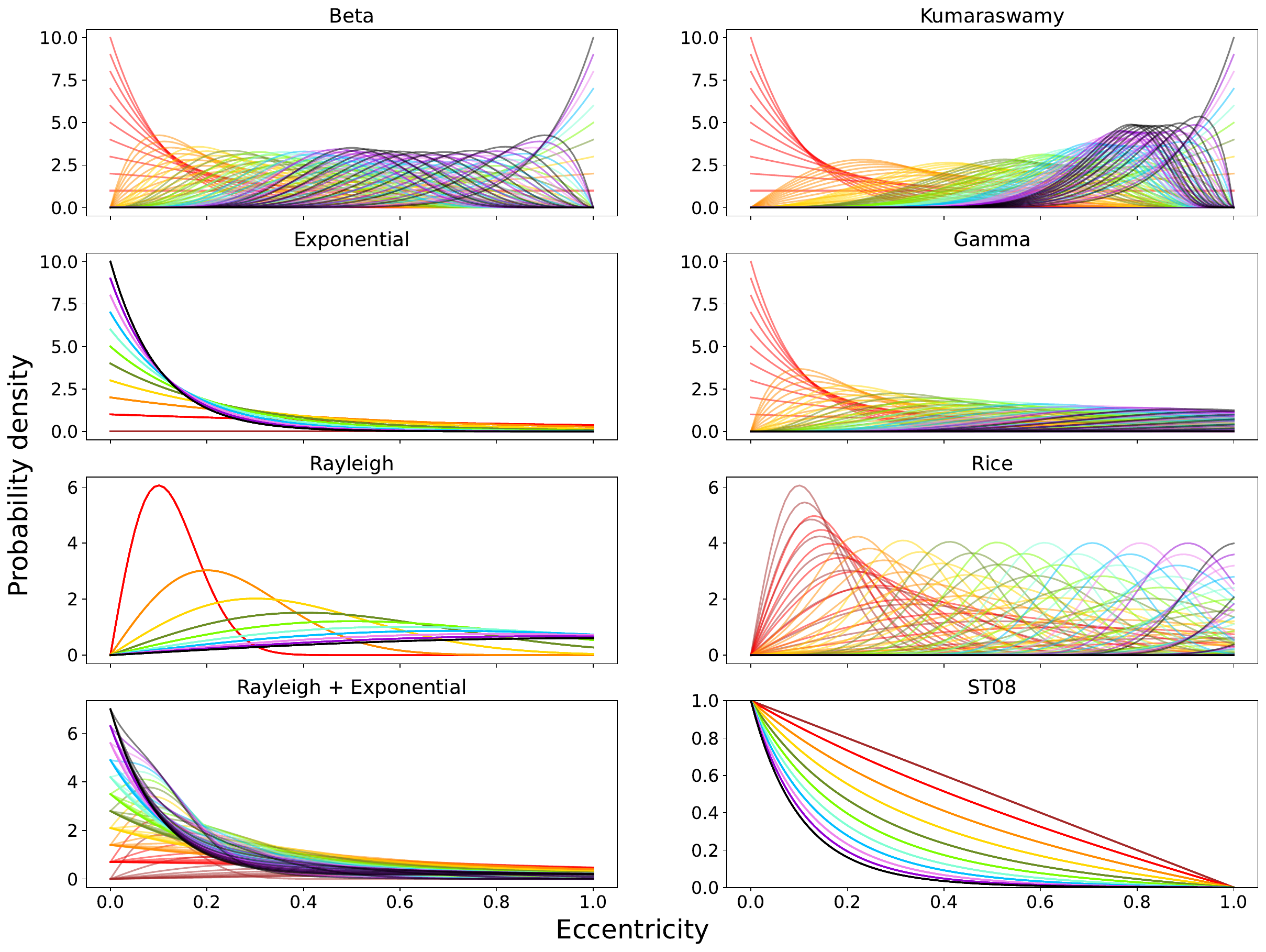}
    \caption{Visualisation of each PDF described in Section~\ref{sec:functions}, highlighting the flexibility of each distribution as parameters are varied. In a similar manner to \citetalias{kipping2013}, we vary one parameter from 0 to 10 in unity steps (associated with each colour, with increasing value from red through to purple and black), and then plot ten curves for the second parameter, also between 0 and 10. For the exponential, Rayleigh, and ST08 distributions, each only has one parameter, and therefore only 10 curves are drawn per plot. The Rayleigh+Exponential mixture panel is plotted with $\alpha=0.70$ (chosen to match Table~\ref{tab:all_ecc_results}) to avoid over-complicating the plot.}
    \label{fig:pdfs}
\end{figure*}

\subsection{Shen \& Turner 2008}
In a study of biases and uncertainties in the exoplanet eccentricity distribution, \citet[][henceforth \citetalias{ShenTurner08}]{ShenTurner08} explored processes that could change the distribution shape by analysing simulated data. They drew eccentricity samples from a PDF model peaking at $e=0$, that diminishes to zero at $e=1$:
\begin{equation}
    P_{\textrm{ST08}}(e;a) = \frac{1}{k}\bigg(\frac{1}{(1+e)^{a}} - \frac{e}{2^{a}}\bigg);
    \label{eqn:ST08pdf}
\end{equation} 
\noindent a family of these PDF curves are plotted in Figure~\ref{fig:pdfs}. We can then calculate the CDF of this function through integration, 
\begin{align}
        C_{\textrm{ST08}} (e;a) &= \int_{0}^{e} \frac{1}{k}\bigg(\frac{1}{(1+e')^{a}} - \frac{e'}{2^{a}}\bigg)~de' \\
        &= \frac{1}{k}\bigg[\frac{1}{(1-a)(1+e')^{a-1}} - \frac{e'^{2}}{2^{a+1}}\bigg]_{0}^{e}, 
\end{align}
\noindent where in these equations $a$ is the shape parameter and $k$ is simply a normalisation factor.

\citetalias{kipping2013} stated that this model is not normalised over 0 to 1 (therefore includes hyperbolic orbits), nor is it uniquely defined for $0\leq e \leq 1$. The correct form of Equation~\ref{eqn:ST08pdf} (taken from \citetalias{ShenTurner08}) is however constrained between 0 and 1, as seen in Fig.~\ref{fig:pdfs}. To account for the $k$ factor, we can normalise the \citetalias{ShenTurner08} CDF for regression to the empirical distribution function of eccentricities:
\begin{equation}
   C_{\textrm{ST08, }\{e\,\in\,\mathbb{R}~:~0\,\leq\,e\,\leq\,1\}} = \frac{C_{\textrm{ST08}}}{\max(C_{\textrm{ST08}})}.
\end{equation}

\subsection{Rayleigh + Exponential}\label{sub:RE}
\citetalias{kipping2013} also considered a mixture between a Rayleigh and an Exponential distribution, created to fit the empirical distribution function. Advantages to this model include claimed physical motivation, as the exponential component reflects effects of tidal dissipation \citep[e.g.][]{Rasio1996,RasioFord1996}, and the Rayleigh distribution is thought to reflect the impact that planet-planet scattering has on observed eccentricities (\citealt{Juric2008}; \citetalias{kipping2013}; \citealt{Weldon2024})

The equations describing these distributions follow. The exponential model (Equations~\ref{eqn:expon_PDF} and~\ref{eqn:expon_CDF}) takes one shape parameter $\lambda$, as does the Rayleigh distribution (Equations~\ref{eqn:rayleigh_PDF} and~\ref{eqn:rayleigh_CDF}), $\sigma$.

\begin{equation}
    P_{\textrm{exponential}}(e;\lambda) = \lambda \exp{[-\lambda e]},~\forall~e\geq 0,
    \label{eqn:expon_PDF}
\end{equation}
\begin{equation}
    C_{\textrm{exponential}}(e;\lambda) = 1 -\exp{[-\lambda e]},~\forall~e\geq 0,
    \label{eqn:expon_CDF}
\end{equation}

\begin{equation}
    P_{\textrm{Rayleigh}}(e;\sigma) = \frac{e}{\sigma^{2}} \exp{\bigg[-\frac{e^{2}}{(2\sigma^{2})}\bigg]},~\forall~e\geq 0,
    \label{eqn:rayleigh_PDF}
\end{equation}
\begin{equation}
    C_{\textrm{Rayleigh}}(e;\sigma) = 1 - \exp{\bigg[-\frac{e^{2}}{(2\sigma^{2})}\bigg]},~\forall~e\geq 0.
    \label{eqn:rayleigh_CDF}
\end{equation}

\noindent The exponential and Rayleigh distribution PDFs are plotted in Fig.~\ref{fig:pdfs}, along with a mixture model PDF.

Combining these distributions creates an appealing mixture model that is grounded in underlying physical processes. However, \citetalias{kipping2013} noted that a major problem is that the resulting distribution would assign hyperbolic orbits ($e\geq 1$) a non-zero probability, at odds with signs of periodic planetary modulation in RV data. To regress the Rayleigh+exponential distribution (henceforth `$\mathcal{RE}$') to the eccentricity EDF, we adjust the CDF equations to ensure that the CDF reaches unity at $e=1$, and account for normalisation factors:
\begin{equation}
    C_{\mathcal{RE}} = \alpha \bigg(\frac{C_{\textrm{exponential}}}{C_{\textrm{exponential}}(e=1)} \bigg)  + (1-\alpha)\bigg(\frac{C_{\textrm{Rayleigh}}}{C_{\textrm{Rayleigh}}(e=1)}\bigg),
\end{equation}
\noindent where $\alpha$ here governs the relative contributions of each distribution. An $\alpha = 1$ would be a pure exponential, and $\alpha = 0$ a pure Rayleigh distribution.

\subsection{Gamma}\label{sub:gamma}
Noticing the peaked nature of the KDE probability density function in Fig.~\ref{fig:ecc-hist} directed us towards the Gamma distribution, which has a similarly shaped PDF. This distribution also has a flexible shape, and can resemble the exponential or chi-squared distributions under special conditions. There are two commonly used parameterisations of the Gamma distribution, ($k,\theta$) or ($\alpha,\beta$), with the latter being usually preferred in Bayesian statistics. We therefore adopt this convention. The PDF is given as

\begin{equation}
    P_{\Gamma} (e;\alpha,\beta) = \frac{e^{\alpha-1}\exp{[-\beta e]}~\beta^{\alpha}}{\Gamma(\alpha)}~\textrm{for}~e>0~~\alpha,\beta>0,   
\end{equation}
\noindent where $\Gamma(\alpha)$ is the gamma function. The CDF is therefore
\begin{equation}
    C_{\Gamma} (e;\alpha,\beta) = \frac{\gamma(\alpha,\beta e)}{\Gamma(\alpha)}, 
\end{equation}
\noindent where $\gamma(\alpha,\beta e)$ is the lower incomplete gamma function. Additionally, we make a similar modification as with other functions, to ensure that the CDF is normalised at the maximum eccentricity of $e=1$, 
\begin{equation}
    C_{\Gamma}~(\textrm{normalised}) = \frac{C_{\Gamma}}{C_{\Gamma}(e=1)}.
\end{equation}
We have plotted the Gamma distribution PDF for a range of ($\alpha,\beta$) in Fig.~\ref{fig:pdfs}.

\subsection{Rice}\label{sub:rice}
The Rice distribution is the probability distribution of a circularly-symmetric bivariate Gaussian random variable, that may have a non-zero mean \citep{Rice}. The distribution is described by PDF
\begin{equation}
    P_{\textrm{rice}}(e; \nu,\sigma) = \frac{x}{\sigma^{2}} \exp{\bigg[\frac{-(e^{2} + \nu^{2})}{2\sigma^{2}}\bigg]}~I_{0}\bigg(\frac{e\nu}{\sigma^{2}}\bigg),
\end{equation}
\noindent with shape parameter $K = \nu^{2}/2\sigma^{2}$ and scale parameter $\Omega = \nu^{2} + 2\sigma^{2}$. The symbol $I_{0}(...)$ denotes the modified Bessel function of the first kind with order zero. Upon inspection of this PDF equation (and plots in Figure~\ref{fig:pdfs}), it is clear that when $\nu=0$, the Rice distribution simplifies to the Rayleigh distribution of Section~\ref{sub:RE}. The CDF is then given as 
\begin{equation}
    C_{\textrm{rice}}(e; \nu,\sigma) = 1- Q_{1}\bigg(\frac{\nu}{\sigma},\frac{e}{\sigma}\bigg),
\end{equation}
\noindent where $Q_{1}$ is the Marcum Q-function \citep{Marcum1960}.

The Rice distribution has previously been suggested to model eccentricity densities of the inner Solar System planets, well-representing the effects of chaotic diffusion \citep{Hara2019}. \citet{Laskar2008} studied how the eccentricities of chaotic planetary orbits in the Solar System would evolve with time and further perturbation (e.g. between Mercury and Jupiter at perihelion). They found that the resulting eccentricities of the terrestrial planets after numerical integration was very well approximated by a Rice PDF. As such, other authors have modified the Rice distribution for creation of eccentricity priors \citep*[cf.][]{Mahmud2023}

The match between Rice and Laskar's results provides another appealing physical interpretation for an eccentricity prior. If the Rice distribution models chaotic diffusion of the inner system planets \citep{Mogavero2017}, it is worth investigation as an eccentricity prior for exoplanets. Chaotic diffusion is though to to cause orbital instabilities in planetary systems once the disk is disrupted, and to be a natural consequence of irregular motion \citep*{MCB2016}. When resonances caused by planetary orbits overlap, instabilities (and chaotic diffusion) will come into play. Multi-resonant configurations will be a likely outcome of migration through interaction with the protoplanetary disk \citep{MCB2016}.

\section{EDF Regression}\label{sec:EDF}

To regress each model to the data EDF, we used the \textsc{multinest} package \citep{FerozHobson2008,multinest} following the procedure adopted by \citetalias{kipping2013}. Our code employs the Python implementation, \textsc{pymultinest} \citep{pymultinest}. Nested sampling (NS) explores the posterior distributions, and also computes the Bayesian evidence of each model, allowing us to easily compare and select the optimal model \citep[e.g.][]{Skilling2004}. The number of additional parameters is penalised in the evidence calculation, ensuring that the simplest, best-fitting model is chosen. For all parameters we chose to use uniform priors for simplicity, from 0 to 10 (apart from $\alpha$ in $\mathcal{RE}$, which is only defined from 0 to 1). \citetalias{kipping2013} elected to use modified Jeffrey's\footnote{The modified Jeffrey's prior is uniform below the knee/inflexion point, and log-uniform beyond it.} priors for most parameters, placing an inflexion point at unity. We performed trials with these priors, finding that the fitted distribution parameters were identical in either case, and the evidences were consistent to within $1\sigma$ uncertainties (typically $\Delta\ln{Z}\sim0.10$). In situations with small sample sizes, priors can have a large effect \citep[e.g.][]{Nagpal2023}. Our large sample (EDF) effectively constrains the parameters of the fitted distribution, meaning that informative (or biasing) priors are not necessary. To perform the most simplistic treatment possible, we did not constrain prior space further and opt to use uniform priors on the distribution parameters.

\begin{figure}
    \centering
    \includegraphics[width=0.9\linewidth]{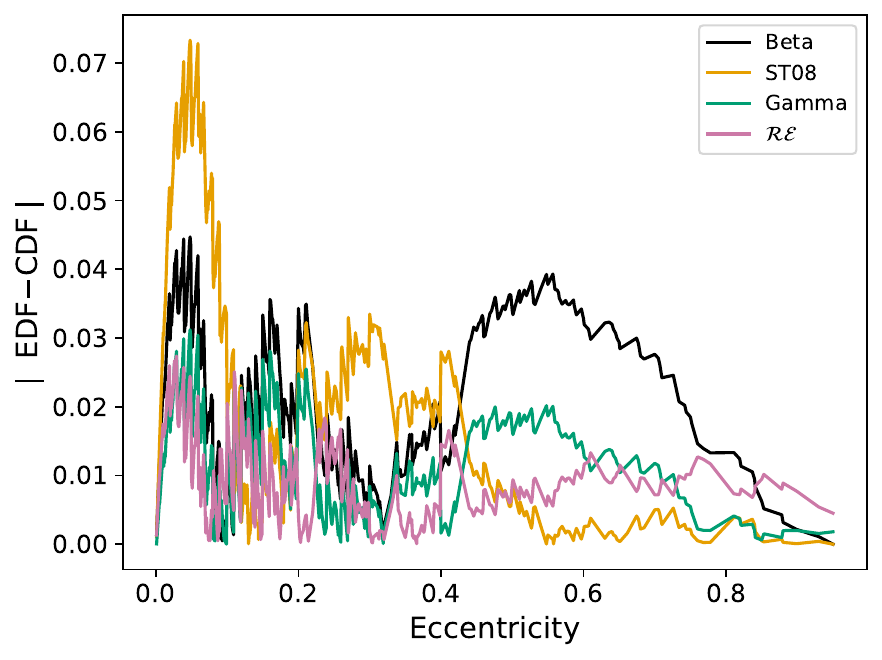}
    \caption{The absolute value of the residuals between regressed distributions (CDFs) of Table~\ref{tab:all_ecc_results}, and the eccentricity EDF of the entire sample. A plot of the fits themselves does not successfully illustrate the differences between the distributions, as all are a relatively close fit to the EDF. 
    Kumaraswamy is not included in this figure, as it is extremely similar to Beta.
    To improve visibility of the lines, we have adopted a colour-blind friendly colour-map in this plot \citep{Wong2011}.}
    \label{fig:EDF_residuals}
\end{figure}

We report the results of our NS regression for the Beta, Kumaraswamy, ST08, Gamma, and Rayleigh + Exponential distributions in Table~\ref{tab:all_ecc_results}. Residuals for each regression are plotted in Fig.~\ref{fig:EDF_residuals}, to compare how successfully each distribution fits the EDF. A purely Rician CDF does not successfully model our sample EDF, so has been excluded from the remainder of the analysis. As it is a more general version of the Rayleigh distribution, it may be of use in mixture models, though this will increase the number of parameters even further so is not explored.

It can be seen that $\mathcal{B}$ and $\mathcal{K}$ are very similar, with $\Delta \ln{Z}\sim 3.63$, less than the required threshold for strong evidence in favour of one model over the other \citep[$\Delta\ln{Z}>5$;][]{Kass1995,Trotta2008}. We now find the highest evidence is attributed to the $\mathcal{RE}$ model, in disagreement with \citetalias{kipping2013}. The ratio of evidences, or Bayes factor (BF), for this model over the next best (Gamma) is equal to a significance of $4.63\sigma$. This corresponds to an odds-ratio of $\sim7800:1$. Here and throughout this work, we convert BFs to `sigmas' to provide a more digestible measure of significance. This is done following the prescription of \citet{Trotta2008}, adapted from implementations by \citet{MacDonald2023} and \citet{Taylor2023}.

Our enlarged sample reveals that other distributions now provide a better match to the eccentricity EDF than Beta/Kumaraswamy. $\mathcal{RE}$ is favoured over Beta by $\Delta\ln{Z} = 77.18$ (odds-ratio of $\sim3\times10^{33}:1$), or $12.64\sigma$ significance. The Gamma distribution provides the closest fit for a non-mixture model, however the mixture of Rayleigh and Exponential distributions allows improved flexibility and is favoured by the evidence here even though it includes an additional parameter over other models.

In these results, and throughout the remainder of this paper, we observe a consistently lower evidence score for ST08 regression to the EDF (e.g. Table~\ref{tab:all_ecc_results}). This distribution is the least favoured option, and we can attribute this to the shape of the sample. As seen in Fig.~\ref{fig:ecc-hist}, the peak of the sample distribution has moved away from zero due to the treatment of planets with eccentricity fixed to zero. The ST08 function was designed to peak at an eccentricity of zero based on the sample at the time \citepalias{ShenTurner08}, so it is no surprise that this distribution is the least successful at modelling the updated EDF.

\renewcommand{\arraystretch}{1.4}
\begin{table*}
    \centering
        \caption{Statistics for regression to the entire usable RV sample eccentricity \textbf{EDF}, when varying model choice. Evidence values (favoured models being more positive) are shown along with the best-fit parameters, where \textsc{corner} plots have been used to determine the median value and 1-$\sigma$ error bounds on the posteriors output by \textsc{pymultinest}.}
    \begin{tabular}{lllll}
    \hline
         Distribution & Evidence ($\ln{Z}$) &  Parameter 1 & Parameter 2 & Parameter 3\\
         \hline
        Beta $(a,b)$ & $775.08 \pm 0.10$  & $1.08\pm 0.02$ & $4.10^{\,+0.09}_{-0.08}$ &  \\
        Kumaraswamy $(\alpha,\beta)$ & $771.45 \pm 0.10$ & $1.05\pm 0.01$ & $4.10 \pm 0.11$ &  \\
        ST08 $(a)$ & $735.49 \pm 0.08$ & $4.74\pm0.04$ &  &  \\
        Rayleigh + Exponential $(\alpha,\lambda,\sigma)$ & $852.26 \pm 0.11$ & $0.70 \pm 0.02$ & $3.39^{\,+0.09}_{-0.10}$ & $0.111 \pm 0.003$ \\
        Gamma $(\alpha,\beta)$ & $843.29 \pm 0.09 $  & $1.30\pm 0.02$ & $5.88^{\,+0.14}_{-0.13}$ &  \\
        \hline
    \end{tabular}
    \label{tab:all_ecc_results}
\end{table*}

\subsection{Constraining prior space}

In our analysis of the eccentricity distribution, we have so far considered uniform priors on function parameters. However, we can see from Fig.~\ref{fig:pdfs} that a subset of parameter choices represent unrealistic scenarios where the PDF increases towards maximum near or at $e=1$. This does not matter for parameter estimation, but could bias our model comparison. Requiring the Beta distribution to have positive skewness only ($a<b$) halves the size of prior space, and may therefore affect the evidence calculation. 

To investigate the effect that changing the prior has on the evidence, we implemented the restriction that $a<b$ in the Beta distribution. Performing the same regression to the EDF as above, the evidence value does not change significantly. Restricting the prior space does not therefore significantly affect our model comparison. The \textsc{multinest} sampler does make half the number of likelihood calculations when we restrict prior space, speeding up computation, but the effect is not overly noticeable given the already fairly rapid execution times ($\mathcal{O}\sim1$\,minute).

\subsection{Fitting to multiple EDFs}
We can split the population of RV exoplanets up further, and regress multiple local functions to the data, instead of using a single global function. This was done by \citetalias{kipping2013} for $\mathcal{B}$, where they split the sample into short- and long-period exoplanets (at the median period), and found the two samples came from different underlying distributions (at $11.6\sigma$ significance). \citet*{2024AJ....168..115B} recently analysed a population of RV orbits for $P<200$\,d. They found that for giant planets, the upper eccentricity envelope is strongly influenced by orbital period, but low-mass planet eccentricities do not show a dependence on orbital period. 
The massive planets' eccentricity distribution's upper envelope is period dependent \citep{2024AJ....168..115B}. This is similar to the behaviour of the eccentricity distributions of spectroscopic binaries \citep*[e.g.][]{2023MNRAS.522.1184B} and brown dwarfs \citep[][]{Stevenson2023-BDs,Shin2024}. As \citet{2024AJ....168..115B} confirm that both period and mass are important in shaping the eccentricity distribution, we have created subsamples based on planetary periods and masses to assess how it changes the parameters of the fitted distributions.

The null hypothesis used for comparison here is a single distribution fitted to the two local EDFs. Whilst seeming counter intuitive, this allows the similarity between EDFs to be tested. If a single global fit has higher evidence than local regression, the two EDFs are similar enough that doubling the number of fitted parameters provides no improvement to the models, and we can conclude that there are not multiple underlying populations combining to dictate the shape of the observed eccentricity distribution. With our enlarged sample we can explore this similarity across various subsamples.

\subsubsection{Short- \& long-period}\label{subsub:short-long}

We replicate the analysis of \citetalias{kipping2013}, splitting the sample into two equally-sized subsamples (at the median orbital period, $371.74$~d here). We also find very decisive evidence that the two distributions are drawn from different underlying parent distributions. The average $\ln{Z}$ difference between the two hypotheses ($\mathcal{H}_{1}$ and $\mathcal{H}_{2}$), across all fitted distributions, is in the range of 300--400 (Table~\ref{tab:short_vs_long}). Note that for global fits, the evidences and inferred parameters differ from Table~\ref{tab:all_ecc_results}. This arises from the binning procedure used in creation of these EDFs, where the number of unique eccentricities defines the maximum resolution we can achieve \citepalias[][]{kipping2013}. The entire sample EDF (used for results in Table~\ref{tab:all_ecc_results}) has the highest resolution and improved evidence, and is therefore more reliable.

The highest evidence multi-local-function fit here is now achieved with the Gamma distribution. The individual EDFs are better approximated by Gamma distributions than the next best ($\mathcal{RE}$), but when combining the sub-sets back into a single dataset, fitting a single Gamma does not provide the same flexibility that the $\mathcal{RE}$ mixture model can offer. $\mathcal{RE}$ provides the highest global evidence in Table~\ref{tab:all_ecc_results}, and in the rows relating to the single global model hypothesis in Table~\ref{tab:short_vs_long}, perhaps indicating that when multiple physical effects are driving the eccentricity population, a mixture model can do a better job at `bridging the gap' and accounting for the blend of individual EDF shapes. Either distribution provides very decisive evidence over the remainder of the trial functions, and we can rule out the Beta/Kumaraswamy and ST08 functions here.

Using local function regression, we have attempted to learn about the underlying physical processes causing observed demographics. Tidal circulation would be stronger for short-period planets than for long-period, distant planets. It has been postulated that this process is best-modelled by an exponential \citep[e.g.,][]{Rasio1996}. We investigate this by fitting a mixture model consisting of only an exponential for sub-median-periods, and $\mathcal{RE}$ for long periods. For samples split at the median period, we find the evidence becomes $\ln{Z} = 904.97$ (see $\mathcal{E}_{\textrm{s}}, \mathcal{RE}_{\textrm{l}}$ model in Table~\ref{tab:short_vs_long}). This is $\Delta\ln{Z} \sim 40$ ($9.2\sigma$) lower evidence than using two $\mathcal{RE}$ mixture distributions, for short and long periods, indicating that for the shorter periods, it is still informative to use the mixture, and an exponential alone provides a worse fit. This suggests that below the median $P=371.74$, tidal dissipation is not the only factor sculpting the demographics.

\begin{figure}
    \centering
    \includegraphics[width=0.9\linewidth]{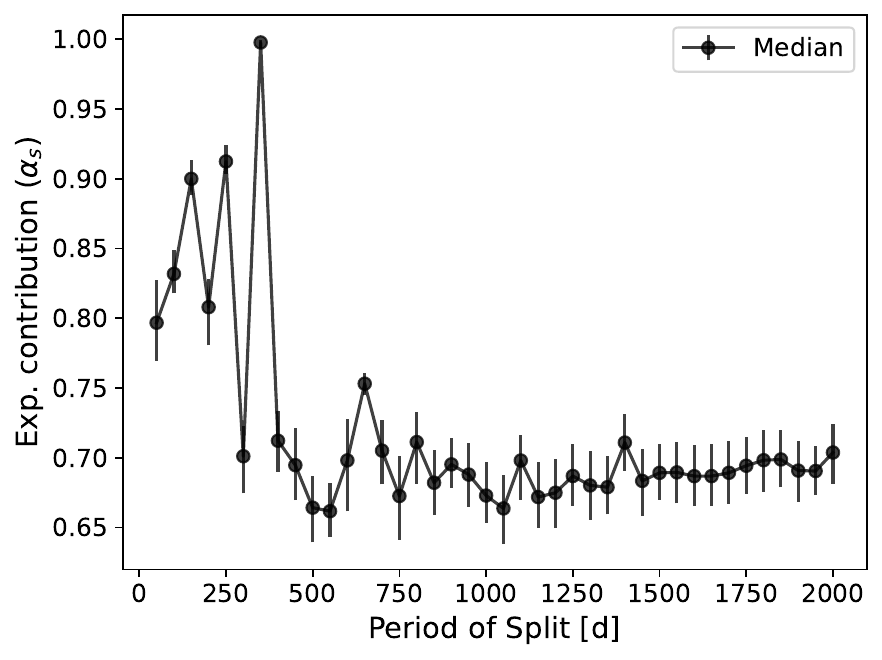}
    \caption{Exponential contributions for `short-period' planets in the case of two local $\mathcal{RE}$ regressions, when varying the period split taken to split the sample `short' and `long' sub-samples. As this period increases, the exponential component eventually becomes less dominant and stabilises. This provides us with an informative location to perform such a period split, at roughly 400 days, which does in-fact happen to be close to the median orbital period.}
    \label{fig:alphaS_evolution}
\end{figure}

\begin{figure*}
    \centering
    \includegraphics[width=0.85\linewidth]{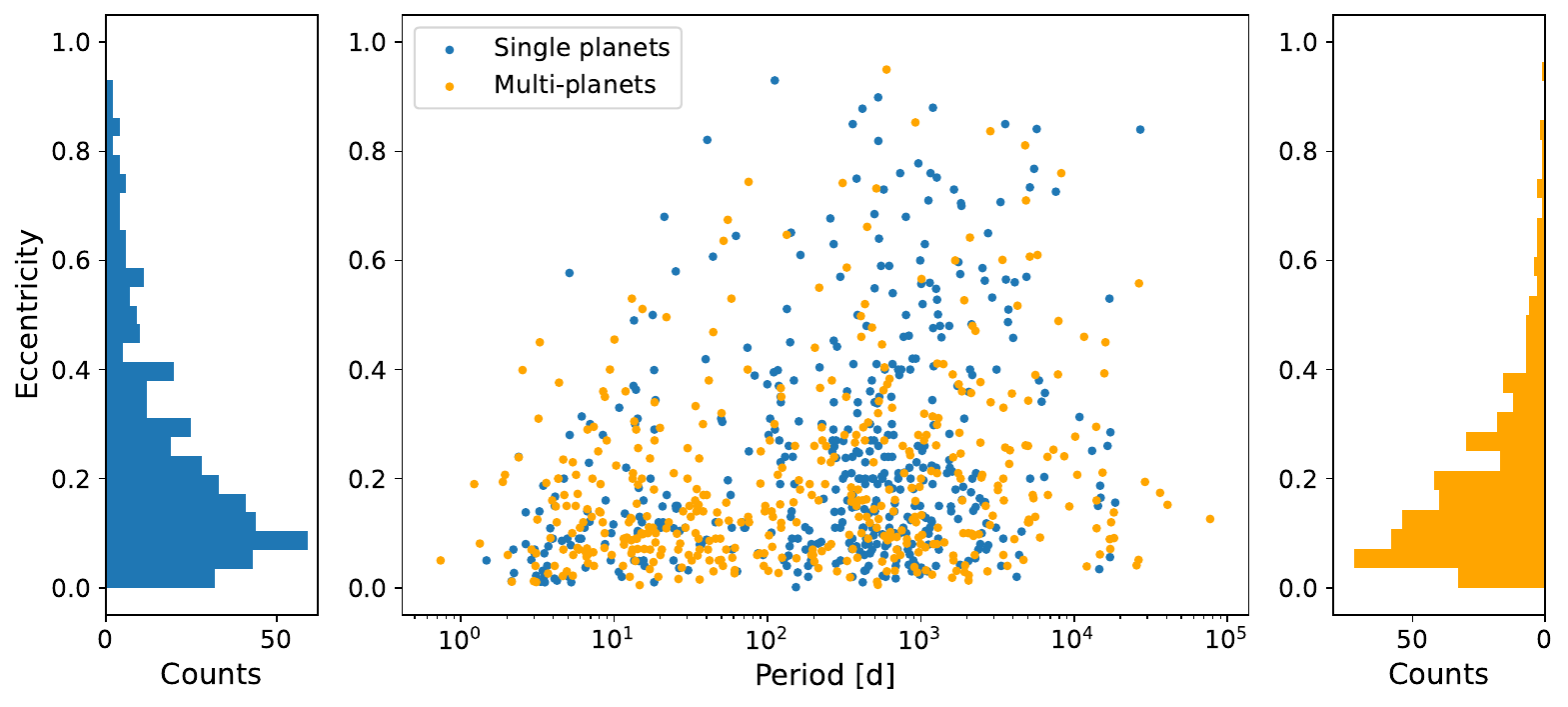}
    \caption{The period--eccentricity plane for all exoplanets in our sample, showing that circularisation has a stronger influence on the short-period planets. Points are coloured based on if the system contains a single planet, or more than one planet. Histograms for eccentricities of either single or multi-planet systems are plotted beside the main panel.}
    \label{fig:P-vs-e}
\end{figure*}

We explored varying the orbital period where the sample is divided, thus examining
if a purely exponential short-period fit is ever justified. This could reveal the orbital period below which tidal circulation alone appears to be driving the eccentricities. We find that the exponential component is larger, more variable, and can come close to unity, for period splits up to roughly $400$\,d (Fig.~\ref{fig:alphaS_evolution}). Planets on close-in orbits will more readily circularise through the loss of orbital angular momentum via frictional losses in either the outer layers of the stellar atmosphere, or within the planetary interior itself \citep{Mathis2015}. Beyond this period, including successively longer orbital period planets causes a consistent departure from any potential $\alpha_{\textrm{s}} \sim 1$ behaviour, as these orbits are less likely to be circularised (cf. Fig.~\ref{fig:P-vs-e}). Therefore, if one is interested in a population of short-period planets, it would be worth bearing the exponential distribution in mind to parametrise the eccentricity space, and including it in model comparisons. Additional study of the short period planet demographics in the future would be worthwhile to investigate any possible connection between exponential behaviour and tidal circularisation.

There may also be some observational bias at play. Lower mass, rocky planets will be less subject to tidal circularisation than gas giant exoplanets, as the circulation rate due to tides raised on a planet is inversely proportional to $R_{\rm p}^{\,5}$ \citep{Jackson2008}. However, the tidal circularisation rate is affected by the planet's composition. For rocky planets, this is partly compensated by the lower modified tidal quality factor \citep[e.g.][]{1966Icar....5..375G}.
These low-mass, rocky planets will also be more difficult to detect at long periods, as the RV amplitude depends on both mass and proximity to the host star \citep[e.g.][]{Haswell2010}. A large number of long-period, low-mass planets may be missing, which are even less likely to be tidally circularised. Results, such as those seen in Figure~\ref{fig:alphaS_evolution}, may be impacted for periods beyond $\sim1$~yr, as this regime will mostly comprise detections of massive gas giant planets. With the ever-improving detection capabilities of high-precision spectrographs, this region will become more populated in the future, and will be worth revisiting.

\renewcommand{\arraystretch}{1.4}
\begin{table*}
    \centering
        \caption{Results when regressing two models to two \textbf{EDFs}, built from planets above and below the median orbital period. Parameter subset `s' denotes short-period, subset `l' denotes long-period planets. Global fits with one model are also included, denoted `g'. We also tabulate regression results where the short-period EDF is fitted with a purely exponential distribution, and long-period EDF with a Rayleigh + Exponential ($\mathcal{E}_{_{\textrm{s}}}\,\&\,\mathcal{RE}_{_{\textrm{l}}}$).}
        \label{tab:short_vs_long}
    \begin{tabular}{llllllll}
    \hline
         Distribution & Evidence ($\ln{Z}$) &  Param.~1 & Param.~2 & Param.~3 & Param.~4 & Param.~5 & Param.~6\\
         \hline
         $\mathcal{H}_{1}$ -- Two local models\\
        Beta $(a_{\textrm{s}},b_{\textrm{s}},a_{\textrm{l}},b_{\textrm{l}})$ & $911.28 \pm 0.13$  & $1.19\pm 0.04$ & $5.82^{\,+0.24}_{-0.22}$ & $1.07 \pm 0.03$ & $3.27 \pm 0.11$ &  & \\
        ST08 $(a_{\textrm{s}},a_{\textrm{l}})$ & $875.79 \pm 0.08$ & $6.12\pm0.08$ & $3.50^{\,+0.07}_{-0.08}$ & &  &  & \\
        $\mathcal{RE}$ $(\alpha_{\textrm{s}},\lambda_{\textrm{s}},\sigma_{\textrm{s}},\alpha_{\textrm{l}},\lambda_{\textrm{l}},\sigma_{\textrm{l}})$ & $945.79 \pm 0.16$ & $0.68^{\,+0.03}_{-0.02}$ & $4.41^{\,+0.18}_{-0.15}$ & $0.098^{\,+0.005}_{-0.004}$ & $0.70^{\,+0.03}_{-0.02}$ & $2.60^{\,+0.13}_{-0.11}$ & $0.132^{\,+0.005}_{-0.004}$ \\
        $\mathcal{E}_{\textrm{s}}\,\&\,\mathcal{RE}_{\textrm{l}}$ $(--,\lambda_{\textrm{s}},--,\alpha_{l},\lambda_{\textrm{l}},\sigma_{\textrm{l}})$ & $904.97 \pm 0.12$ & $--$ & $5.19\pm 0.06$ & $--$ & $0.68\pm 0.03$ & $2.53\pm 0.16$ & $0.13\pm0.01$  \\
        Gamma $(\alpha_{\textrm{s}},\beta_{\textrm{s}},\alpha_{\textrm{l}},\beta_{\textrm{l}})$ & $951.72 \pm 0.12 $  & $1.38 \pm 0.04 $ & $7.81^{\,+0.30}_{-0.31}$ & $1.33 \pm 0.04$  & $4.96^{\,+0.20}_{-0.18}$ & &  \\
        \hline
        $\mathcal{H}_{2}$ -- One global model\\
        Beta $(a_{\textrm{g}},b_{\textrm{g}})$ & $522.36 \pm 0.09$  & $1.04\pm 0.02$ & $3.86\pm 0.10$ & & &  & \\
        ST08 $(a_{\textrm{g}})$ & $520.31 \pm 0.08$ & $4.60\pm0.05$  & &  &  &  & \\
        $\mathcal{RE}$ $(\alpha_{\textrm{g}},\lambda_{\textrm{g}},\sigma_{\textrm{g}})$ & $581.90 \pm 0.12$ & $0.70\pm0.02$ & $3.14^{\,+0.12}_{-0.10}$ & $0.103^{\,+0.004}_{-0.003}$ &  &  &  \\
        Gamma $(\alpha_{\textrm{g}},\beta_{\textrm{g}})$ & $570.76\pm 0.09 $  & $1.24\pm 0.03$ & $5.48\pm 0.16$ &  \\
        \hline
    \end{tabular}
\end{table*}

\subsubsection{Three period regimes}\label{sec:3pr}

Testing the differences between `short' and `long' period exoplanets by splitting at the median orbital period (into two equally-sized subsamples) is overly-simplistic. The eccentricities of the exoplanetary population are evidently strongly affected by orbital period, so testing further sub-groupings may reveal features sculpted by, for example, tidal dissipation. We split the sample into three period intervals (to retain a substantial number of samples in each bin): 1--10, 10--300, and $>$300\,d. These ranges closely match the current definitions of `hot', `warm' and `cold' planets \citep[similar to the divisions used in e.g.][and others]{BihanBanerjee2024}.

We split the sample into these period bins and performed both local and global regression to the created EDFs. The local parameterisations are again preferred by $\Delta\ln{Z}\sim300$, very strong positive evidence \citep{Trotta2008}. Fitting a $\mathcal{RE}$ distribution to these EDFs shows that for 1-10\,d and 10-300\,d period ranges, the distribution is almost entirely exponential ($\alpha\sim1$). This is perhaps expected based on the results shown in Figure~\ref{fig:alphaS_evolution}, where `short' period exoplanet eccentricities are more likely to follow an exponential distribution than `long' period planets. This is thought to be related to the tidal circularisation for short period planets, as mentioned above.

Additionally, the 1--10\,d period planets are found to be far more circular than for other period ranges. They have mean $\overline{e}=0.132$, and are all $<0.6$. The other ranges extend up to $e\sim0.9$, with means of $0.197$ and $0.259$, for 10--300\,d and $>$300\,d periods, respectively. 
Proximity to the star circularises the orbit of a planet, as the circularisation time scales as $(M_{\rm p}/M_{\rm s}) \times (a/R_{\rm p})^{5}$ \citep[e.g.][]{Adams2006}. A significant fraction of planets in the 1--10 and 10--300\,d period bins will exhibit eccentricities that have been substantially impacted by tidal forces.
When considering the eccentricity distributions of exoplanets, and using this to inform our choice of a prior, it would make sense to use one that reflects the possible eccentricities for expected system architectures.

\subsubsection{Three mass regimes}\label{sec:3m} 

The eccentricity distribution of exoplanets is also inherently linked to the masses of planets themselves. Small planets in multiple systems are likely to have smaller eccentricities \citep[e.g.][]{VEA2015,Xie2016,VE19}, whereas giant planets in wide orbits overwhelmingly exhibit larger eccentricities \citep*[][and references therein]{Bitsch2020}. To investigate the distinction between low (super-Earths), moderate (Neptune-type planets), and high-mass planets (gas giants, like the gaseous giant planets of the Solar System), we have split our sample into three mass bins.

Many studies have investigated mass-radius relations to find transition points between `types' of planets \citep{HatzesRauer2015,ChenKipping2017,Otegi2020,Edmondson2023,Muller2024}. There is no definitive answer, so we have chosen approximate\footnote{This selection criteria does not need to be exact, given that the Neptune-mass part of the mass distribution is under-populated, caused in part by the Neptune desert \citep*[e.g.][]{Mazeh2016}.} regions to provide a broad picture of how planetary mass affects the eccentricity distribution. We place the boundary between rocky and volatile-rich planets at $20$~M$_{\earth}$ \citep[close to the value found by][]{Otegi2020}, and the boundary defining the most massive giants at $100$~M$_{\earth}$ \citep{HatzesRauer2015}.

We therefore split our sample into bins of 
$1$-$20$~M$_{\earth}$, $20$-$100$~M$_{\earth}$, and $>100$~M$_{\earth}$. EDFs are again created for the three sub-samples, and distributions fitted with the same method as previous sections. The evidence is compared with the null hypothesis of fitting a single over-arching distribution. It is worth noting at this point that the majority of these masses will obviously be minimum masses, given that the planets were all identified using the radial velocity technique. We have used the `best' available masses quoted on the NASA exoplanet archive (\texttt{pl\_bmass}), tabulating the true mass if available, and the $m\sin{i}$ if not. Many masses will be fractionally increased, by an average factor of $\sim1.19$ -- based on the expectation value of an inclination vector distributed randomly over the surface of a sphere \citep{HoTurner2011,LJ2012}.

The increase in evidence massively favours local parameterisations accounting for different planetary masses: $\Delta\ln{Z}~\sim100$--$200$, depending on the chosen distribution. 
This aligns with what one would expect when calculating the average reported eccentricity for planets in each mass bin: $\overline{e}=0.164,0.191,0.245$, in order of increasing mass. The preference for local fits provides overwhelming evidence that mass is also a key factor sculpting the eccentricity distribution. This will however also be coupled to the period distribution, due to the detection sensitivity of our instruments and the currently known catalogue of exoplanets showing clusters of detections in the cold-Jupiter or hot/warm super-Earth regimes (see Section~\ref{subsub:period-mass} and Figure~\ref{fig:3m3p-plane}).

As with previous sections, we have searched for physical meaning in the fitted distributions by inspecting the parameters of the Rayleigh and Exponential mixture model.  For the most massive planets (the Jovian-type planets, and the low number of planets with $20\, {\textrm{M}}_{\earth} < M < 100\, {\textrm{M}}_{\earth} $), the mixture fraction $\alpha$ is extremely close to unity, and perhaps indicates that tidal circularisation (and dependence on orbital period) governs the eccentricity behaviour. For low-mass planets however, $\alpha=0.38$, the lowest value for $\alpha$ found anywhere throughout this paper. This is the only time the $\mathcal{RE}$ distribution is weighted in favour of the Rayleigh distribution, which is perhaps as expected. 

The Rayleigh distribution has been theorised to reflect the eccentricity distributions of planets that have undergone significant scattering due to other bodies in the system. The lower the mass of the planet, the weaker the perturbation needs to be to achieve a significant level of scattering. The above result is in agreement with \citet{2024AJ....168..115B}, who found that small planets were circularised less effectively -- and that the cause was likely to be dynamical interactions from other planets in the system. Additionally, due to their lower masses, small planets are more likely to remain stable in higher-multiplicity planetary systems \citep[considering, for example, the angular momentum deficit;][]{Laskar2017}, allowing for configurations that would exhibit a high degree of interaction between the constituent planets.

\subsubsection{Period and Mass}\label{subsub:period-mass}
\begin{figure}
    \centering
    \includegraphics[width=0.99\columnwidth]{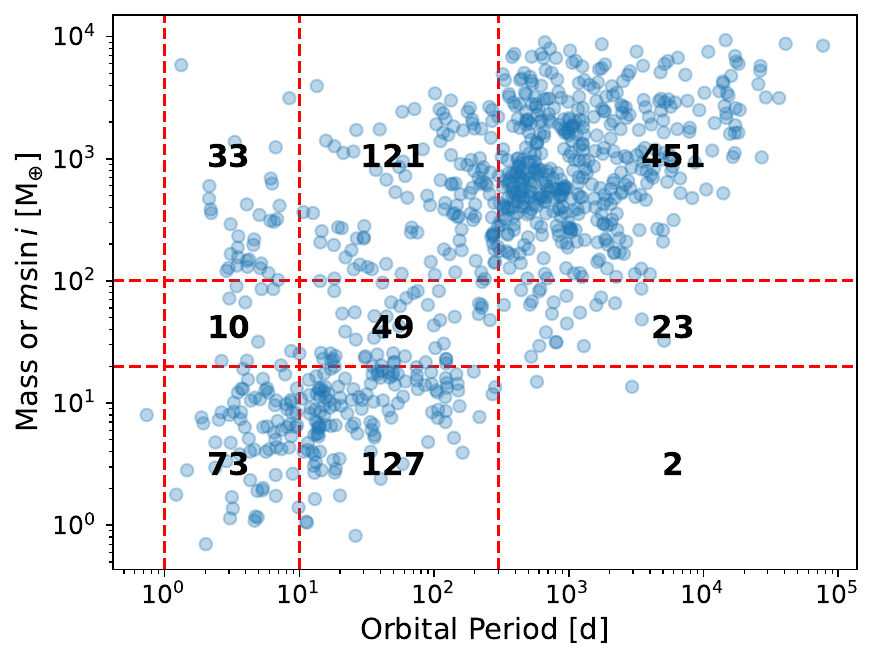}
    \caption{
    The period-mass plane for RV-detected exoplanets in our sample. Dashed lines are drawn on to indicate splits in period (hot, warm, and cold planets) and mass (super-Earth, Neptune-mass, and giant planets). For each bin bounded by the dashed lines, the number of planets in that bin is written on the plot.
    }
    \label{fig:3m3p-plane}
\end{figure}

The two sections above confirm that both period \textit{and} mass are very influential in shaping the eccentricity distribution of RV-detected exoplanets. Clearly one could divide the sample into the 9 regions delimited in Figure~\ref{fig:3m3p-plane}, which shows the period-mass plane with lines indicating the divisions used in Sections~\ref{sec:3pr} and~\ref{sec:3m}.  However, this would require 9 EDFs to be created, and the eccentricity binning would become more influential due to the small populations in most regions. In particular there are too few `hot Neptunes' or `cold super-Earths' (due to the Neptune desert, and our detection sensitivity, respectively) to allow the creation of a useful EDF, or an effective fitting of a prior distribution. For $P-M$ bins containing a sufficient number of eccentricities, the parameters of regressed distributions are reported in Table~\ref{tab:PMplane_fits} to provide a comparison between the populations.

Due to the significant evidence in favour of local parametrisations in Sections~\ref{sec:3pr} and~\ref{sec:3m}, it would be advisable to restrict the sample in both period and mass when deriving a prior distribution for a particular type of planet, where possible. For example, there are plentiful RV detections of `cold Jupiter' type planets, and these could be used to create a prior that would be employed when recovering the signal of such a planet from RV data. This will be the subject of a further study, forming an extension to this work where we will also use HBMs to infer the underlying eccentricity distribution through use of posterior distributions (see Section~\ref{sub:e_errors}).

\begin{table}
    \centering
        \caption{ 
        Parameters for a selection of distributions regressed to EDFs created by splitting the RV-planet sample in both period and mass.
        Of the nine bins in Figure~\ref{fig:3m3p-plane}, four contain a sufficient number of planets (and unique eccentricities) to fit a distribution.}
        \label{tab:PMplane_fits}
    \begin{tabular}{lccccc}
    \hline
    Distribution & $\ln{Z}$ & Param.~1 & Param.~2 & Param.~3 \\
    \hline 
    \multicolumn{5}{c}{\underline{$1<P<10$\,d; $1<M_{\rm p}<20$~M$_{\oplus}$ (\emph{`hot super-Earths'})}}  \\
    Beta $(a,b)$ & $57.48\pm0.07$ & $1.56^{\,+0.11}_{-0.15}$& $9.02^{\,+0.70}_{-1.01}$ & \\
    $\mathcal{RE} (\alpha,\lambda,\sigma)$ & $54.49 \pm0.08$ & $0.41^{\,+0.12}_{-0.12}$ & $6.09^{\,+2.35}_{-1.56}$ & $0.11^{\,+0.02}_{-0.01}$ \\
    Gamma $(\alpha,\beta)$ & $55.61\pm0.07$  & $1.48^{\,+0.07}_{-0.10}$ & $9.47^{\,+0.40}_{-0.73}$ & \\
    \hline
    \multicolumn{5}{c}{\underline{$10<P<300$\,d; $1<M_{\rm p}<20$~M$_{\oplus}$ (\emph{`warm super-Earths'})}} \\
    Beta $(a,b)$ & $101.67\pm0.08$ & $1.78^{\,+0.08}_{-0.11}$& $9.43^{\,+0.41}_{-0.70}$ & \\
    $\mathcal{RE} (\alpha,\lambda,\sigma)$ & $100.79 \pm0.09$ & $0.24^{\,+0.08}_{-0.06}$ & $2.27^{\,+1.27}_{-1.27}$ & $0.11^{\,+0.01}_{-0.01}$ \\
    Gamma $(\alpha,\beta)$ & $98.35\pm0.08$  & $1.67^{\,+0.04}_{-0.05}$ & $9.78^{\,+0.17}_{-0.33}$ & \\
    \hline
    \multicolumn{5}{c}{\underline{$10<P<300$\,d; $M_{\rm p}>100$~M$_{\oplus}$ (\emph{`warm giants'})}}  \\
    Beta $(a,b)$ & $131.86\pm0.08$ & $0.86^{\,+0.07}_{-0.07}$& $3.10^{\,+0.33}_{-0.30}$ & \\
    $\mathcal{RE} (\alpha,\lambda,\sigma)$ & $134.40 \pm0.07$ & $0.96^{\,+0.03}_{-0.05}$ & $4.52^{\,+0.46}_{-0.29}$ & $4.66^{\,+3.71}_{-3.73}$ \\
    Gamma $(\alpha,\beta)$ & $133.09\pm0.08$  & $1.00^{\,+0.09}_{-0.08}$ & $4.16^{\,+0.52}_{-0.48}$ & \\
    \hline
    \multicolumn{5}{c}{\underline{$P>300$\,d; $M_{\rm p}>100$~M$_{\oplus}$ (\emph{`cold giants'})}}  \\
    Beta $(a,b)$ & $500.42\pm0.09$ & $1.08^{\,+0.03}_{-0.03}$& $3.31^{\,+0.12}_{-0.11}$ & \\
    $\mathcal{RE} (\alpha,\lambda,\sigma)$ & $540.24 \pm0.11$ & $0.65^{\,+0.03}_{-0.03}$ & $2.41^{\,+0.16}_{-0.17}$ & $0.13^{\,+0.01}_{-0.01}$ \\
    Gamma $(\alpha,\beta)$ & $532.96\pm0.09$  & $1.35^{\,+0.04}_{-0.04}$ & $5.05^{\,+0.19}_{-0.19}$ & \\
    \hline
    \end{tabular}
\end{table}

\subsubsection{Single or multiple planets}\label{subsub:single-multi}

The eccentricity distribution for transiting \textit{Kepler} planets was recently analysed by \citet{VE19} for FGK stars, with M-dwarfs being included by \citet{Sagear2023} later. Both of these works find that there are clearly different underlying parent eccentricity distributions in cases of either a single transiting planet, or systems containing multiple transiting planets. Both works also conclude that single-planets are more eccentric, whereas multi-planets are closer to circular, in agreement with \citet{Xie2016}, who found that single transiting planets had mean eccentricity $\overline{e}=0.3$, with multiple transiting planets showing a far smaller $\overline{e}=0.04$.

For the RV planets considered herein, we find a very similar result: single and multi-planets have an extremely significant evidence increase ($\Delta\ln{Z}\gg5$) in favour of two local parameterisations, as opposed to a stand-alone global model (Table~\ref{tab:single_vs_multi}). Hence, they are clearly drawn from two separate underlying distributions. Also in agreement with findings from other works, the multis tend to have slightly lower eccentricity, with the EDF increasing faster with eccentricity than it does for single planets. Compact multi-planet systems, like the multiple transiting planets in \textit{Kepler} studies, will have the eccentricity moderated by planet-planet interactions \citep{Juric2008} -- and will also be more resistant to perturbations from external bodies. However, as RV searches can detect planets regardless of their orbital orientation with respect to the observer and are sensitive to longer orbital periods, there may be misaligned outer giants accompanying close in planets included in our sample. These systems could end up contributing larger eccentricities as the inner planet(s) may have their eccentricity pumped by dynamical perturbations, and far out giant planets are less likely to be circular if there has been scattering in the systems history \citep{Carrera2019}. This must be a fairly uncommon architecture within the sample, as we find multi-planets are more circular than single planets. It could also be indicative that for such systems, tidal circularisation of the inner planet happens on a more rapid timescale than the outer planet can drive an eccentric orbit -- which in itself may be dependent on the masses of each interior planet.

The Gamma distribution also provides the highest evidence local function fits for these EDFs, and the $\mathcal{RE}$ comes in at a (relatively) close second. All other functions are far less successful at recreating the shape of the EDFs. On the other hand, when combining the sub-sets back into a single dataset, fitting a single Gamma does not provide the same flexibility that the $\mathcal{RE}$ mixture model can offer, and consequently demonstrates slightly lower evidence.

As in the above study of short and long period planets, fitting a $\mathcal{RE}$ mixture distribution can potentially inform us of the physics behind the demographics. As stated in Section~\ref{sub:RE}, the exponential component is theorised to reflect the effects of tidal dissipation, and the Rayleigh distribution describes the impact of planet-planet scattering \citep{Rasio1996,Juric2008}.

We can see from Table~\ref{tab:single_vs_multi} that the single-planet fit for the $\mathcal{RE}$ distributions is getting close to entirely exponential ($\alpha_{\textrm{s}} \sim 0.9$). If a Rayleigh distribution really does describe the outcome from planet-planet interaction, many single-planets perhaps exist without undergoing a high degree of dynamical interaction, and the dominant forces sculpting observed eccentricities are tides. The multi-planet systems clearly exhibit a greater Rayleigh contribution, potentially indicating scattering occurring in these systems.

In a similar manner to before, we tried fitting a purely exponential model to the single-planet eccentricity EDF. In Table~\ref{tab:single_vs_multi}, the $\mathcal{E}_{\textrm{s}},\mathcal{RE}_{\textrm{m}}$ model displays very similar evidence to the double--$\mathcal{RE}$ fit ($\Delta\ln{Z}=0.05$; $1.13\sigma$). There is therefore insufficient evidence to separate these models, and both are equally suitable for parametrising the single-planets eccentricity EDF. The Rayleigh component can perhaps be disregarded for single-planets, which agrees with the physical interpretation if this model reflects eccentricities modulated by planet-planet scattering \citep{Rasio1996,Juric2008}.

\renewcommand{\arraystretch}{1.4}
\begin{table*}
    \centering
        \caption{Similar to Table~\ref{tab:short_vs_long}, this table lists information for regression to two \textbf{EDFs}, created from either single- or multi-planet systems. Parameter subset `s' denotes singles, and subset `m' denotes multis. Subscript `g' denotes the global model regressed to two combined EDFs.}
        \label{tab:single_vs_multi}
    \begin{tabular}{llllllll}
    \hline
         Distribution & Evidence ($\ln{Z}$) &  Param.~1 & Param.~2 & Param.~3 & Param.~4 & Param.~5 & Param.~6\\
         \hline
         $\mathcal{H}_{1}$ -- Two local models\\
        Beta $(a_{\textrm{s}},b_{\textrm{s}},a_{\textrm{m}},b_{\textrm{m}})$ & $911.81 \pm 0.13$  & $0.97\pm 0.03$ & $3.15\pm0.11$ & $1.26^{\,+0.04}_{-0.03}$ & $5.62^{\,+0.20}_{-0.18}$ &  & \\
        ST08 $(a_{\textrm{s}},a_{\textrm{m}})$ & $876.95 \pm 0.08$ & $3.91\pm0.08$ & $5.75\pm 0.08$ &  & &  & \\
        $\mathcal{RE}$ $(\alpha_{\textrm{s}},\lambda_{\textrm{s}},\sigma_{\textrm{s}},\alpha_{\textrm{m}},\lambda_{\textrm{m}},\sigma_{\textrm{m}})$ & $931.92 \pm 0.16$ & $0.91\pm0.02$ & $3.63^{\,+0.07}_{-0.06}$ & $0.15\pm 0.02$ & $0.63\pm0.02$ & $4.05^{\,+0.14}_{-0.15}$ & $0.113^{\,+0.005}_{-0.004}$\\
        $\mathcal{E}_{\textrm{s}}\,\&\,\mathcal{RE}_{\textrm{m}}$ $(--,\lambda_{\textrm{s}},--,\alpha_{\textrm{m}},\lambda_{\textrm{m}},\sigma_{\textrm{m}})$ & $ 931.87 \pm 0.12 $ & $--$ & $3.66\pm 0.05 $ & $--$ & $0.61\pm 0.03$ & $3.85^{\,+0.21}_{-0.20} $ & $0.110^{\,+0.005}_{-0.004}$  \\
        Gamma $(\alpha_{\textrm{s}},\beta_{\textrm{s}},\alpha_{\textrm{m}},\beta_{\textrm{m}})$ & $950.36 \pm 0.12  $  & $1.18\pm 0.03$ & $4.63^{\,+0.18}_{-0.19}$ & $1.48 \pm 0.04 $  & $7.75^{\,+0.29}_{-0.28} $ & & \\
        \hline
        $\mathcal{H}_{2}$ -- One global model\\
        Beta $(a_{\textrm{g}},b_{\textrm{g}})$ & $765.36 \pm 0.09$  & $1.10\pm 0.02$ & $4.24 \pm 0.11$ & & &  & \\
        ST08 $(a_{\textrm{g}})$ & $736.76 \pm 0.08 $& $4.78\pm0.05$ &   &  &  & \\
        $\mathcal{RE}$ $(\alpha_{\textrm{g}},\lambda_{\textrm{g}},\sigma_{\textrm{g}})$ & $833.28 \pm 0.09$ & $0.66\pm 0.02$ & $3.26\pm0.13$ & $0.108\pm 0.003$ &  &  &  \\
        Gamma $(\alpha_{\textrm{g}},\beta_{\textrm{g}})$ & $816.14\pm 0.09$ & $1.31 \pm 0.03 $  & $6.01\pm 0.17$ &  \\
        \hline
    \end{tabular}
\end{table*}

\subsubsection{Stellar \& Massive outer companions}\label{sub:outer-companions}

We also chose to investigate the effects a companion star has on the eccentricity distribution of RV exoplanets. A massive companion may excite perturbations for an inner planet, driving evolution and impacting the eccentricity of the orbits, via processes such as the eccentric Kozai-Lidov mechanism \citep[see e.g.][for a review]{Naoz2016}. All planets in systems where the NASA Exoplanet Archive indicates the presence of more than one star star can be selected from the sample by selecting those with `\texttt{sy\_snum}' greater than one. 741 planets exist in a system with a single star, whereas 149 planets are in systems with $>1$ star (in `S-type' circumstellar orbits).

Following an identical methodology to the previous sections, we test whether local regression is favoured over a single global fit to the two EDFs, to test the similarity between them. Here though, two local regressions provide no evidence improvement (e.g.~$\mathcal{RE}$: $\Delta\ln{Z}=-3.66$; $3.18\sigma$). This evidently suggests that these two distributions are inherently drawn from a common underlying sample.

Many systems with one star will however have external giant planets, which 
could impact the eccentricities of smaller, closer-in planets in the same system. These giant planets may even have a larger perturbing effect than a binary companion
if there is very wide separation between the two stars. We therefore relax our sub-sample from \textit{stellar} companion, to any type of \textit{massive outer companion}. This selection is performed by taking two cuts on the RV planet archive from Section~\ref{sec:sample}:
\begin{enumerate}
    \item systems with more than one star, as above;
    \item multi-planet systems, where when sorted by orbital period, the outermost planet is more massive than Saturn ($0.3$\,M$_{\textrm{Jup}}$), and is also more massive than the innermost planet.
\end{enumerate}

This results in 280 planets in systems where there is a massive outer companion, and 610 where there is not. Regressing two local $\mathcal{RE}$ functions to the constructed EDFs is $\Delta\ln{Z}=-5$ worse than fitting a single global model. Just as before, we accept the null hypothesis with $3.6\sigma$ significance that these two sub-samples are drawn from the same parent distribution, and posit that this points towards outer companions perhaps not being a main driver for the overall eccentricity demographics we observe. However, \citet{Weldon2024} recently found that Kozai-Lidov oscillations induced by a stellar perturber will be hindered by planetary companions. Therefore, it is worth noting that the multiplicity of these systems could be affecting the results in this section.

\subsection{Host star}\label{sub:host-star}

Recently, \citet{Sagear2023} studied the eccentricity distribution for transiting planets orbiting M-dwarf stars using the so-called ``photoeccentric effect''. The resulting distributions for systems with either a single transiting planet, or multiple transiting planets, were found to be qualitatively similar in comparison with those derived for FGK transiting planets \citep{VE19}. 

To investigate if this also holds true for RV-detected exoplanets, we wanted to split our sample up by spectral type (SpT) of the host star. For a large majority of hosts in our sample, the NASA archive provides no SpT information. We therefore have elected to split the sample up by stellar mass, as 882 of 891 total usable RV exoplanets have a tabulated host mass. This however raises the question of where to place the mass boundaries to approximate a spectral type for each star. Different sources provide different ranges, and there is no one-size-fits-all answer. 
We could have used bounds from works such as \citet{PecautMamajek2013}, though found that many stars classified as spectral types M or G would instead fall in the K-type mass bin.
Without analysing each star individually, and to place our own bounds on the mass bins, we use data from the \textit{Gaia} DR3 astrophysical parameters database \citep{Creevey2023}.

To find these bounds we search for a sample of 50000 stars\footnote{This number was chosen arbitrarily to provide a large sample size, but still be computationally tractable and achievable with synchronous queries to the TAP interface of \href{https://gaia.aip.de/}{gaia.aip.de}.} at each of the F, G, K, and M spectral types (with A-types also included solely for placing an upper constraint on F-type stars). We aim to ensure these stars are main sequence by requiring \texttt{evolstage\_flame} $<420$. This is an integer parameter indicating the evolutionary phase of a star, where values greater than $420$ typically indicate sub-giant stars, and $>490$ are giant stars \citep[more information can be found in][and associated \textit{Gaia} documentation]{Creevey2023}. We extract the spectral type (\texttt{spectraltype\_esphs}) and estimated FLAME mass of each star (\texttt{mass\_flame}), along with the effective temperature (\texttt{teff\_gspphot}).

\begin{figure}
    \centering
    \includegraphics[width=0.99\linewidth]{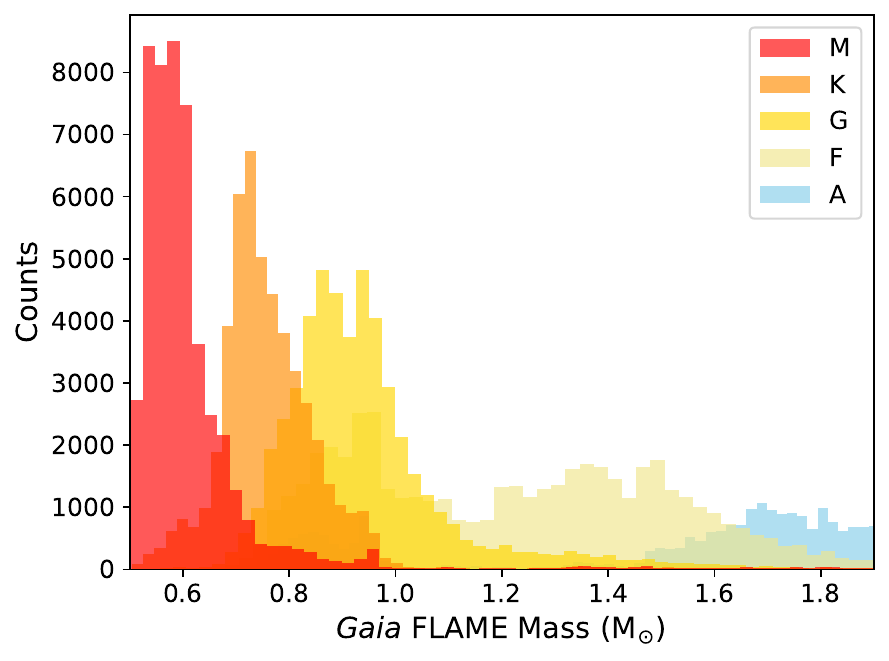}
    \caption{Histograms of stellar masses for each spectral type. We accessed 50000 random stars of each SpT with \texttt{evolstage\_flame} $<420$ from \texttt{gaiadr3.astrophysical\_parameters} table \citep{Creevey2023}. The regions where histograms overlap give approximate bounds for masses associated with each SpT (Tables~\ref{tab:SpT},~\ref{tab:SpT_transit} and~\ref{tab:SpT_Kepler}), to use in the following analysis of the eccentricity distribution for different classes of host stars.} 
    \label{fig:gaia_mass_SpT}
\end{figure}

Plotting mass histograms for each stellar spectral type (Fig~\ref{fig:gaia_mass_SpT}), we see that there is significant overlap, and that spectral type is obviously not clear-cut in mass space. For the purposes of our study, we are therefore assuming an `approximate' spectral type, purely to aid interpretation. The mass boundaries are designated at locations where the decreasing envelope of one histogram intersects with the increase of the next.

Grouping host stars by these masses gives us the samples shown in Table~\ref{tab:SpT}. We have 115 planets hosted by 84 M-type stars, in comparison with \citet{Sagear2023}, who had 163 planets across 101 systems. Dividing the sample into this many sub-samples will result in fewer unique eccentricities and subsequently lower resolution EDFs (and larger DKW error bounds on each bin, scaling with $n^{-\frac{1}{2}}$).

\begin{table}
    \centering
        \caption{Statistics for our RV-detected planet sample from \href{https://exoplanetarchive.ipac.caltech.edu/}{NASA Exoplanet Archive}, when split by stellar masses into corresponding spectral types. Multi-planet fraction here is the fraction of host stars listed in the archive with \texttt{sy\_pnum}, the number of planets in the system, greater than 1 (regardless of detection method).}
        \label{tab:SpT}
    \begin{tabular}{lcccc}
    \hline
    Mass ($\textrm{M}_{\sun}$) & Approx. SpT & Planets & Host Stars & Multi-planet frac. \\
    \hline
    $1.1-1.6$ & F & 256 & 218  & 0.25 \\
    $0.85-1.1$ & G & 288 & 204 & 0.45 \\
    $0.65-0.85$ & K & 125 & 87 & 0.46 \\
    $<0.65$ & M & 115 & 84 & 0.37 \\
        \hline
    \end{tabular}
\end{table}

\begin{figure}
    \centering
    \includegraphics[width=0.90\linewidth]{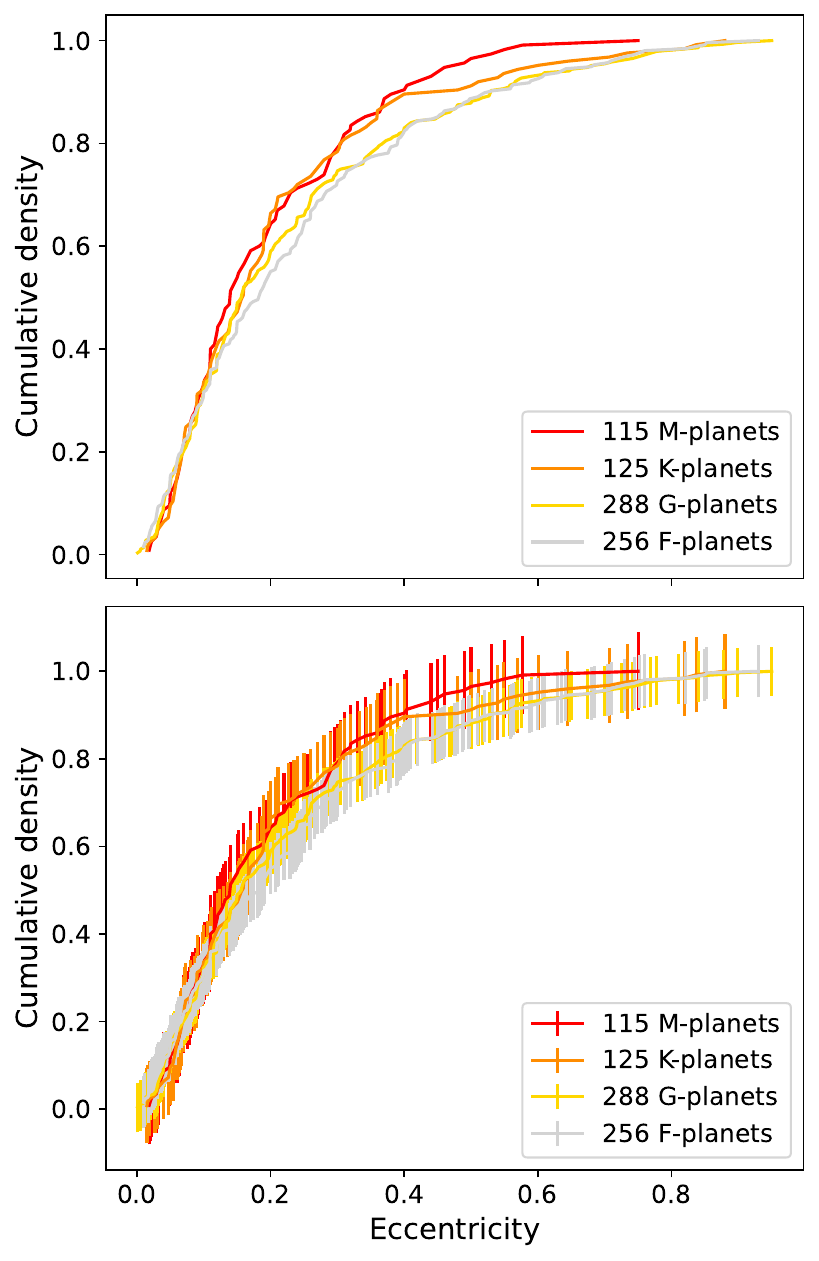}
    \caption{Eccentricity EDFs computed for each inferred spectral type (Table~\ref{tab:SpT}). The EDFs alone are shown in the \textbf{top} panel, and with DKW uncertainties included in the \textbf{bottom} panel.}
    \label{fig:SpT_EDFs}
\end{figure}

For each spectral type in Table~\ref{tab:SpT}, the computed EDFs are displayed in Fig.~\ref{fig:SpT_EDFs}. A qualitative visual inspection casts doubt upon the idea that planets hosted by lower and higher-mass stars have the same eccentricity distribution.

To quantify this, we perform similar tests as in previous sections, computing two EDFs and comparing local function regression against a singular global fit. The tests are performed with the Rayleigh and Exponential mixture distribution, due to its increased flexibility and high evidence ($\ln{Z}$) throughout the preceding parts of this work.

Perhaps as anticipated, a global fit is preferred for G- and F-type stars, indicating they share a common eccentricity distribution. The $\Delta\ln{Z}$ is $\sim8$ ($4.4\sigma$) in favour of a single global model. This evidence change will be due to the additional parameters needed for the local models (twice as many), penalised in evidence calculations -- without improving the actual fit to the EDFs. 

Upon deriving that G- and F-hosted planets share an eccentricity distribution, we can compare either of these with different spectral types, selecting just one to ensure the number of bins (and data points) are approximately comparable between the constructed SpT EDFs. We have chosen to repeat the process with G and M-type stars. This comparison results in lower evidence for $\mathcal{RE}$ distribution regression in the local parameter case, with $\Delta\ln{Z} \sim 5~(3.6\sigma)$ in favour of a single global model.

One would perhaps think that the EDFs in the top panel of Figure~\ref{fig:SpT_EDFs} look clearly different, though the evidence suggests the opposite. In the bottom panel, the DKW error bounds are seen to overlap for all spectral types, providing reinforcement that a single model provides an adequate fit given the uncertainty in the EDF shape. Any minor improvement in fitting is outweighed by doubling the number of free parameters in the model. 

To be consistent with other works on this subject (albeit ones that use transit samples), we can group F, G, and K stars together, and compare the resulting planet eccentricity distribution to that of M-dwarf hosted planets. Stellar structure changes to fully convective in the M-dwarf regime, and consequently, may have an impact on the observed exoplanet eccentricities. Although, as mentioned above, our spectral types are loosely based on approximate masses, this provides a way to compare the least massive stars in the sample with those of higher, almost `solar-type' masses. Regressing two $\mathcal{RE}$ functions to the M or FGK EDFs provides a clearly disfavoured evidence compared to a global fit, with $\Delta\ln{Z} = -5$ ($3.6\sigma$ significance). The results of this fit are shown in Table~\ref{tab:FGKvsM_CDF}.

\begin{table*}
    \centering
        \caption{Regression statistics for \textbf{EDFs}
        created after splitting our exoplanet sample by host spectral type. We report the results of our comparison between FGK and M-type hosted planets, as well as the comparison between FG and KM hosted planets. Note that the global (`g') fits will differ in this table: these are created by adding together two different EDF combinations (FGK + M, FG + KM), and will therefore have different binning and slightly different fitted parameters.
        In each test, two local parameterisations are considered as an alternative to a single global fit.
        }
        \label{tab:FGKvsM_CDF}
    \begin{tabular}{lccccccc}
    \hline
    Case & Evidence ($\ln{Z}$) & Param.~1 & Param.~2 & Param.~3 & Param.~4 & Param.~5 & Param~6. \\
    \hline 
    \multicolumn{8}{c}{\underline{FGK vs M}}  \\
    Global $\mathcal{RE} (\alpha_{\textrm{g}},\lambda_{\textrm{g}},\sigma_{\textrm{g}})$ & $815.21 \pm 0.11$ & $0.72 \pm 0.02$ & $3.33^{\,+0.11}_{-0.12}$ & $0.110\pm 0.004$ \\
    Two $\mathcal{RE}(\alpha_{\textrm{FGK}},\lambda_{\textrm{FGK}},\sigma_{\textrm{FGK}},\alpha_{\textrm{M}},\lambda_{\textrm{M}},\sigma_{\textrm{M}})$ & $809.48 \pm 0.14 $ & $0.72 \pm 0.02$ & $3.30^{\,+0.12}_{-0.11}$  & $0.111 \pm 0.004$ & $0.988^{\,+0.008}_{-0.017}$ & $4.88^{\,+0.21}_{-0.20}$ & $6.63^{\,+2.21}_{-2.90}$\\
    \hline
    \multicolumn{8}{c}{\underline{FG vs KM}}  \\
    Global $\mathcal{RE} (\alpha_{\textrm{g}},\lambda_{\textrm{g}},\sigma_{\textrm{g}})$ & $820.31 \pm 0.11$ & $0.73 \pm 0.03$ & $3.34\pm0.13$ & $0.109\pm 0.004$ \\
    Two $\mathcal{RE}(\alpha_{\textrm{FG}},\lambda_{\textrm{FG}},\sigma_{\textrm{FG}},\alpha_{\textrm{KM}},\lambda_{\textrm{KM}},\sigma_{\textrm{KM}})$ & $828.57 \pm 0.16 $ & $0.89 \pm 0.02$ & $3.87^{\,+0.051}_{-0.050}$  & $0.138 \pm 0.010$ & $0.53^{\,+0.04}_{-0.05}$ & $3.48^{\,+0.24}_{-0.38}$ & $0.111\pm0.004$\\
        \hline
    \end{tabular}
\end{table*}

The evidence does not provide any support for two local parameterisations, so we would assume this to mean that FGK and M-hosted planets share a common eccentricity distribution \citep[in agreement with studies such as][]{Sagear2023}. However, in Table~\ref{tab:FGKvsM_CDF}, we see the regression parameters are not consistent, and most notably, the M-planet CDF looks to be almost fully exponential (and consequently has a very poor fit to the $\sigma_{M}$ parameter). The EDF for FGK stars (or F/G alone) will have a far greater number of unique eccentricities (EDF data points) than M-type, and these could be getting drowned out. The global fit will be driven by the FGK EDF, and local parameterisation will not change much besides adding three additional free parameters into the model. It would be worth investigating this with a larger M-hosted planet sample, or one more homogeneous in sizes between different spectral types. 

\begin{figure}
    \centering
    \includegraphics[width=0.99\linewidth]{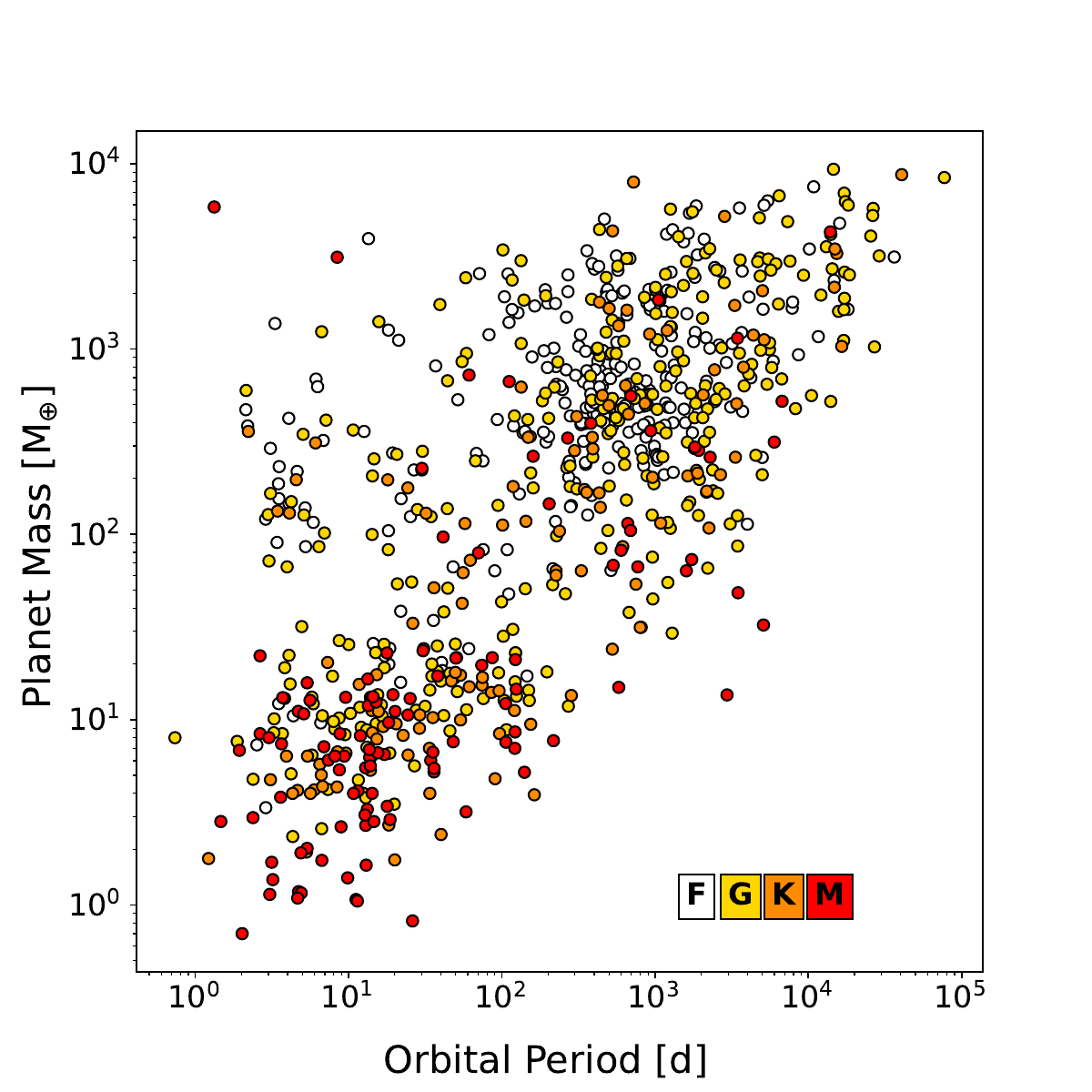}
    \caption{Figure showing the demographics of RV detected exoplanets across the Period-Mass plane. Colours represent mass/approx. SpT of the host (see Table~\ref{tab:SpT}). It is immediately apparent that lower-mass stars tend to host lower-mass planets, typically on shorter period orbits, with a definite lack of external giant planets.}
    \label{fig:mass-period-SpT}
\end{figure}

Potential similarity between populations across differing spectral types may also be due to the types of planets hosted by these stars. In this section we have broadly considered all types of planets. The planet population orbiting M-dwarfs is dominated by low-mass, short-period planets (see Figure~\ref{fig:mass-period-SpT}), and the work in \citet{Sagear2023} also primarily focuses on these planets. If we were to consider only giant planets, they may show a different eccentricity distribution as fewer of these planets form, and are thus less likely to scatter onto high-eccentricity orbits due to interactions between planets. One should therefore also take the mass and period of planets into account, as discussed in Section~\ref{subsub:period-mass}. However, there are currently too few RV-detected giant planets orbiting M-dwarfs for a quantitative comparison. 
Only $18$~per~cent of M-dwarf hosted RV planets are gas giants more massive than $100$~M$_{\earth}$, whereas the fraction increases massively for RV gas giants orbiting G-type stars ($\sim67$~per~cent). An unbiased spectral type comparison is worth pursuing as the subject of future studies.

We also tested the similarity between combined F\&G vs K\&M SpT hosted planet eccentricity EDFs, as these look the most similar in Figure~\ref{fig:SpT_EDFs}. The sample sizes and number of bins for these two joint EDFs is more similar than FGK vs M. We see a preference for local parameterisation, with the two $\mathcal{RE}$ model hypothesis preferred by evidence of $\ln{Z}\sim8~(4.4\sigma)$. As KM-type planet eccentricities seem to be distinct from FG-type planet eccentricities, we now outline any potential reasons the distributions may be dissimilar.

It is well documented that planets orbiting M-type dwarf stars are less likely to be accompanied by an external giant planet, which could perturb inner planet(s) and excite eccentricity or inclination changes in the system. These stars have fewer external giants compared to more massive stars \citep{Cumming2008,Johnson2010,GaidosMann2014}, and also have a lower binarity fraction \citep[see table~1 in][for example]{DucheneKraus2013}. However, our results from Section~\ref{sub:outer-companions} tentatively suggest that the systems where an external perturber is present do not show a quantitatively different distribution of eccentricities to planets without such an outer companion.  

As can be seen from Fig.~\ref{fig:mass-period-SpT}, the RV-detected exoplanets hosted by M-type stars are clustered towards lower-masses and shorter-periods, whereas FGK-planets are scattered more evenly throughout the $P-M$ plane. This highlights potential biases that may affect the shape of the regressed eccentricity distribution. RV searches will be more sensitive to low-mass planets when orbiting lower-mass stars, as these two factors are crucial for determining the scale of the semi-amplitude we observe. Planetary system formation processes could also cause the low-mass stellar host planet distributions to look superficially different from those of higher-mass stars. Massive stars will form from higher-mass disks, and consequently will have a greater amount of mass available for subsequent use in planetary formation \citep*[e.g.][and references therein]{Hinkel2024}. Protoplanetary environment properties scale strongly with stellar mass \citep*{Mulders2015}, such as disk mass, dust evolution, and inner disk radius. The inner edge of the protoplanetary disk changes location depending on the co-rotation radius and the dust sublimation radius \citep{Mulders2015}. For lower-mass stars, these distances are much closer to the star, and hence permit closer in-situ formation, or will trap migrating planets on smaller orbits than an equivalent planet orbiting a more massive star. As M-dwarfs are predisposed to host smaller, closer in planets, the eccentricity distribution may inherently be skewed. We see in Section~\ref{subsub:short-long} that for the sample composed of all spectral types, short period planets are typically less eccentric, so this could play a part in shaping the observed EDF for M-planets.

Planets that orbit close to their host on short orbital periods are likely to circularise over time, due to tidal dissipation of orbital energy in either the host star, or planet itself \citep{Mathis2015}. Dissipation will impact the orbital configurations \citep{Rasio1996}, and could feasibly vary depending on the host star. The tidal energy absorbed due to friction in the convective envelopes of stars may vary with mass, as the fraction of the stellar material forming the convective envelope will increase towards lower masses. The M-dwarfs (below masses of $\sim 0.35$~M$_{\sun}$) have a fully convective structure  \citep{CB1997}. \citet{Mathis2015} found that for stars on the main sequence, the tidal quality factor (Q) of the convective envelope was at a minimum for 0.5--0.6~M$_{\sun}$ stars, where a lower Q means dissipation is stronger and evolution is more rapid. If low-mass stars are able to dissipate orbital energy more efficiently in their outer envelopes, the system architecture of short-period planets may be affected in a diverse manner across the host-mass continuum.

\section{PDF Regression}\label{sec:PDF}

This method has been implemented in \textsc{pymc} to draw posterior samples from given priors via computation of each distribution's likelihood in light of the eccentricity data. For the likelihoods, we use the in-built distributions for Beta, Gamma, and exponential, and a special case of \texttt{Weibull.dist($\alpha$=2, $\beta=\sqrt{2} * \sigma$)} which is exactly equivalent to the Rayleigh distribution.

To calculate the evidence and facilitate model comparison in an MCMC framework, we use the Sequential Monte Carlo sampler functionality of \textsc{pymc}. This transforms prior into posterior through steps of annealed sequences, and a creates the marginalised likelihood, or evidence, as a by-product.

\subsection{Results}
Estimating distribution parameters directly from the PDFs in light of the measured data gives us another way to check which best fits the observed eccentricity population. Using the Beta, Rayleigh+Exponential, and Gamma distributions, we estimate the evidence for each fit to find which function is preferred. 
Note that we do not include either Kumaraswamy or ST08 distributions in this section: Kumaraswamy is almost identical to Beta (for our purposes), and the ST08 distribution provides consistently low evidence throughout previous sections of this work, so is excluded for simplicity. Fitting our sample, the results are similar to Section~\ref{sec:EDF}: Beta is the least preferred, and Rayleigh+Exponential and Gamma are the most appropriate for this sample. However, Gamma now has the highest global evidence, out-performing the $\mathcal{RE}$ mixture distribution, by $\Delta\ln{Z}=4.16$: $3.35\sigma$ significance (Table~\ref{tab:all_ecc_results_PDF}). This constitutes a moderate, non-definitive evidence increase \citep[on the scale of][]{Trotta2008}, yet is the preferred model, as it benefits from additional simplicity with fewer parameters and no mixture element. 

Any potential differences between the two methods and why Gamma becomes the highest global evidence model will be discussed in Section~\ref{sub:EDFvsPDF}.
For the remainder of this Section however, we shall adopt Gamma for our analysis, and investigate how splitting the sample up and fitting local functions, as in previous sections, could help differentiate between origin/evolution of exoplanet sub-populations.

\renewcommand{\arraystretch}{1.4}
\begin{table*}
    \centering
        \caption{Similar to Table~\ref{tab:all_ecc_results}, this table reports statistics for eccentricity regression to function \textbf{PDFs}, when varying model choice. Evidence values (favoured models being more positive) are shown along with the best-fit parameters, where \textsc{arviz} (a Python package for analysis of Bayesian models) summaries are used on the \textsc{pymc} chains to determine values.}
    \begin{tabular}{lllll}
    \hline
         Distribution & Evidence ($\ln{Z}$) &  Parameter 1 & Parameter 2 & Parameter 3\\
         \hline
        Beta $(a,b)$ & $ 448.16 \pm 0.04  $ & $1.06 \pm 0.04 $ & $ 3.54\pm 0.17$ &  \\
        $\mathcal{RE}~(\alpha,\lambda,\sigma)$ & $ 474.20 \pm 0.10 $ & $0.68  \pm 0.05 $ & $ 3.32 \pm 0.25 $ & $ 0.11 \pm 0.01 $ \\
        Gamma $(\alpha,\beta)$ & $ 478.36 \pm 0.06 $  & $ 1.44\pm 0.06 $ & $6.50\pm 0.33$ &  \\
        \hline
    \end{tabular}
    \label{tab:all_ecc_results_PDF}
\end{table*}

In Table~\ref{tab:altmethod_similaritytests} we report the change in evidence for fitting two local Gamma distributions, over one single local model. In a similar vein to Section~\ref{sec:EDF} and the EDF method, we find that short and long period planets (split at the median) are significantly distinct, as are the distributions of planets in single and multi-planet systems (with evidence improvement here just below the $\Delta\ln{Z}=5$ strong-evidence threshold). These populations may require independent parameterisations as it clear they are not drawn from the same parent eccentricity distribution, quantified by the Bayes factors. We also find, similar to before, that splitting the sample up by those with either a stellar or massive outer planetary companion in the system provides no improvement, and that the presence of a known massive outer perturbing object does not impact the shape of the distribution. For different combinations of host spectral types, all planets again seem to be drawn from a common distribution. The low number of M-dwarf-planets compared to more massive spectral types may influence the results, and will require future study.

\renewcommand{\arraystretch}{1.4}
\begin{table}
    \centering
        \caption{Results for model comparison between a single, global Gamma distribution fitted to observed eccentricities, and two local functions used on sub-sets of the sample, split as indicated below and in Section~\ref{sec:EDF}. The Gamma distribution was adopted here as it provides the highest evidence fit. Change in evidence/BF is listed, where positive values favour two local parameterisations, and negative values indicate preference for a single global model.}
    \begin{tabular}{lc}
    \hline
         Sub-sets & Evidence for `local' fits ($\Delta\ln{Z}$) \\
         \hline
         Short \& long-period & $16.04$  \\
         Single \& multi-planets & $4.42$   \\
         With \& without Massive companion & $-3.36$   \\
         Three period regimes & 20.49 \\
         Three mass regimes & 18.00  \\
         FGK \& M-hosted & $0.19$ \\
         FG \& KM-hosted & $0.90$ \\
        \hline
    \end{tabular}
    \label{tab:altmethod_similaritytests}
\end{table}

\section{Discussion}\label{sec:discussion}

\subsection{EDF vs PDF}\label{sub:EDFvsPDF}
In the two preceding sections, we have investigated two methods for regressing functional forms to the exoplanet eccentricity distribution. On the whole, the results are fairly consistent. In both methods, Bayesian evidence model comparison no longer prefers the Beta distribution. In detail there are however differences between the methods.

Both methods agree that the Gamma distribution, or Rayleigh + Exponential mixture distribution are by far the two `best' models to use when performing a global fit. The EDF method favours use of the $\mathcal{RE}$ by $\Delta\ln{Z}=8.97$ (strong evidence, $4.63\sigma$), whereas PDF method favours Gamma by $\Delta\ln{Z}=4.16$ (just below the strong evidence threshold, $3.35\sigma$ significance). Performing multiple local regressions, the Gamma distribution is always preferred, irrespective of sub-sample cuts or regression method (EDF or PDF). We also find, that for some of the distributions, the fitted parameters are not consistent. Comparing between Tables~\ref{tab:all_ecc_results} and~\ref{tab:all_ecc_results_PDF}, we see that only the  $\mathcal{RE}$ distributions are fully consistent to within $1\sigma$ uncertainties on the derived parameters.

The EDF method preferring $\mathcal{RE}$ for global regression, but Gamma for local regression is interesting. As we briefly mention in Section~\ref{subsub:single-multi}, this may be down to how we split the sample and then create the EDF. When dividing the sample into subsamples that have a good physical reasoning (e.g., the distribution of short and long period planets is likely to be different), the Gamma distribution performs better at modelling each individual EDF. When combined into a single EDF, and we have two competing groups with their own physical drivers, the $\mathcal{RE}$ distribution provides greater flexibility due to being a mixture model, and may provide higher evidence when modelling the merged populations that are each sculpted by distinct processes. 

There will be inherent differences due to the representation of the eccentricities as an EDF, compared to using them directly in the PDF method. However, there may be other reasons why the methods can have slight disagreement. We tested the recovery of known distributions for each method by generating five sets of random samples from a fixed Beta distribution (with the same number of observations as the number of observed eccentricities), and fitting each sample with either EDF or PDF regression. The EDF method always gives far tighter posteriors, but they can be offset from the `true' parameter values used to generate the distribution by a significant amount, potentially indicating that the EDF uncertainties are too small. The PDF method parameter posteriors tend to be far broader, and always include the true values. 

This is probably related to the error treatment. The PDF method does not yet incorporate any errors, but the EDF method does, in a slightly indirect way: the DKW error interval we used as an update to the Poisson errors originally used by \citetalias{kipping2013}. As the DKW error is equal for every EDF bin, it seems that the regressed CDF can not change shape much, but instead is relatively free to move up and down within the error bars. These distributions will be nearly the same, but shifted, leading to small variations in the parameters. The optimal method is yet to be determined, but we note that the PDF method provides larger parameter errors, including the true value more often in our above tests.

It will therefore be justified to re-run the analysis in this paper, including any cuts on the eccentricity sample for a specific problem, and make judgements upon which distribution would be best to use as a prior on a case-by-case basis. For this reason, we make our code available to the community. 

\subsection{Bespoke functions for the eccentricity prior in RV analysis}

At high significance, we have shown that period, (minimum) mass, and multiplicity all strongly affect the underlying eccentricity distributions of exoplanets. This has prompted us to reconsider the functions we use as a prior in our RV exoplanet parameter inference techniques. If, for instance, one cannot use a broad period prior due to sampling/time-span constraints, or instability zones created by other bodies (e.g. for planets in either circum-primary or circum-binary systems: \citealt{Barnes2020,Stevenson2023a,Standing2023,Sairam2024}), it would be best to also use an eccentricity prior with parameters informed by demographics of an exoplanet sub-population. As can be seen in \citetalias{kipping2013} and our analysis above, it is statistically significant to perform these local fits, and will not always suit the problem to use a distribution describing the entire population. Therefore, we have developed a Python code, \textsc{eccentriciPy}, to not only model up-to-date global properties of the eccentricity distribution (downloading the most recent version of an exoplanet archive), but to take user defined prior information on specific planetary systems. The code will then produce all statistics and comparisons needed, allowing the user to select the optimal eccentricity prior in light of any constraints on their data.

\subsection{Implementation in \textsc{kima}}
To test the effects of a different eccentricity parameterisation, compared to the traditional use of \citetalias{kipping2013}'s Beta, we performed trials with the \textsc{kima} code \citep{kima-joss,kima-ascl}, comparing the posteriors when recovering known planets from RV data. For these tests, we chose to use the mixture of Rayleigh + Exponential distributions because this is the only distribution that has consistent parameters between regression methods. We adopt ($\alpha, \lambda, \sigma$) from Table~\ref{tab:all_ecc_results_PDF}, as the PDF method typically provides parameter estimates with more conservative errors.

The $\mathcal{RE}$ mixture distribution is now included with the \textsc{kima} package (version 6.0.20 onwards), and can be employed when defining planetary eccentricity priors with \texttt{ExponentialRayleighMixture($\alpha$,scale=$1/\lambda$,$\sigma$)}, from the \texttt{kima.distributions} module.

Additionally, we have also tested the posterior response for these systems when employing a uniform prior. Exoplanetary literature has moved away from uniform priors in favour of those that represent what we may expect to find \citep[e.g.][]{Faria2016,Standing2022}, though it is worth assessing what happens when we do not apply constraining prior knowledge. This is discussed in Appendix~\ref{app:uniform_prior}.

\subsubsection{Case study \uppercase\expandafter{\romannumeral1\relax}: 51 Pegasi b}
To test the $\mathcal{RE}$ distribution as a prior, we first considered a simple example, to assess how the prior could change the posterior distribution. We use the example included with the original \textsc{kima} package \citep{kima-joss} of 51 Peg b, where 256 Hamilton Spectrograph RVs \citep{Butler2006} are used to demonstrate the ability to re-discover the first exoplanet found orbiting a Sun-like star \citep{MayorQueloz1995}. We use identical parameters to the example, but modify the eccentricity prior to be either $\mathcal{RE}\,(\alpha=0.68,\lambda=3.32,\sigma=0.11)$, or a Kumaraswamy with parameters chosen to match \citetalias{kipping2013}'s Beta distribution ($\alpha=0.881,\beta=2.878$, see Section~\ref{sec:functions}). These are kept constant for all studies in this section. We also increase the number of samples from 5000 to 50000 to improve the resolution of the final posterior. 

\begin{figure}
    \centering
    \includegraphics[width=0.80\linewidth]{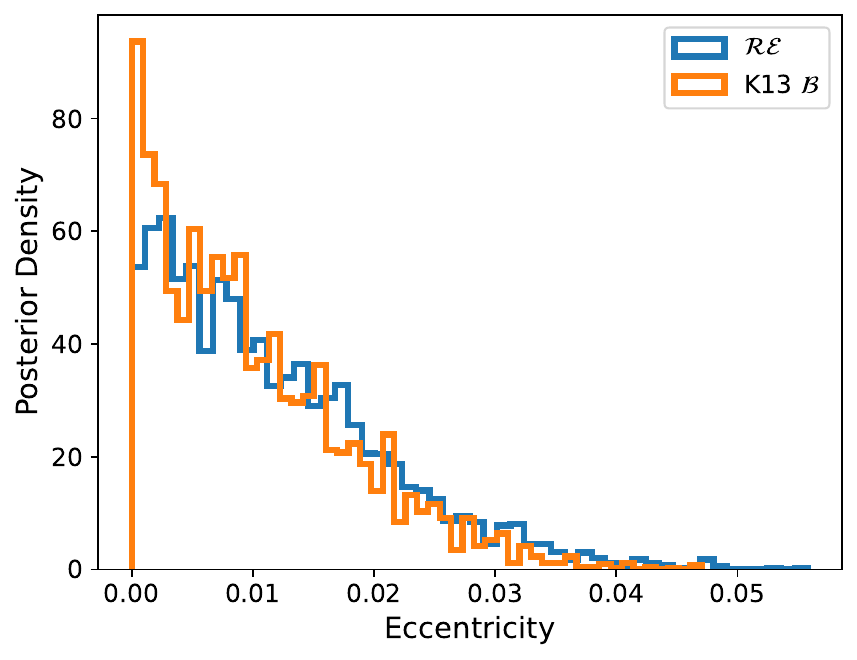}
    \caption{
    Eccentricity posterior histograms for \textsc{kima} runs of 51 Peg b, with two different eccentricity priors.
    }
    \label{fig:51Peg_ecc}
\end{figure}

The histograms of posterior samples are shown in the top panel of Figure~\ref{fig:51Peg_ecc}, for either choice of prior. It can be seen that for a planet on an approximately circular orbit, the choice of prior has a very small effect. Taking the median and $1\sigma$ credible bounds from \textsc{corner} \citep{corner}, we find $e=0.0098^{\,+0.0108}_{-0.0071}$ and $e=0.0082^{\,+0.0099}_{-0.0063}$ for $\mathcal{RE}$ and Beta respectively -- fully compatible well within $1\sigma$.
The two \textsc{kima} runs had comparable execution times, confirming that changing the prior does not drastically alter the computational cost.

\subsubsection{Case study \uppercase\expandafter{\romannumeral2\relax}: DMPP-3}
To further test the effect a new prior has on the posterior, we can apply the above method to a radial velocity detection with a clearly defined high eccentricity, rather than a near-circular-orbit planet. 

DMPP-3 (HD\,42936) is a planet-hosting binary star system consisting of K-type primary (A) and secondary (B) that is just massive enough to fuse hydrogen \citep[][]{Barnes2020,Stevenson2023a}. The binary orbit has a very well-constrained eccentricity, and an extremely high signal-to-noise ratio, suiting the purposes of this study. 

\begin{figure}
    \centering
    \includegraphics[width=0.80\linewidth]{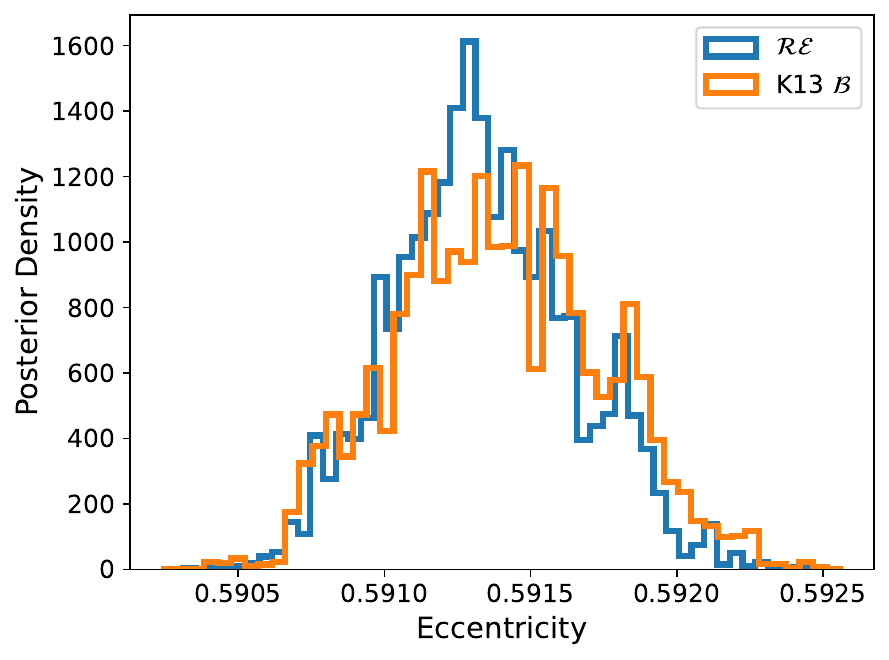}
    \caption{
    Eccentricity posterior histograms for \textsc{kima} runs of DMPP-3, with two different eccentricity priors. 
    }
    \label{fig:DMPP3_ecc}
\end{figure}

We use a dataset of 115 HARPS RVs, reduced with the \textsc{s-bart} pipeline \citep[see][]{S-BARTpaper}. The \textsc{kima} code is run with either $\mathcal{RE}$ or Beta prior for 100k samples, and resulting posterior histograms for the eccentricity of the DMPP-3B orbit are shown in Figure~\ref{fig:DMPP3_ecc}. Median values and one-sigma uncertainties are as follows: $e_{\mathcal{RE}}= 0.591319^{\,+0.000336}_{-0.000287}$; $e_{\mathcal{B}} = 0.591387^{\,+0.000392}_{-0.000349}$. The Beta prior results in a posterior that is ever-so-slightly wider, with a noisier distribution peak.
Most importantly, these results are extremely consistent when using either prior, the ideal outcome when recovering well-resolved eccentricities. Introducing this new prior does not seem to negatively affect the values inferred by orbit fitting routines.

\subsubsection{Case Study \uppercase\expandafter{\romannumeral3\relax}: Multi-super-Earth system}

The use of an alternate eccentricity prior may have consequences for systems with multiple, low-amplitude signals, where eccentricity is not well constrained and posterior sampling is more likely to return the prior. 

We have chosen to investigate a published compact multi-super-Earth system, HD\,39194, which was fully characterised by \citet{Unger2021}. These planets are on `mostly circular' orbits, with one-sigma eccentricity upper limits of $e<0.105,0.078,0.174$ for planets b, c, and d, respectively. In the analysis by \citet{Unger2021}, planet d's eccentricity posterior was found to not significantly deviate from the input prior distribution. The planetary orbital periods are $P=5.64,14.03,33.91$, and minimum masses are $M_{\textrm{p}}\sin{i}=4.0,6.3,4.0~\textrm{M}_{\earth}$.

In a \textsc{kima} run of 100000 samples, we use 288 HARPS RVs, reduced with \textsc{s-bart}. 
Summary statistics (median and one-sigma bounds) for the resulting eccentricities are listed in Table~\ref{tab:HD39194_result}. It seems that the $\mathcal{RE}$ prior consistently provides higher eccentricities. However, when such a posterior is bounded up against zero, these statistics do not provide as much information and are skewed due to the hard boundary \citep{KippingWang2024}.

\begin{table}
    \centering
        \caption{Median and one-sigma eccentricity uncertainties for \textsc{kima} runs of HD\,39194, when varying the choice of prior.}
        \label{tab:HD39194_result}
    \begin{tabular}{lcccc}
    \hline
    Planet & Beta prior eccentricity & $\mathcal{RE}$ prior eccentricity  \\
    \hline
    HD\,39194\,b & $0.053^{\,+0.059}_{-0.040}$ & $0.066^{\,+0.061}_{-0.043}$ \\
    HD\,39194\,c & $0.032^{\,+0.039}_{-0.024}$ &  $0.045^{\,+0.039}_{-0.029}$ \\
    HD\,39194\,d & $0.087^{\,+0.109}_{-0.065}$ & $0.101^{\,+0.082}_{-0.064}$ \\
        \hline
    \end{tabular}
\end{table}

\begin{figure}
    \centering
    \includegraphics[width=0.80\linewidth]{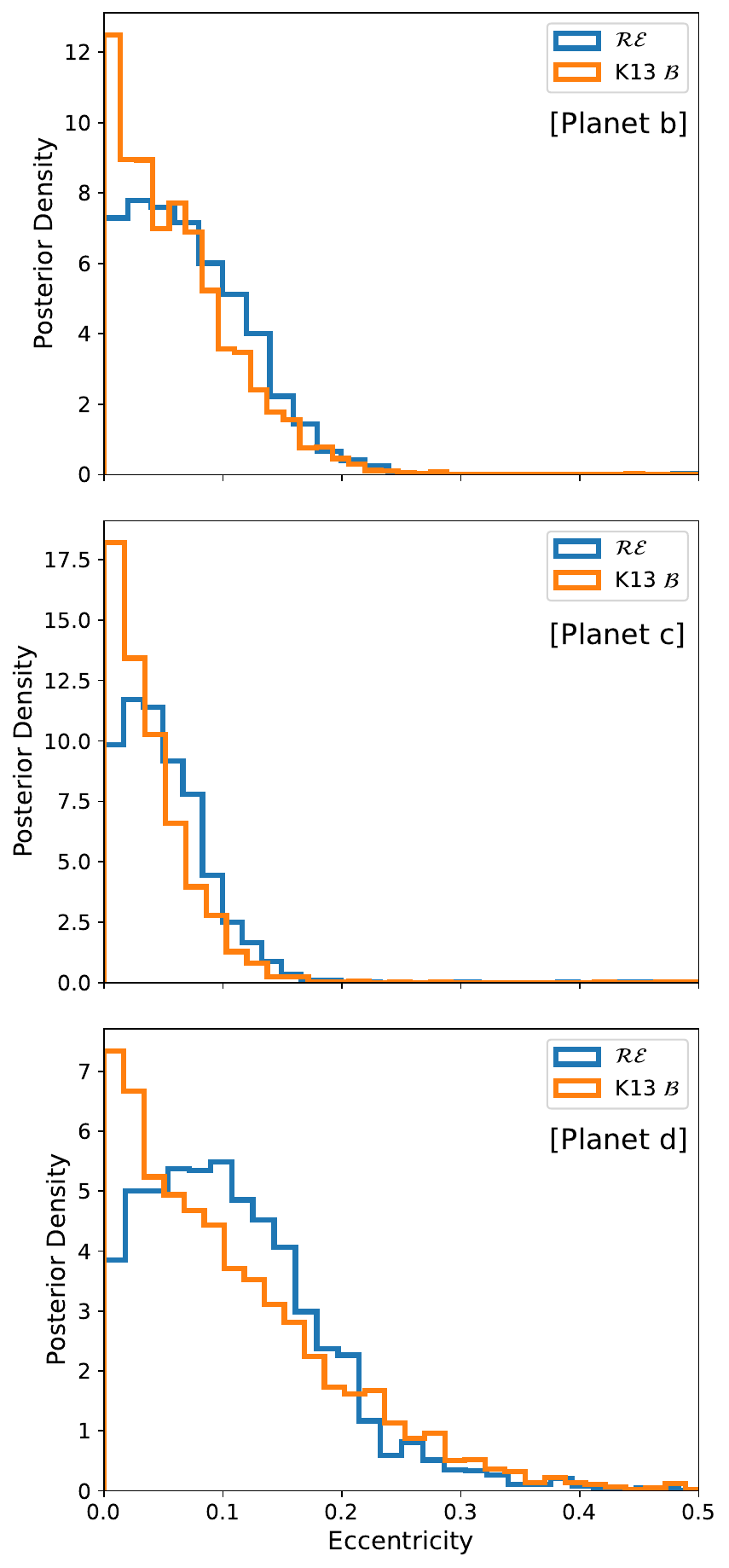}
    \caption{
    Eccentricity posterior histograms for \textsc{kima} runs of HD\,39194, with two different eccentricity priors. \textbf{Top:} Eccentricity posteriors for planet b, with $P=5.64$~d. \textbf{Middle:} Eccentricity posteriors for planet c, with $P=14.03$~d. \textbf{Bottom:} Eccentricity posteriors for planet d, with $P=33.91$~d \citep{Unger2021}.
    }
    \label{fig:HD39194_ecc}
\end{figure}

Looking directly at the posterior distributions plotted in Figure~\ref{fig:HD39194_ecc}, we can see differences starting to emerge depending on choice of prior. Again, the $\mathcal{RE}$ prior leads to distributions shifted towards higher eccentricity values, and the distributions also appear slightly wider. These posteriors' peaks are slightly offset from zero, rather than exhibiting a peak at zero and continual decrease seen with the Beta prior. This hints at the first steps towards resolving small but non-zero eccentricities, a possible improvement over the previous best practice of quoting upper limits. This is probably due to the updated parameterisation, where the population peaks away from zero (Figure~\ref{fig:ecc-hist}), and due to our removal/updating of planets where eccentricity was fixed to zero, thus reducing bias towards circular orbits. Even with many data points, the prior may have a significant impact on the posterior distribution. When interested in small but non-zero eccentricities, it therefore emphasises the need to use the best available priors.

\subsubsection{Recovery of injected eccentricity: 51 Peg}

The tests in these three case studies, particularly that for HD\,39194, show that using the $\mathcal{RE}$ prior results in a slightly different recovered posterior. However, when analysing observed data we do not know the actual underlying eccentricity or distribution that we wish to reproduce or extract. In this section, we perform analysis on simulated data. We assess the posterior response for each prior when we arbitrarily define the eccentricity. For the 51 Peg and HD\,39194 systems, RVs were simulated for the Keplerian signals with specified eccentricities, at the same observation epochs as the original datasets (256 RVs from \citealt{Butler2006} for 51\,Peg, and 288 RVs from the ESO archive for HD\,39194).
For eccentric orbits, the argument of periastron is simulated using the value obtained from the maximum likelihood solution of the \textsc{kima} runs on the real datasets.
Error bars are informed by the datasets, and Gaussian white noise scatter is added to each simulated RV set, with standard deviation equal to the error bar size. We repeated each simulation and recovery test with multiple sets of randomly generated Gaussian noise applied to the simulated RVs.

\begin{figure*}
    \centering
    \includegraphics[width=0.33\linewidth]{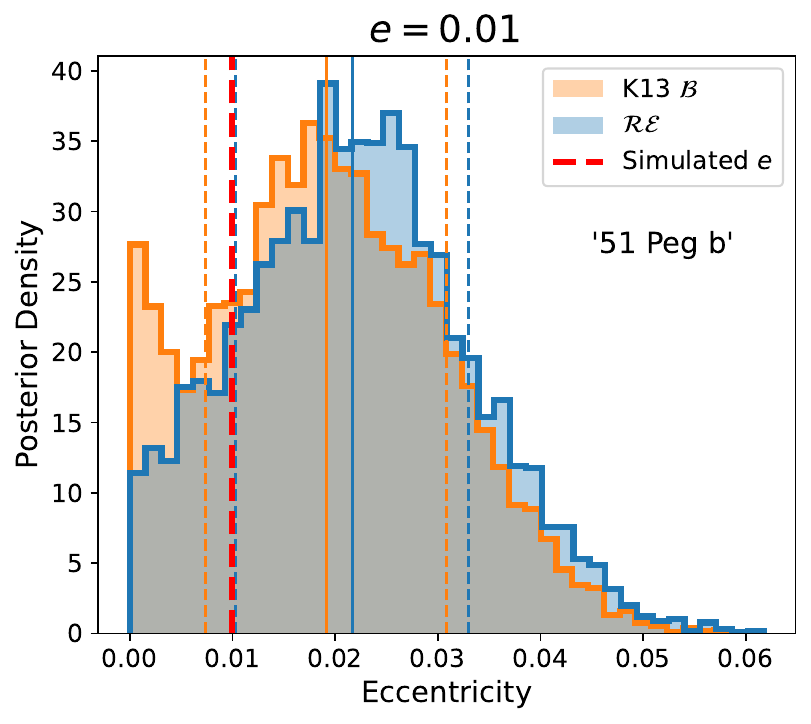}\includegraphics[width=0.33\linewidth]{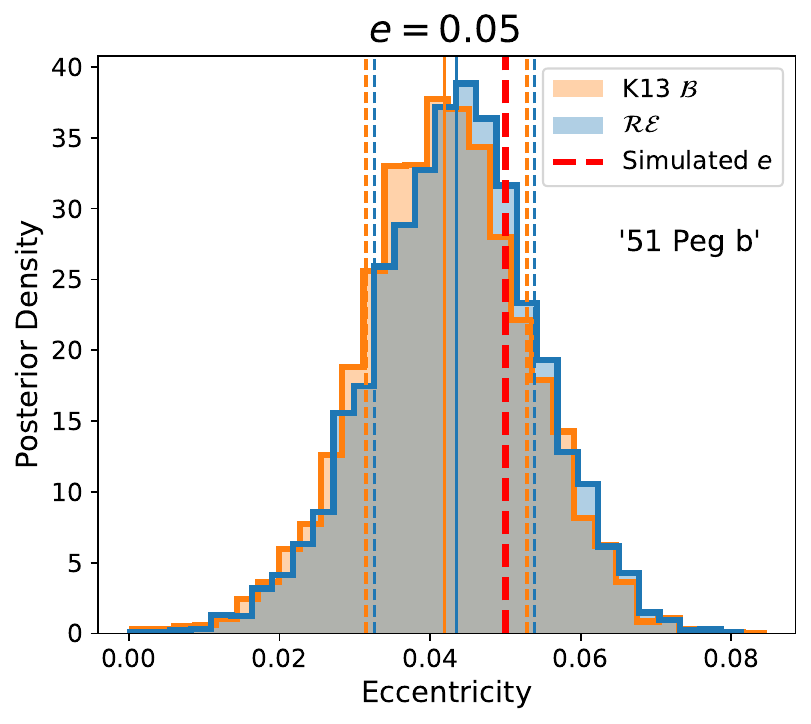}\includegraphics[width=0.33\linewidth]{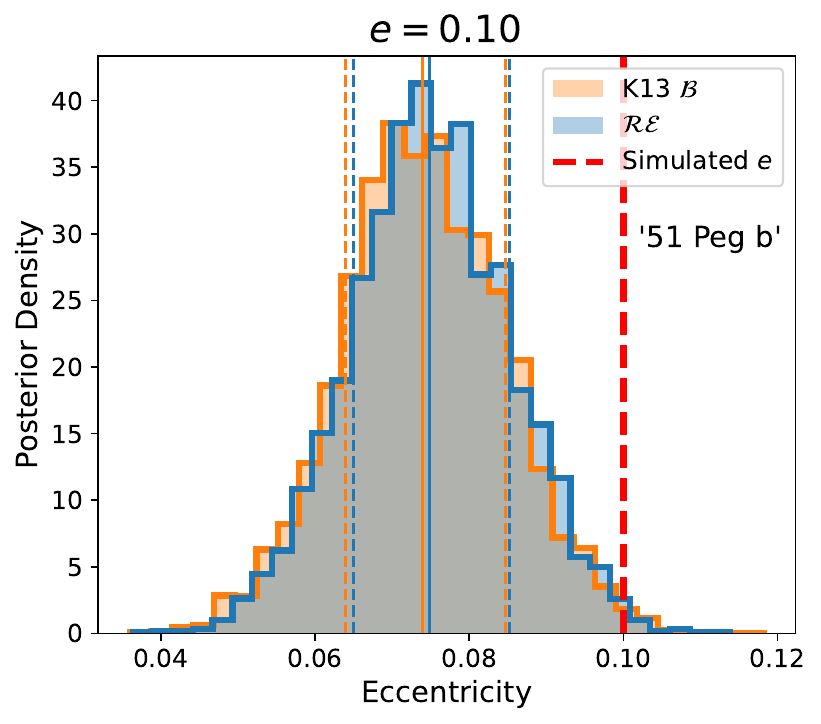}
    \caption{
    Eccentricity posterior histograms for \textsc{kima} runs of 51~Peg\,b, with two different priors. RVs are simulated in an attempt to recover ground truth eccentricities. Results are shown for signals with $e=0.01,0.05,0.10$  (\textbf{left to right}).
    }
    \label{fig:51pegSims}
\end{figure*}

Starting with 51 Peg b, we simulated various small eccentricity values, increasing until the posterior peaks were well resolved and recovery allowed clear distinction from a circular orbit. Some of the posteriors for trialled values are shown in Figure~\ref{fig:51pegSims} ($e=0.01,0.05,0.10$). In each case, the median and $1\sigma$ bounds of the distributions are mostly consistent, though the shape is not always. In the leftmost panel of Fig.~\ref{fig:51pegSims}, we see that when using the $\mathcal{RE}$ prior, the posterior peak seems to be moved further from zero, with fewer samples indicating a completely circular orbit. Both posteriors overestimate the simulated value ($e=0.01$), though using the $\mathcal{RE}$ prior 
gives slightly less weight to a circular orbit interpretation, with fewer samples at $e=0$ -- despite being generally similar.
This posterior shape may warrant investigation, for further study into consequences of a non-circular orbit \citep[e.g. tidal heating;][]{Sing2024,Welbanks2024}. However, the ground truth eccentricity lies just outside the $1\sigma$ error estimate for the $\mathcal{RE}$-posterior, and just within for the Beta-posterior. 

In some simulations, the scatter we apply to simulated RVs, to make them more realistic than a perfect Keplerian signal, can make a very slightly eccentric orbit (e.g. $e=0.01$) appear circular. Different random scatter initialisations causes this to fluctuate, but in all cases the $\mathcal{RE}$ posterior provides larger upper limits as well as being consistent with the overall interpretation from the Beta posterior. The $\mathcal{RE}$ is therefore not drastically influencing the shape when there is no reason to.

The middle panel of Fig.~\ref{fig:51pegSims} ($e=0.05$) shows little difference between prior choice, and both distributions provide a good estimate of $e$. The right panel ($e=0.10$) is included as it shows both posteriors are very similar, yet are consistently offset from the ground truth value, by far more than one sigma. This is due to the precision and random scatter of the simulated measurements. The amount the median is offset from the truth is not constant, and varies for different random noise initialisations. When we re-simulated the RVs for all three eccentricity cases (e=0.01, 0.05 and 0.10) as if they were all taken by ESPRESSO (optimistic 0.1~m\,s$^{-1}$ error bar, and correspondingly small white noise added), the eccentricities become fully aligned with the ground truth to within very narrow $1\sigma$ confidence intervals.
 
\begin{figure}
    \centering
    \includegraphics[width=0.80\linewidth]{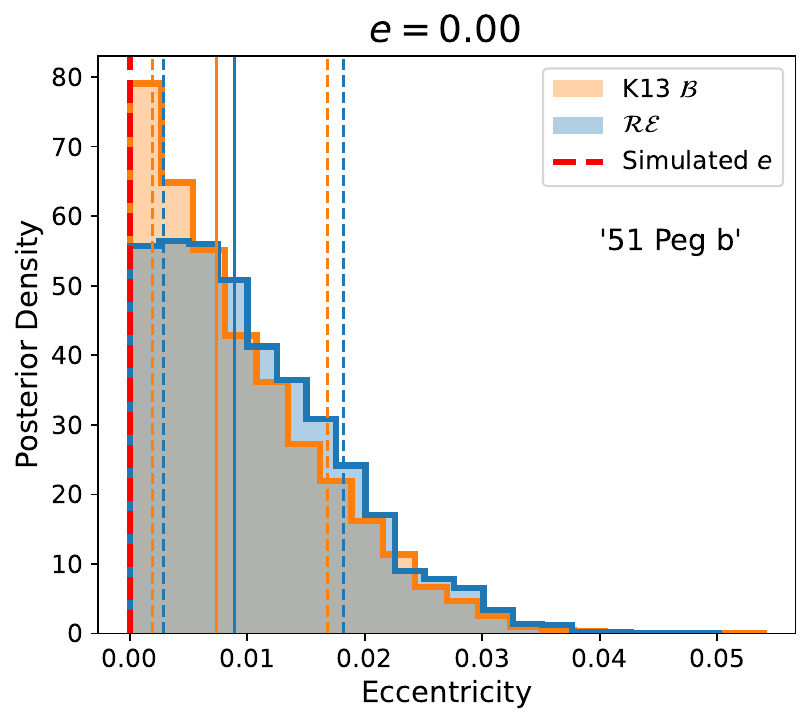}
    \caption{
    As for Figure~\ref{fig:51pegSims}, but with RVs simulated for an exactly circular 51~Peg\,b Keplerian. Neither posterior shows significant difference from the circular orbit interpretation, and shows that the new prior is not overly biasing results away from $e=0$.
    }
    \label{fig:51peg_circ_Sims}
\end{figure}

This does however raise a question. What happens if the orbit is exactly circular? Our updated parameterisation with the $\mathcal{RE}$ prior reflects the current exoplanet population, where the peak in eccentricity is shifted from zero (cf. Fig.~\ref{fig:ecc-hist}). The previously used Beta distribution prior has a peak at $e=0$ (fig.~2 in \citetalias{kipping2013}), so the two priors may treat truly circular planets differently. Accordingly, we repeat our simulations for 51~Peg\,b, but generate RVs for a Keplerian with $e=0$, to assess any non-zero bias. The simulated data retains error and scatter dictated by the mean error on the existing data set. Fig.~\ref{fig:51peg_circ_Sims} shows the posteriors for the aforementioned simulated results, using either prior. The Beta-posterior still peaks at zero as expected, whereas the $\mathcal{RE}$-posterior is almost flat in density at the lowest eccentricities, before dropping off. The shape is broadly similar, and neither rules out the circular-orbit interpretation -- either quantitatively or qualitatively. The main functional difference between these posteriors is that the $\mathcal{RE}$ prior gives a slightly higher one-sigma upper limit estimate. Depending on how the simulated random noise falls within the RV time-series, it can emulate a very slightly eccentric orbit, and the \textsc{kima} recovery posteriors peak away from zero. The response from either prior is very similar, with both $\mathcal{RE}$ and Beta agreeing on a non-zero peak. This highlights the importance of precise observations, and reducing scatter on observations by accurately modelling noise sources such as stellar activity.

In the regime of poor quality/ uninformative data, replicating the prior may lead to an over-estimated eccentricity through use of the global sample $\mathcal{RE}$ prior. Returning the prior would prompt one to take the eccentricity results with a pinch of salt regardless, and will always justify further investigation to break free of the low information regime. Additionally, slightly higher eccentricity estimates err on the side of caution from a planetary system stability consideration. If we do not know that eccentricities truly are zero, then assessing the stability of prior-biased presumed circular orbits may permit configurations that are in-fact unrealistic. 

\subsubsection{Recovery of injected eccentricity: HD\,39194}

\begin{figure*}
    \centering
    \includegraphics[width=0.305\linewidth]{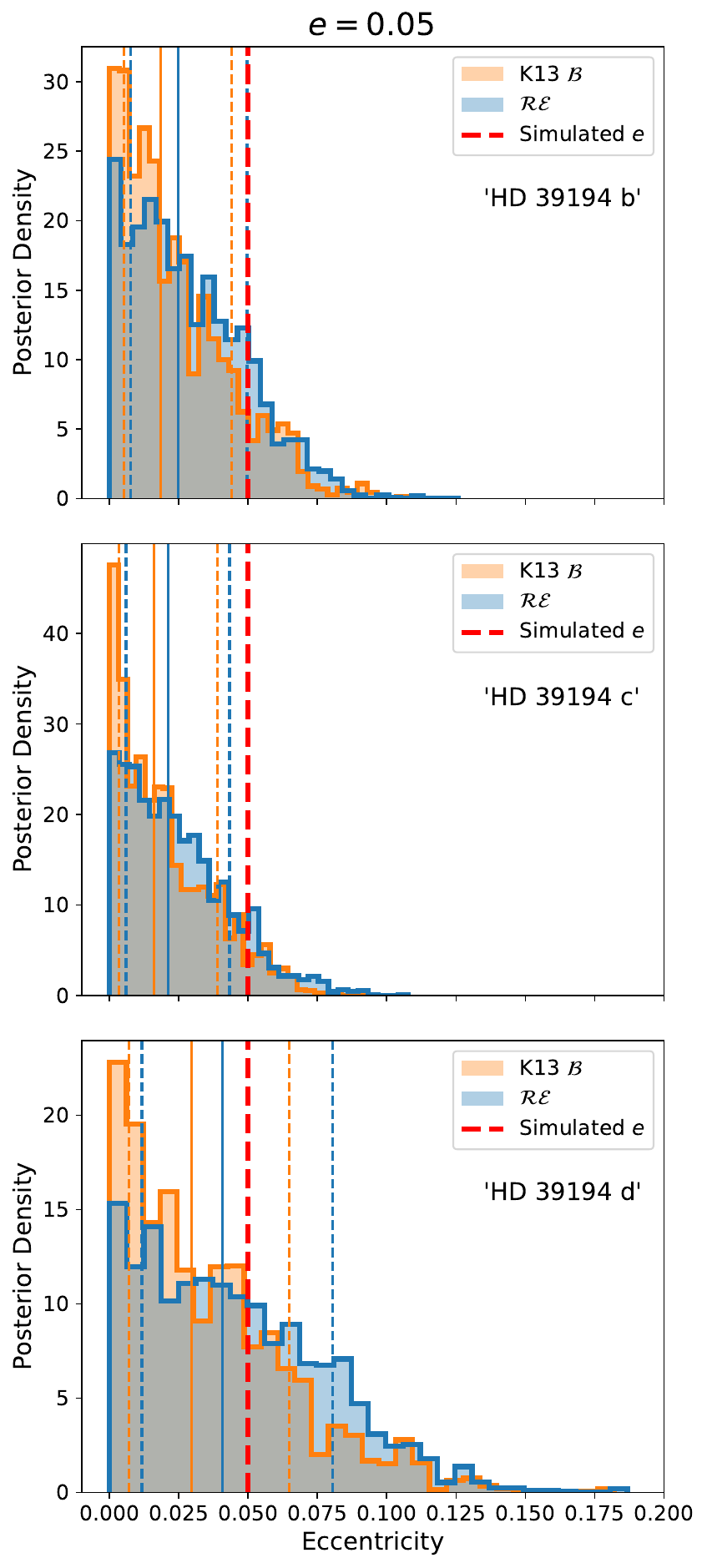}\includegraphics[width=0.30\linewidth]{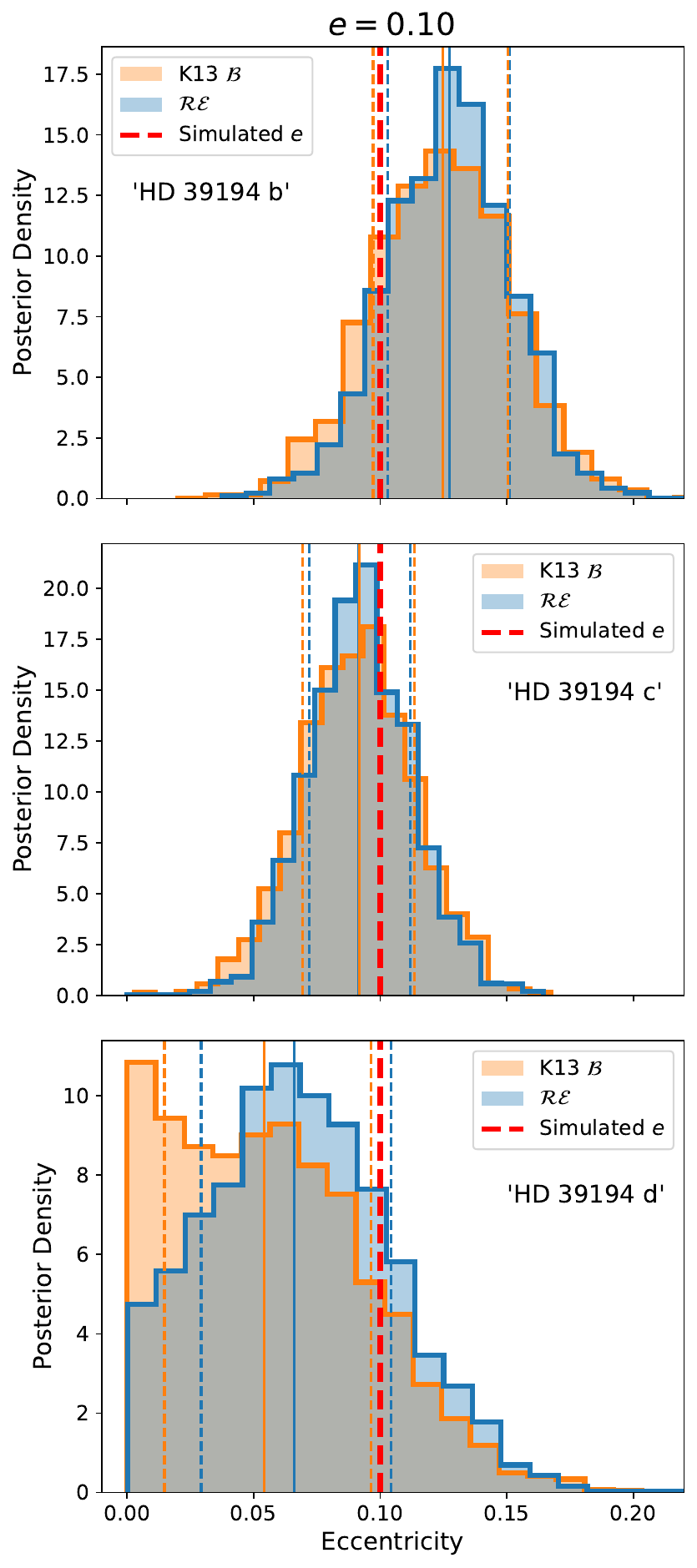}\includegraphics[width=0.31\linewidth]{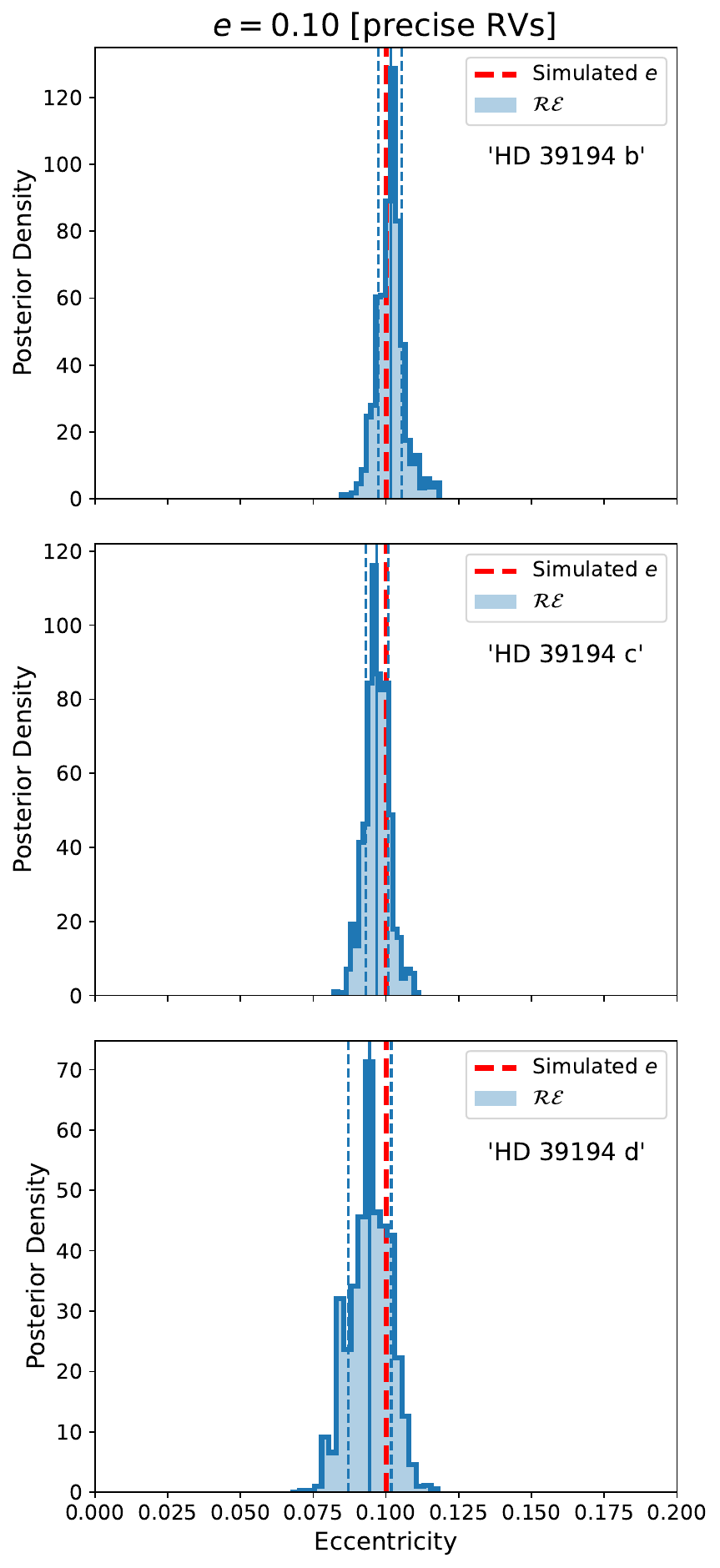}
    \caption{
    Eccentricity posterior histograms for \textsc{kima} runs of HD\,39194, with two different priors. RVs are simulated in an attempt to recover ground truth eccentricities. Results are shown for signals with $e=0.05$ and $e=0.10$  (\textbf{left, middle}, respectively). All planets in the system are simulated with the same value of eccentricity (and marked by the red dashed lines). \textbf{Right:} this column shows the excellent recovery of true values if all simulated RVs were recorded with the highest precision (0.1~m\,s$^{-1}$ uncertainties) and had correspondingly small white noise applied.
    To simplify the plots, only one posterior is shown for each planet, as the $\mathcal{B}$ (K13) and $\mathcal{RE}$ priors produce posteriors with almost identical medians and $1\sigma$ uncertainties.
    }
    \label{fig:HD39194Sims}
\end{figure*}

Similarly, we then assessed the posterior results from \textsc{kima} for simulated data with specific eccentricities in the HD\,39194 system (where we impose that the three planets share a common eccentricity value for simplicity). The RVs are simulated with error bars and standard deviation of Gaussian white noise equal to $0.51$~m\,s$^{-1}$, the mean uncertainty on the 288 archival RVs. The posterior distributions for both priors are shown in Figure~\ref{fig:HD39194Sims}. In the left panels, results are shown for an $e=0.05$ simulation. At low eccentricity, the posteriors are fairly unconstrained, for either prior. Both posteriors are a similar shape, rising to the bound at $e=0$. The median clearly underestimates this value (and the issue with the median is discussed further in Section~\ref{sub:biases}), but neither prior shows significant difference from a circular orbit.  
From these plots (and circular orbit simulations in Fig.~\ref{fig:HD39194CircSims}), we see the result of weakly-informative data. We can conclude that, despite occasionally peaking just away from zero, the $\mathcal{RE}$ prior is not biasing the posterior to a particular value when the data cannot constrain the eccentricity well. The new prior is not forcing the shape away from a preference for $e=0$.

In the middle column of Fig.~\ref{fig:HD39194Sims}, the injected eccentricity has been increased to $e=0.10$. We see that for simulated planets b and c, the posterior peak is well resolved and significantly different from zero. The median $e_{\textrm{(planet b)}}$ is offset from the true value for either prior, but this is due to the precision and random scatter on simulated data, as mentioned above. For these planets, the simulated data would be informative enough for us to correctly rule out circular orbits, irrespective of our choice of prior.

However, for HD\,39194\,d simulated with $e=0.10$, the story is different. Whilst medians of both posteriors underestimate the `true' value, the Beta-posterior shape shows fairly strong consistency with zero, peaking at $e=0$ then a relatively smooth decrease (within the noise).  
The $\mathcal{RE}$-posterior shows an almost-Gaussian-shaped peak that is, at least qualitatively, less compatible with a circular orbit. 
This posterior comes closer to recovering the properties of the injected Keplerian signal, where the mean lies less than $1\sigma$ away from the true eccentricity.
A result like this on real data, such as the similarly-shaped eccentricity posterior for \textit{actual} HD\,39194\,d data (Fig.~\ref{fig:HD39194_ecc}), suggests that the orbit is in fact not fully consistent with $e=0$, that use of the Beta prior would have one believe. Compact multi-planetary systems are typically difficult to characterise \citep{HD28471paper}, with the further complication that the eccentricities of the planets are likely to vary due to planet-planet gravitational interactions \citep[e.g.][]{Staab2020,Ross2024}.

If we instead simulate very precise data, at the same observational epochs as the HARPS RVs, we are able to recover the injected small eccentricities ($e=0.10$) very successfully (Fig.~\ref{fig:HD39194Sims}, right). Our original simulations had error bars and standard deviation on Gaussian white noise equal to $0.51$~m\,s$^{-1}$, whereas this has now been reduced to $0.10$~m\,s$^{-1}$. This is a slightly optimistic estimate for data recorded by ESPRESSO, but is not entirely improbable for a bright nearby star. Provided that any stellar activity and instrumental effects are well-modelled, and non-Keplerian scatter on the data is therefore minimised, ultra-precise observations can greatly aid in resolving small non-zero eccentricities for low-mass planets in compact systems. Further observations with improved precision would be warranted to reveal the true nature of the HD\,39194 system, and others like it.

\subsection{Inherent biases of an RV sample}\label{sub:biases}

There may be biases from observations of RV exoplanets, or the inference methods to detect these planets, which will impact the eccentricity distribution and our presented results. We discuss potential biases underlying our analysis below.

\subsubsection{Inference biases}
When estimating eccentricity with parameter inference methods like Markov Chain Monte Carlo (MCMC) techniques, we face the issue that eccentricity is a positive definite quantity, and has a hard bound at $e=0$, as negative eccentricities are nonphysical. This means that an MCMC walker, stepping around in parameter space, will come up against this hard boundary, and be biased towards taking steps towards larger rather than smaller eccentricities when $e\sim0$ \citep{KippingWang2024}. Due to the positive-definite bias, \citet{LucySweeney1971} found that to be 95\,per~cent confident an eccentricity is non-zero, it must be $2.45\sigma$ above zero, rather than the naive $2.0\sigma$ \citep{Eastman2024}. Nested sampling \citep{Skilling2004} can circumvent this problem, but then another arises from our use of summary statistics to describe posterior distributions \citep{KippingWang2024}. Median and one-sigma credible intervals of the posterior distributions will be skewed due to the $e=0$ boundary.

To counter the sub-optimal use of summary statistics, \citet*{Zakamska2011} suggest using the mode of the marginalised posterior eccentricity distribution instead of the median. However, an all-together more appropriate solution would be to use HBMs, discussed earlier in Section~\ref{sub:e_errors} \citep[see figure~2 in][for a informative diagram on the method]{Hogg2010}. When one has access to individual eccentricity posterior distributions, using these directly will circumvent the issue of calculating skewed medians and error estimates. 

Some planets in the sample have had their eccentricities fixed to be zero in the discovery papers, when $e$ is found to be consistent with zero (or different at low significance). In some situations this may be a justified assumption, but prevents any errors being calculated. Small but significant eccentricities should in principle not be excluded \citep{Bonomo2017}.

There will also be different biases in how the discovery papers perform fitting procedures, with varying choices of priors used and the implementation of additional jitter terms. Non-Gaussian noise can have an impact on signal characterisation, and the wrong global $\chi^{2}$ minimum will be reached if the error bars are incorrectly accounted for \citep{Hara2019}. It has become the norm to include such a jitter term, as without one, the eccentricity can either be underestimated, or in a perhaps worse outcome, overestimated if the solution is forced to fit outlying data points \citep{Bonomo2017};  although high eccentricity models may be penalised in a Bayesian framework and disfavoured over a lower eccentricity model \citep{Hara2019}. Where white noise is present, including jitter and estimating the eccentricity from posteriors can give acceptable retrieval of input parameters, verifying the approach \citep{Zakamska2011}.

\subsubsection{Observational biases}
Eccentricity can be a notoriously difficult parameter to characterise in RV studies \citep{ShenTurner08, HaddenLithwick2017,Wittenmyer2019,VE19,Stock2023,GilbertPetiguraESS}. For a fixed number of observations per orbital period, detection efficiency was found to steeply drop off at higher eccentricities \citep{Cumming2004}. Sparse sampling could easily miss the brief RV peaks when a highly eccentric planet is near periastron \citep{Cumming2004,Zakamska2011}. Additionally, there exists a degeneracy between solutions for an eccentric planet or two resonant, circular orbit planets, requiring precise strategic observations to discriminate between the two scenarios \citep*{anglada2010}.

Naively, we might expect more eccentric planets to be easier to detect as $K$ is larger \citep{Haswell2010}. \citetalias{ShenTurner08} found there was a positive bias in detecting eccentric planets based on the SNR and time span of the observations, two controlled parameters of an observing campaign. Recently, \citet{KippingWang2024} investigated how detections of eccentric exoplanets were biased by the SNR of the signal. The probability of detecting a signal (completeness rate) was originally assumed to depend on this SNR, which in turn would depend upon parameters of the system, making some surveys more complete to certain regions of orbital-parameter space. \citet{KippingWang2024} calculate an idealised SNR of detecting a given signal, and determine general and analytic formulas to identify basic trends that would cause RV-method-specific biases. Instead however, they find the detection SNR is remarkably flat with eccentricity. The ratio between SNR of an eccentric signal and circular signal tends towards $0.9$ as $e\to1$, therefore only being noticeable at very high eccentricity. As SNR dependency marginalised over all orientations is close to being flat, \citet{KippingWang2024} conclude that the observed eccentricity distribution for RV exoplanets is approximately unbiased.

\citet{KippingWang2024} also study the SNR required to break the degeneracy between a single eccentric planet and two circular planets in resonance \citep[e.g.][]{anglada2010}. They find, for eccentricities up to $e=0.316$, that precise observations are required to break the degeneracy. The SNR of the discrimination term ($\propto Ke^{2}$) is typically far less than a tenth of the SNR of the measured RV signal.

Finally, \citet{KippingWang2024} also find that orbits with the semi-major axis vector pointed towards the observer enjoy a positive detection bias. This corresponds to a maximisation of the radial component of the orbital motion.

\section{Conclusions}\label{sec:conclusion}

We have performed a re-analysis of the RV exoplanet population, using a similar methodology to \citetalias{kipping2013} but with a sample size more than twice as large. This has been extended to incorporate an additional method, allowing us to compare EDF to PDF (likelihood) regression, and investigate differences between the results. We studied how an updated sample size changes which functions best describe observed eccentricity distributions. The eccentricity distribution is suitable to use as a prior
in exoplanet detection and parameter inference with nested sampling codes such as \textsc{kima}. Our results in summary are:

\begin{enumerate}
    \item Using an EDF method similar to \citetalias{kipping2013}, we find that a mixture model of Rayleigh + Exponential distributions now provides a much closer match to the data than Beta/Kumaraswamy distributions, which have previously been extensively used as the optimal choice of prior.
    \item A more direct PDF parameterisation method favours the use of a Gamma distribution to describe the eccentricity population, with moderate evidence improvement over the Rayleigh and Exponential mixture model. All other models are heavily disfavoured by Bayesian evidence comparison.
    \item In agreement with \citetalias{kipping2013}, we find that short and long period planets show different eccentricity distributions. Using the Rayleigh + Exponential mixture, we investigate where a split in planetary orbital period between `short' and `long' should be positioned. The relative strength of each distribution (the mixture fraction $\alpha$) changes with period, informing us where a transition may lie. As these distributions are often used to model different physical processes, this transition period may represent the regime where, for shorter period planets, tidal circularisation becomes almost entirely dominant in sculpting the observed eccentricities. Splitting the planets into three period regimes, we find that `hot', `warm' and `cold' planets also display distinct eccentricity distributions.
    \item As the masses of planets affect interactions during planetary system formation, this also impacts the eventual eccentricities observed. Dividing our sample (by the \textit{minimum} planetary masses) into groups roughly corresponding to super-Earth, Neptune, and gas giant sized planets, we find that fitting local distributions is preferred over a global fit to a high degree of significance. We also find that the lowest mass planets ($<20~\rm{M}_{\earth}$) are best fit by a Rayleigh-dominated $\mathcal{RE}$ eccentricity distribution, indicating that scattering and perturbations are strongest for these planets, hindering circularisation \citep[in agreement with][]{2024AJ....168..115B}. The masses of planets should be considered when deriving an eccentricity prior to use for inference of RV-planet parameters.
    \item The eccentricity distributions for single planets and multi-planets are still distinct for RV discoveries, in agreement with results derived for transiting \textit{Kepler} planets \citep{VE19,Sagear2023}.
    \item We investigated whether outer massive companions affect the eccentricity distribution. Systems with either a binary companion star or outer giant planet appear to be drawn from the same underlying eccentricity population as those without.
    \item In a similar vein to \citet{Sagear2023}, we find that planets orbiting M-dwarfs seem to share the same eccentricity distribution as those hosted by FGK stars. The inferred parameters appear significantly different however, and this may be due to the relative size of each sample. The M-dwarf planet EDF has far fewer data points than the FGK-type EDF, so regressing a single distribution to the two EDFs will be skewed: by both the number of points, and the smaller DKW error bar weighting each point of the FGK-type EDF. Combining FG and KM-hosted planets together, we find that a local EDF parameterisation is preferred by $\ln{Z}\sim8$. Similar regression with the PDF method does not echo this improvement ($\Delta\ln{Z}\sim0.9$). This requires further study, with a larger, more homogeneous sample.
    \item We have created code, \textsc{eccentriciPy}, to re-run this analysis for an updated sample of RV exoplanets, and take user input to create trial eccentricity priors for planet discovery, depending on features of the data/system being investigated.
    \item Three case studies compared the use of an updated Rayleigh + Exponential prior to the traditionally-used \citetalias{kipping2013} Beta. The strongest influence of the prior is identified for a low-eccentricity, compact multi-planet system of super-Earths, where a Rayleigh + Exponential prior allows the peak of the distribution to be separated from the hard boundary at $e=0$.
    \item On the whole, posterior distributions can be highly dependent on the choice of prior, and this is strongest for low S/N, poorly sampled signals. Using a prior that reflects the underlying population provides a good estimate of the expected planetary distributions, though can sometimes bias the results. One should thoroughly investigate eccentricities for newly-detected RV planets, experimenting with multiple priors and comparing the influence and results. 
    \item Limitations and biases are discussed, with extensions to the study identified. Considering each exoplanet's posterior eccentricity distribution would provide a better method than summary statistics like the median/max posterior \citep{Hogg2010}. In future work, using HBMs will be far more suitable, on condition that posteriors can be derived for all exoplanets in the sample.
\end{enumerate}

Accurately quantifying the small but non-zero eccentricities for close-in exoplanets is vitally important for consideration of their energy balance. Tidal effects arising from non-circular orbits impact atmosphere and envelope loss, atmospheric structure, and for rocky planets may cause volcanism. Exploiting the full information content of the extant RV data is thus important to guide target selection and data interpretation for facilities including JWST. Our case study of 51\,Peg\,b demonstrates that even for massive planets orbiting bright stars, 256 RV epochs are insufficient to precisely determine the eccentricity. Further high precision RV observations may thus be justified for particularly interesting known systems.  

\section*{Acknowledgements}
The authors wish to thank reviewers and editor for their suggestions and input, significantly improving the quality of this article. \\

ATS is supported by an Science and Technology Facilities Council (STFC) studentship. CAH and JRB are supported by consolidated grant ST/X001164/1 from STFC. JKB is supported by an STFC Ernest Rutherford Fellowship (grant ST/T004479/1). MRS acknowledges support from the European Space Agency as an ESA Research Fellow. \\

This research has made use of the \href{https://exoplanetarchive.ipac.caltech.edu/}{NASA Exoplanet Archive}, which is operated by the California Institute of Technology, under contract with the National Aeronautics and Space Administration under the Exoplanet Exploration Program.

Radial velocity data were analysed with the \textsc{kima} package, freely accessible online and described in \citet{kima-joss}.

This work has made use of data from the European Space Agency (ESA) mission {\it Gaia} (\url{https://www.cosmos.esa.int/gaia}), processed by the {\it Gaia} Data Processing and Analysis Consortium (DPAC, \url{https://www.cosmos.esa.int/web/gaia/dpac/consortium}). Funding for the DPAC has been provided by national institutions, in particular the institutions participating in the {\it Gaia} Multilateral Agreement.

The following Python modules have been used in this work: 
\textsc{arviz} \citep{arviz}
\textsc{astropy} \citep{Astropy1,Astropy2,Astropy3}, 
\textsc{chaospy} \citep{chaospy},
\textsc{corner} \citep{corner},
\textsc{matplotlib} \citep{matplotlib},
\textsc{numpy} \citep{numpy}, 
\textsc{pandas} \citep{pandas},
\textsc{pymultinest} \citep{multinest,pymultinest}, 
\textsc{pymc} \citep{pymc},
\textsc{pyvo} \citep{pyvo}, 
\textsc{scipy} \citep{scipy}.

Some data products used in this publication were taken on observing runs where funding was assisted by travel grants from the Royal Astronomical Society (RAS), allowing postgraduate students the opportunity to get in-person experience of observational astronomy that had previously been hindered by the COVID-19 pandemic.

\section*{Data Availability}

All data underlying this article is available from the \href{https://exoplanetarchive.ipac.caltech.edu/}{NASA Exoplanet Archive}. The RV exoplanet sample used is available as journal supplementary material online. The \texttt{eccentriciPy} code and sample are both available at \url{https://github.com/adam-stevenson/eccentriciPy}.


\bibliographystyle{mnras}
\bibliography{main} 



\appendix

\section{Transit detections across stellar masses}
Tables~\ref{tab:SpT_transit} and~\ref{tab:SpT_Kepler} are included here for comparison with Table~\ref{tab:SpT} and Section~\ref{sub:host-star}, to highlight how statistics change with varying stellar mass/approx. spectral type. 
\begin{table}
    \centering
        \caption{Equivalent to Table~\ref{tab:SpT}, but for exoplanets discovered via the transit method.}
        \label{tab:SpT_transit}
    \begin{tabular}{lcccc}
    \hline
    Mass ($\textrm{M}_{\sun}$) & Approx. SpT & Planets & Host Stars & Multi-planet frac. \\
    \hline
    $1.1-1.6$ & F & 748 & 659  & 0.14 \\
    $0.85-1.1$ & G & 1560 & 1350 & 0.17 \\
    $0.65-0.85$ & K & 667 & 530 & 0.21 \\
    $<0.65$ & M & 396 & 301 & 0.23 \\
        \hline
    \end{tabular}
\end{table}

\begin{table}
    \centering
        \caption{Same as Table~\ref{tab:SpT_transit}, but for \textit{Kepler} planets, to see statistics for a more coherent sample with consistent and well-understood biases.}
        \label{tab:SpT_Kepler}
    \begin{tabular}{lcccc}
    \hline
    Mass ($\textrm{M}_{\sun}$) & Approx. SpT & Planets & Host Stars & Multi-planet frac. \\
    \hline
    $1.1-1.6$ & F & 413 & 348  & 0.18 \\
    $0.85-1.1$ & G & 1080 & 945 & 0.16 \\
    $0.65-0.85$ & K & 362 & 315 & 0.14 \\
    $<0.65$ & M & 128 & 93 & 0.23 \\
        \hline
    \end{tabular}
\end{table}

\section{Sample}
A truncated version of our sample is included in Table~\ref{tab:archive}. The entire archive of used RV exoplanets is available as supplementary material online. The column \texttt{updated\_ecc} in the data file contains the eccentricities used in this work, and all other columns are identical to the NASA archive (we have called this ``Eccentricity'' in Table~\ref{tab:archive}).

\input{{table.tex}}

\raggedright
\section{Simulations for circular HD~39194 planets}
\begin{figure*}
    \centering
    \includegraphics[width=0.32\linewidth]{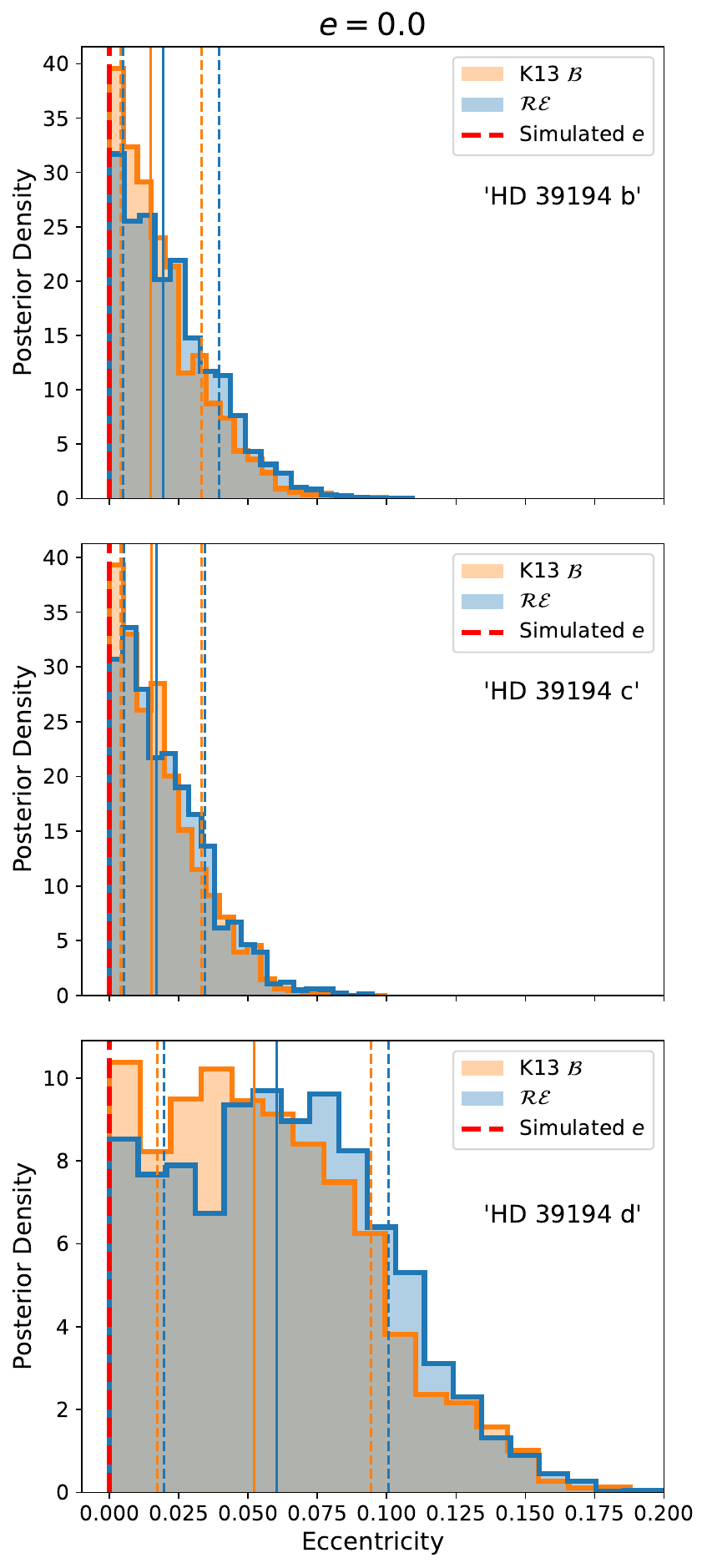}\includegraphics[width=0.32\linewidth]{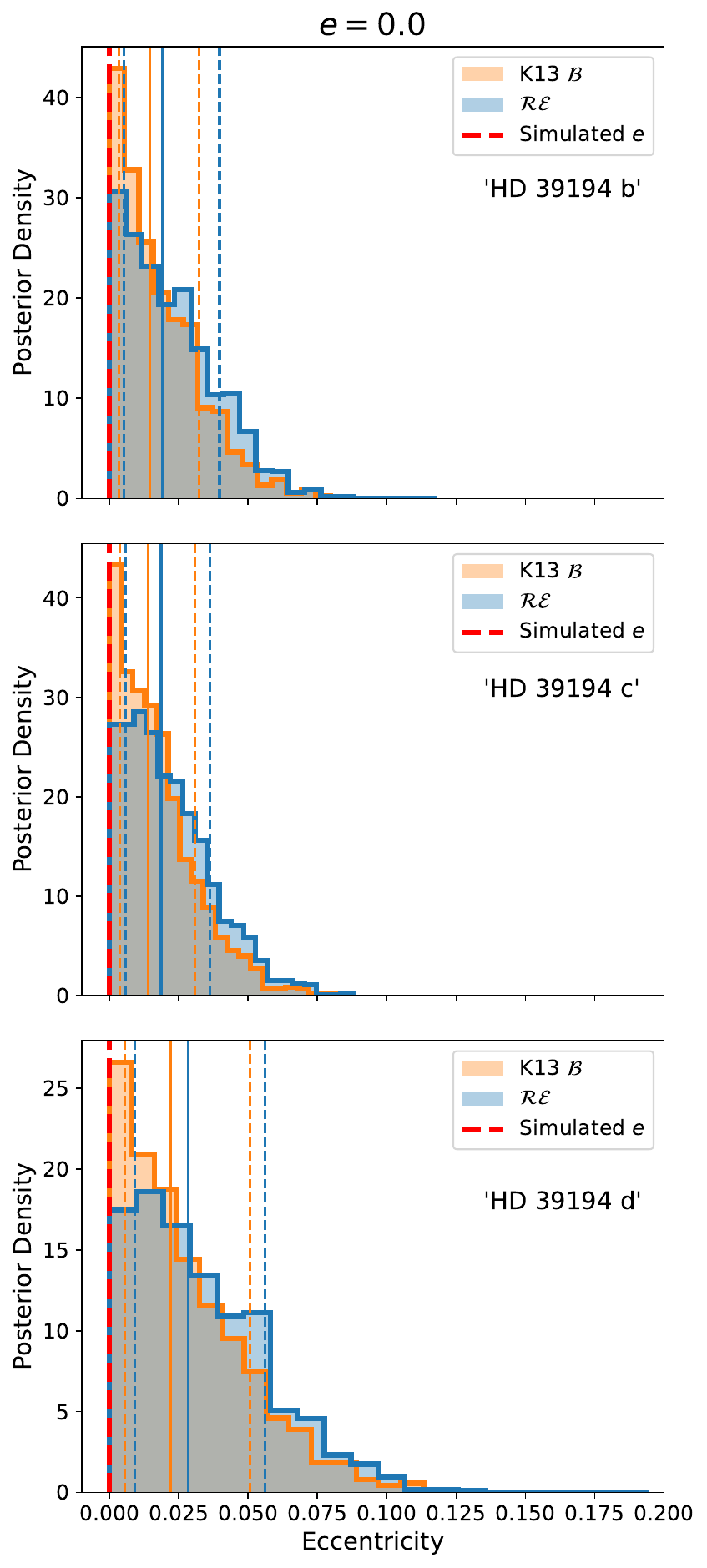}\includegraphics[width=0.32\linewidth]{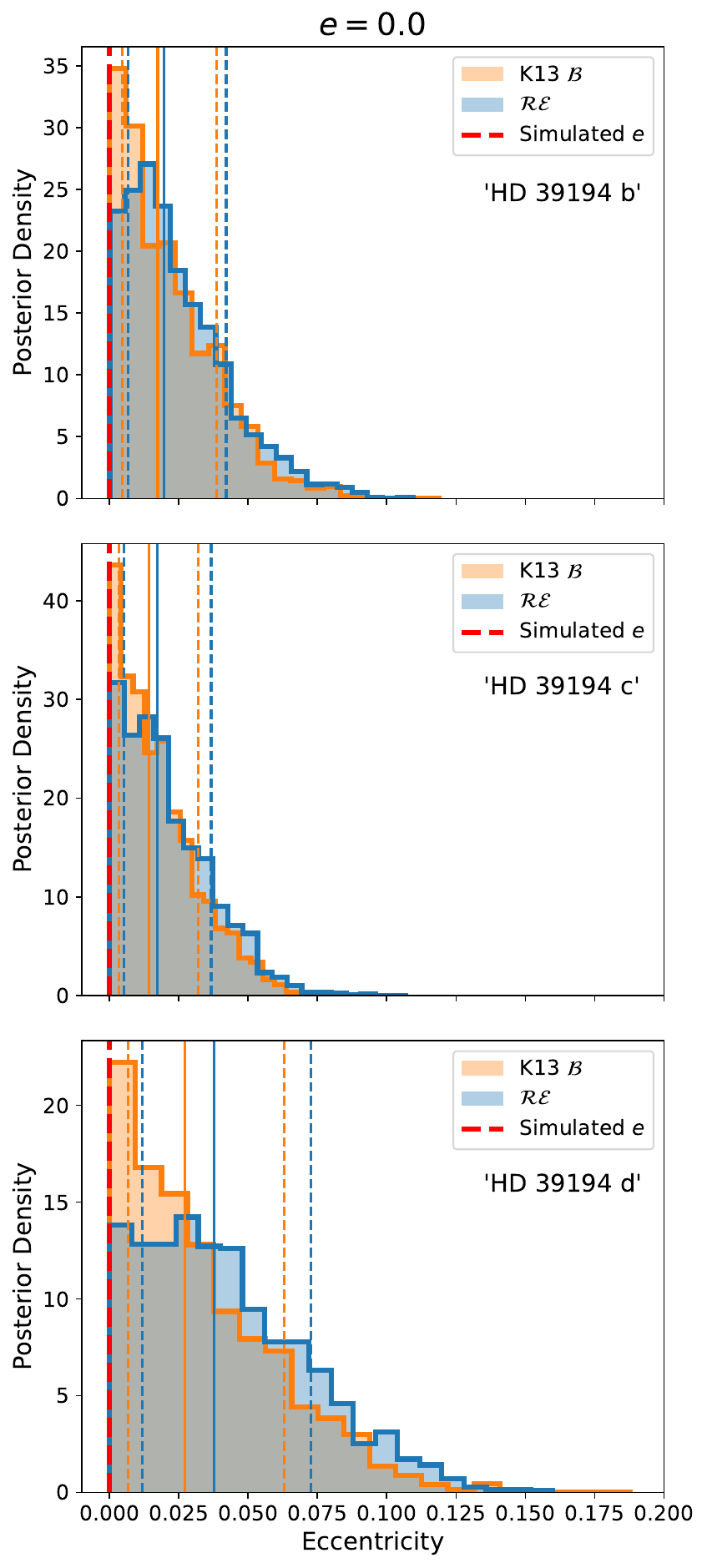}
    \caption{
    Eccentricity posterior histograms for \textsc{kima} runs of HD\,39194, with two different priors. RVs are simulated in an attempt to recover ground truth eccentricities. Each plot shows results for a different random initialisation of constant standard deviation Gaussian white noise scatter on the simulated RVs. All planets in the system are simulated with the same value of eccentricity ($e=0$, marked by the red dashed lines).
    }
    \label{fig:HD39194CircSims}
\end{figure*}

\raggedright
Results for circular simulated data are shown in Figure~\ref{fig:HD39194CircSims}. Each \textsc{kima} recovery run is the same, the only difference being the random noise added to the simulated RVs. This white noise always has a constant scale, but a different random initialisation each time, in case scatter falls such that it mimics truly eccentric Keplerians.

\section{Uniform eccentricity priors}\label{app:uniform_prior}

Eccentricity priors describe our previously-held beliefs about typical eccentricity values for exoplanetary systems. These are informed by our knowledge of currently detected systems, the Solar System, and physical processes which would modulate the increase or decrease of eccentricity in an elliptical orbit.

\begin{figure}
    \centering
    \includegraphics[width=0.80\columnwidth]{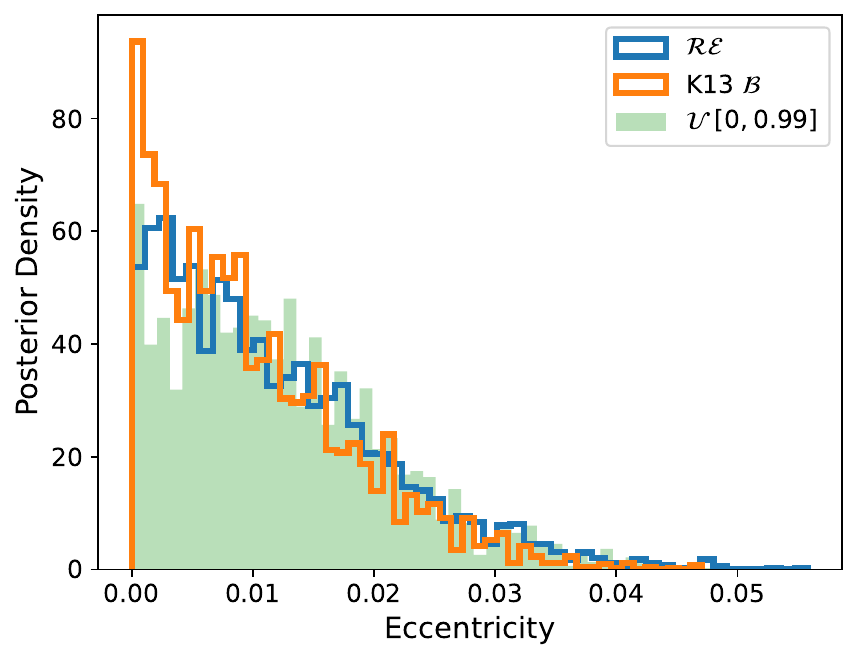}
    \caption{Eccentricity posteriors for \textsc{kima} runs on 51\,Peg RVs. Similar to Figure~\ref{fig:51Peg_ecc}, this plot now includes the posterior created with a uniform ($\mathcal{U}$) prior (green). It is very similar to the $\mathcal{B}$ and $\mathcal{RE}$ posteriors, indicating that the RVs do constrain the solutions fairly well and the prior is not drastically biasing results.}
    \label{fig:uniform:51Peg}
\end{figure}

\begin{figure}
    \centering
    \includegraphics[width=0.80\columnwidth]{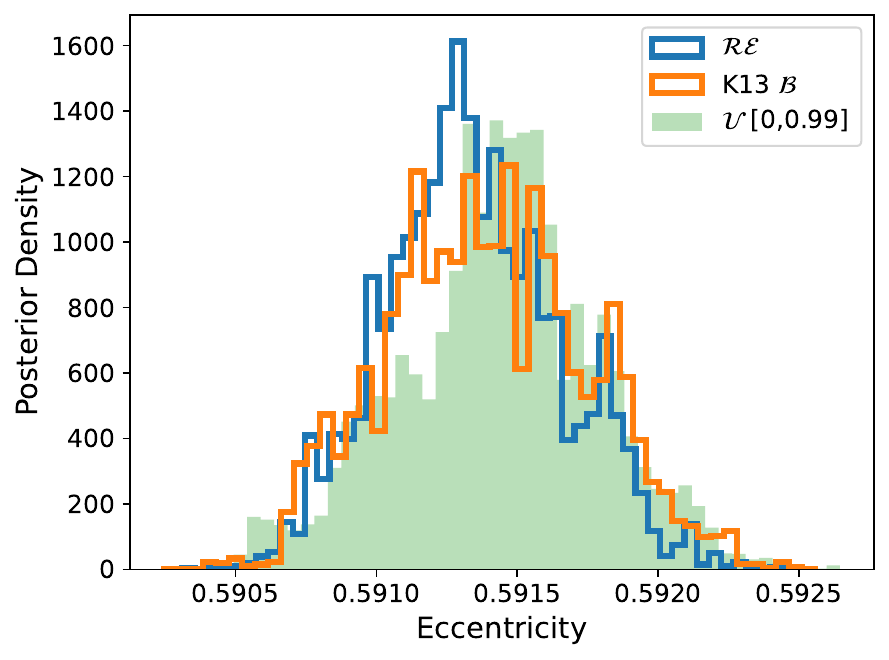}
    \caption{Eccentricity posteriors for \textsc{kima} runs on DMPP-3 RVs. Similar to Figure~\ref{fig:DMPP3_ecc}, this plot now includes the posterior created with a uniform ($\mathcal{U}$) prior (green). It is very similar to the $\mathcal{B}$ and $\mathcal{RE}$ posteriors, indicating that the RVs do constrain the solutions fairly well and the prior is not drastically biasing results.}
    \label{fig:uniform:DMPP3}
\end{figure}

\begin{figure}
    \centering
    \includegraphics[width=0.80\columnwidth]{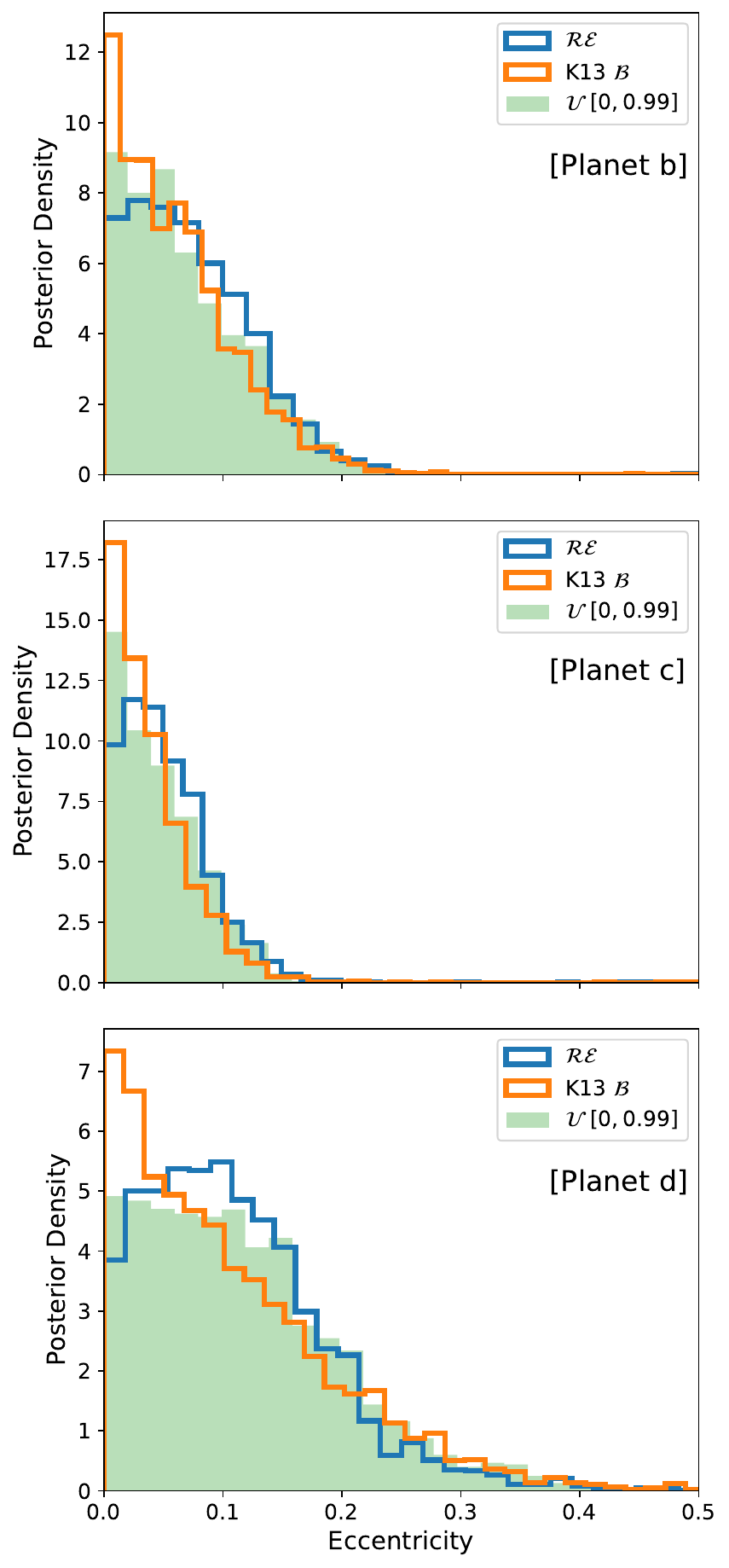}
    \caption{Eccentricity posteriors for \textsc{kima} runs on HD\,39194 RVs. Similar to Figure~\ref{fig:HD39194_ecc}, this plot now includes the posterior created with a uniform ($\mathcal{U}$) prior (green). It is very similar to the $\mathcal{B}$ and $\mathcal{RE}$ posteriors, indicating that the RVs do constrain the solutions fairly well and the prior is not drastically biasing results.}
    \label{fig:uniform:HD39194}
\end{figure}

These priors may however introduce biases, and a more general uniform prior may provide the most un-biased, uninformative distribution to sample from. If signals are strongly detected at sufficient S/N, the effect of the prior will be minimal (evidenced by some of the examples in Section~\ref{sec:discussion}). We have assessed how the posterior response would change when using a uniform $e$ prior (between 0 and 0.99) for the three case studies in Section~\ref{sec:discussion}. Figure~\ref{fig:uniform:51Peg} shows the 51\,Peg\,b posterior, Figure~\ref{fig:uniform:DMPP3} shows the same for DMPP-3B, and Figure~\ref{fig:uniform:HD39194} shows the posterior for HD\,39194.

The posteriors created using a uniform prior (green filled histogram in all plots) show that for these three case studies, the response does not change much depending on use of the prior. Despite not being able to determine a definitive eccentricity value (that is significantly different from zero) for 51\,Peg\,b and the HD\,39194 planets, the RVs do successfully constrain the shape of the resulting posterior. This behaviour is replicated for the posteriors of simulated RVs at given eccentricity values.

\begin{figure}
    \centering
    \includegraphics[width=0.80\columnwidth]{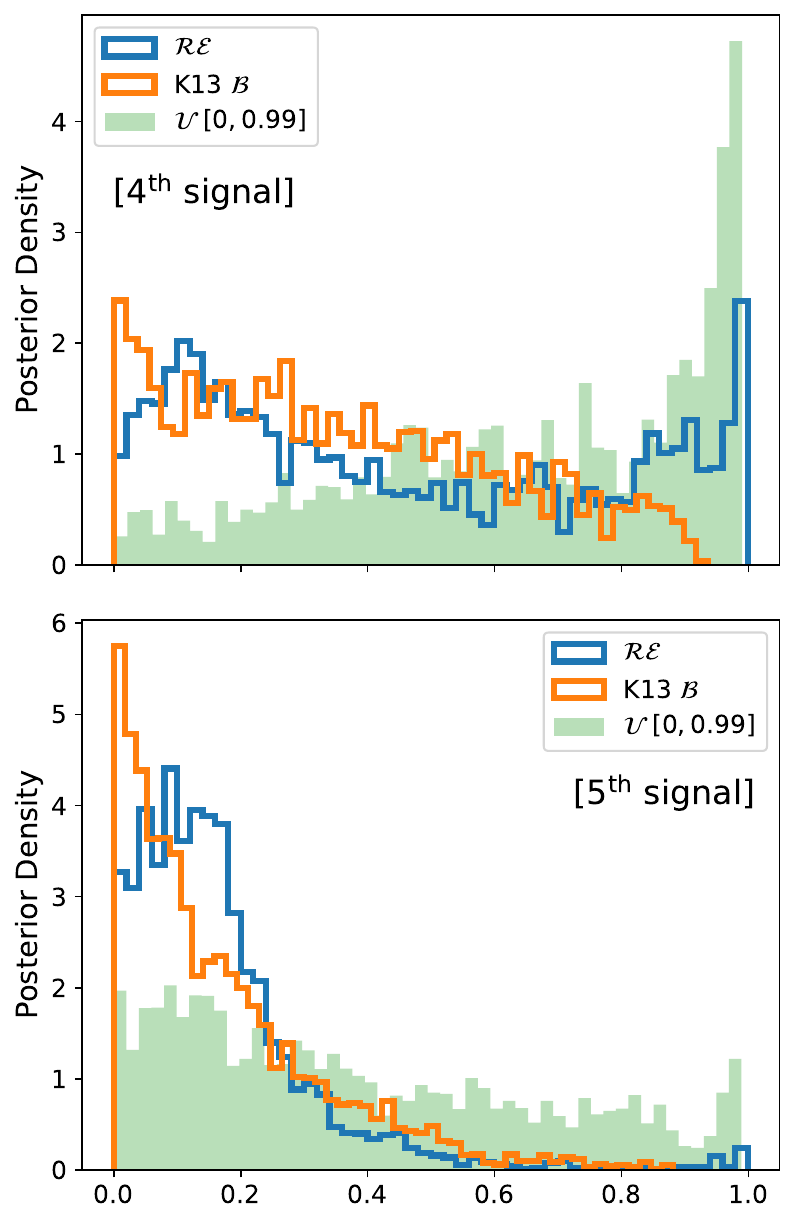}
    \caption{Eccentricity posteriors for additional signals ($P\sim 370~\& \sim 2000$\,d) that are detected in the HD\,39194 system, though are not confirmed as planets \citep{Unger2021}. These periodicities are recovered in \textsc{kima} runs, though the Bayes Factor only allows confirmation of the three known planets b, c, and d \citep{BrewerDonovan2015,Faria2016}. As these signals are poorly modelled, it highlights the response of each trialled prior.}
    \label{fig:uniform:HD39194extra}
\end{figure}

The informative priors are therefore most useful in scenarios where the RVs \textit{do not} effectively constrain the eccentricity of the solutions. Some examples are included in this appendix to illustrate this. Figure~\ref{fig:uniform:HD39194extra} shows the posteriors for two additional periodicities detected in the HD\,39194 system at $\sim370$ and $\sim2000$\,d. These are currently not confirmed to be of planetary origin, with potential causes described further in \citet{Unger2021}. However, they do demonstrate the effect of using alternate priors to study poorly-sampled, low S/N RV signals. In the top panel (labelled ``4$^{\rm th}$ signal'', after planets b, c, and d), we can see the vastly different response. This periodicity is thought to be a calibration error caused by a period stitching effect \citep[again, see][]{Unger2021}, and may in reality not be best modelled by a Keplerian signal. Using the Beta prior here (and the $\mathcal{RE}$ to a lesser extent) biases the behaviour and makes it look more plausible that the eccentricity may be produced by a planet -- confounding the assessment of the signal's nature. For the bottom panel (``5$^{\rm th}$ signal''), the eccentricity posterior almost exactly reproduces the prior as the RVs do not fully sample the periodicity.

\begin{figure}
    \centering
    \includegraphics[width=0.80\columnwidth]{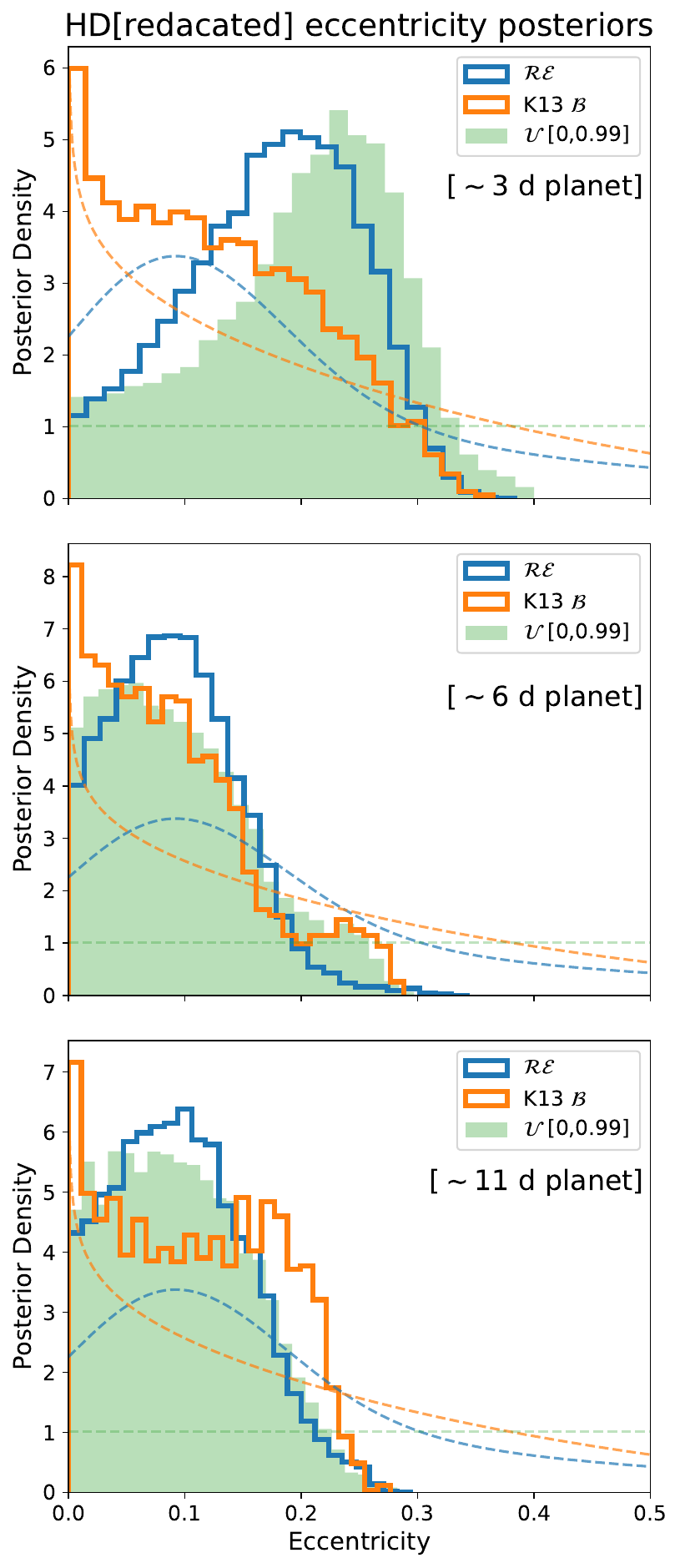}
    \caption{Eccentricity posterior distributions for \textsc{kima} runs of an unpublished compact RV system. Three putative planet eccentricities are shown, for a choice of input priors. These priors are shown with dashed lines. High eccentricities are disfavoured due to stability constraints.}
    \label{fig:uniform:HD28471}
\end{figure}

A further example can be seen when assessing the output of \textsc{kima} runs for an as-yet-unpublished compact multiplanet system \citep{HD28471paper}. The posteriors for putative planets on $\sim3$, $\sim6$ and $\sim11$\,d orbits are shown in Figure~\ref{fig:uniform:HD28471}. For each panel, the upper eccentricity is limited by considerations of the system stability \citep[for further information, see][]{HD28471paper}. In general, all priors produce a similar posterior, though the Beta posterior peaks at zero whereas $\mathcal{RE}$ and Uniform do not. This is most pronounced for the innermost planet (top panel), where we see that the Beta prior is biasing the results towards completely circular orbits. The $\mathcal{RE}$ prior here is less constraining, providing an intermediary between an uninformative Uniform prior, and a biasing Beta prior.

\bsp	
\label{lastpage}
\end{document}

%% file: table.tex
\centering
\begin{table*}\label{tab:archive}
\caption{A truncated version of the archive used in this work, with a selected subset of columns and rows. The full table of 891 RV exoplanets (from 12 June 2024), after taking the cuts described in Section~\ref{sec:sample}, can be found in the supplementary material online. For a description of the columns in the full table, see the \href{https://exoplanetarchive.ipac.caltech.edu/}{NASA Exoplanet Archive}.}
\begin{tabular}{lrrllll}
\toprule
Name & $N_{\textrm{planets}}$ & $N_{\textrm{stars}}$ & Period [d] & Eccentricity & $K$ [m\,s$^{-1}$] & $M_{\star}$~[M$_{\sun}$] \\
\midrule
11 Com b & 2 & 1 & $323.21^{\,+0.06}_{-0.05}$ & $0.24^{\,+0.01}_{-0.01}$ & $288.63^{\,+2.39}_{-2.37}$ & $2.09^{\,+0.64}_{-0.63}$ \\
11 UMi b & 1 & 1 & $516.22^{\,+3.20}_{-3.20}$ & $0.08^{\,+0.03}_{-0.03}$ & $189.70^{\,+7.20}_{-7.20}$ & $2.78^{\,+0.69}_{-0.69}$ \\
14 Her b & 1 & 2 & $1765.04^{\,+1.68}_{-1.87}$ & $0.37^{\,+0.01}_{-0.01}$ & $89.58^{\,+0.96}_{-0.82}$ & $0.91^{\,+0.11}_{-0.11}$ \\
16 Cyg B b & 3 & 1 & $798.50^{\,+1.00}_{-1.00}$ & $0.68^{\,+0.02}_{-0.02}$ & $50.50^{\,+1.60}_{-1.60}$ & $1.08^{\,+0.04}_{-0.04}$ \\
17 Sco b & 1 & 1 & $578.38^{\,+2.01}_{-2.09}$ & $0.06^{\,+0.03}_{-0.02}$ & $92.40^{\,+3.90}_{-3.84}$ & $1.22^{\,+0.13}_{-0.14}$ \\
18 Del b & 2 & 1 & $982.85^{\,+1.06}_{-0.92}$ & $0.02^{\,+0.01}_{-0.02}$ & $114.96^{\,+1.98}_{-0.96}$ & $2.10^{\,+0.13}_{-0.09}$ \\
24 Boo b & 1 & 1 & $30.33^{\,+0.00}_{-0.01}$ & $0.03^{\,+0.04}_{-0.02}$ & $55.67^{\,+2.31}_{-2.66}$ & $1.05^{\,+0.27}_{-0.18}$ \\
24 Sex b & 1 & 2 & $452.80^{\,+2.10}_{-4.50}$ & $0.09^{\,+0.14}_{-0.06}$ & $40.00^{\,+4.90}_{-7.80}$ & $1.54^{\,+0.08}_{-0.08}$ \\
30 Ari B b & 4 & 1 & $335.10^{\,+2.50}_{-2.50}$ & $0.29^{\,+0.09}_{-0.09}$ & $272.00^{\,+24.00}_{-24.00}$ & $1.93^{\,+0.27}_{-0.27}$ \\
4 UMa b & 1 & 1 & $270.27^{\,+0.24}_{-0.09}$ & $0.45^{\,+0.02}_{-0.04}$ & $242.24^{\,+7.45}_{-14.31}$ & $1.23^{\,+0.15}_{-0.15}$ \\
42 Dra b & 2 & 1 & $479.10^{\,+6.20}_{-6.20}$ & $0.38^{\,+0.06}_{-0.06}$ & $110.50^{\,+7.00}_{-7.00}$ & $0.98^{\,+0.05}_{-0.05}$ \\
47 UMa b & 1 & 3 & $1078.00^{\,+2.00}_{-2.00}$ & $0.03^{\,+0.01}_{-0.01}$ & $48.40^{\,+0.80}_{-0.90}$ & $1.06^{\,+0.13}_{-0.14}$ \\
47 UMa c & 1 & 3 & $2391.00^{\,+100.00}_{-87.00}$ & $0.10^{\,+0.05}_{-0.10}$ & $8.00^{\,+1.00}_{-1.00}$ & $1.06^{\,+0.13}_{-0.14}$ \\
47 UMa d & 1 & 3 & $14002.00^{\,+4018.00}_{-5095.00}$ & $0.16^{\,+0.09}_{-0.16}$ & $13.80^{\,+2.20}_{-2.90}$ & $1.06^{\,+0.13}_{-0.14}$ \\
51 Peg b & 1 & 1 & $4.23^{\,+0.00}_{-0.00}$ & $0.01^{\,+0.01}_{-0.01}$ & $55.94^{\,+0.69}_{-0.69}$ & $1.03^{\,+0.17}_{-0.09}$ \\
55 Cnc c & 2 & 5 & $44.40^{\,+0.00}_{-0.00}$ & $0.03^{\,+0.02}_{-0.02}$ & $9.89^{\,+0.22}_{-0.22}$ & $0.91^{\,+0.01}_{-0.01}$ \\
55 Cnc d & 2 & 5 & $5574.20^{\,+93.80}_{-88.60}$ & $0.13^{\,+0.02}_{-0.02}$ & $38.60^{\,+1.30}_{-1.40}$ & $0.91^{\,+0.01}_{-0.01}$ \\
55 Cnc e & 2 & 5 & $0.74^{\,+0.00}_{-0.00}$ & $0.05^{\,+0.03}_{-0.03}$ & $6.02^{\,+0.24}_{-0.23}$ & $0.91^{\,+0.01}_{-0.01}$ \\
... & ... & ... & ... & ... & ... & ... \\
kap CrB b & 1 & 1 & $1253.68^{\,+3.07}_{-6.48}$ & $0.04^{\,+0.01}_{-0.03}$ & $23.78^{\,+0.52}_{-0.99}$ & $1.33^{\,+0.24}_{-0.25}$ \\
mu Leo b & 1 & 1 & $357.80^{\,+1.20}_{-1.20}$ & $0.09^{\,+0.06}_{-0.06}$ & $52.00^{\,+5.40}_{-5.40}$ & $1.50^{\,+0.10}_{-0.10}$ \\
nu Oph b & 1 & 2 & $529.93^{\,+0.08}_{-0.07}$ & $0.13^{\,+0.00}_{-0.00}$ & $287.60^{\,+1.19}_{-0.74}$ & $2.61^{\,+0.32}_{-0.57}$ \\
nu Oph c & 1 & 2 & $3180.60^{\,+5.01}_{-4.52}$ & $0.18^{\,+0.01}_{-0.01}$ & $175.47^{\,+1.57}_{-1.02}$ & $2.61^{\,+0.32}_{-0.57}$ \\
ome Ser b & 2 & 1 & $278.59^{\,+0.80}_{-0.61}$ & $0.18^{\,+0.07}_{-0.13}$ & $23.11^{\,+1.94}_{-3.38}$ & $1.01^{\,+0.07}_{-0.08}$ \\
omi CrB b & 1 & 1 & $187.01^{\,+0.15}_{-0.16}$ & $0.10^{\,+0.04}_{-0.08}$ & $33.56^{\,+1.71}_{-2.16}$ & $1.30^{\,+0.95}_{-0.36}$ \\
omi UMa b & 2 & 1 & $1569.81^{\,+12.76}_{-13.59}$ & $0.15^{\,+0.04}_{-0.08}$ & $33.11^{\,+1.64}_{-2.77}$ & $2.72^{\,+0.14}_{-0.16}$ \\
pi Men d & 1 & 3 & $124.64^{\,+0.48}_{-0.52}$ & $0.22^{\,+0.08}_{-0.08}$ & $1.68^{\,+0.17}_{-0.17}$ & $1.07^{\,+0.04}_{-0.04}$ \\
psi1 Dra B b & 3 & 1 & $3117.00^{\,+42.00}_{-42.00}$ & $0.40^{\,+0.05}_{-0.05}$ & $21.00^{\,+1.00}_{-1.00}$ & $1.19^{\,+0.07}_{-0.07}$ \\
rho CrB c & 1 & 4 & $102.19^{\,+0.27}_{-0.22}$ & $0.10^{\,+0.05}_{-0.05}$ & $4.00^{\,+0.18}_{-0.17}$ & $0.95^{\,+0.01}_{-0.01}$ \\
rho CrB e & 1 & 4 & $12.95^{\,+0.01}_{-0.01}$ & $0.13^{\,+0.05}_{-0.08}$ & $1.14^{\,+0.01}_{-0.01}$ & $0.95^{\,+0.01}_{-0.01}$ \\
tau Boo b & 2 & 1 & $3.31^{\,+0.00}_{-0.00}$ & $0.01^{\,+0.01}_{-0.01}$ & $471.73^{\,+2.97}_{-2.97}$ & $1.32^{\,+0.24}_{-0.18}$ \\
tau Cet e & 1 & 4 & $162.87^{\,+1.08}_{-0.46}$ & $0.18^{\,+0.18}_{-0.14}$ & $0.55^{\,+0.13}_{-0.09}$ & $0.78^{\,+0.01}_{-0.01}$ \\
tau Cet g & 1 & 4 & $20.00^{\,+0.02}_{-0.01}$ & $0.06^{\,+0.13}_{-0.06}$ & $0.49^{\,+0.07}_{-0.11}$ & $0.78^{\,+0.01}_{-0.01}$ \\
ups And b & 2 & 3 & $4.62^{\,+0.00}_{-0.00}$ & $0.02^{\,+0.00}_{-0.00}$ & $70.51^{\,+0.45}_{-0.45}$ & $1.30$ \\
ups And c & 2 & 3 & $241.26^{\,+0.06}_{-0.06}$ & $0.26^{\,+0.01}_{-0.01}$ & $68.14^{\,+0.45}_{-0.45}$ & $1.30$ \\
ups And d & 2 & 3 & $1276.46^{\,+0.57}_{-0.57}$ & $0.30^{\,+0.01}_{-0.01}$ & $56.26^{\,+0.52}_{-0.52}$ & $1.30$ \\
\bottomrule
\end{tabular}
\end{table*}